\documentclass[aps,prl,reprint,groupaddress]{revtex4-2}

\usepackage{amsmath}
\usepackage{amsfonts}
\usepackage{mathtools}

\usepackage[colorlinks=true, pdfstartview=FitV, linkcolor=blue,
citecolor=blue, urlcolor=blue]{hyperref}
\usepackage[capitalise]{cleveref}

\usepackage[resetlabels,labeled]{multibib}
\newcites{S}{SM}

\usepackage{siunitx}
\usepackage{physics} 

\usepackage{graphicx}

\usepackage{xcolor}

\usepackage[T1]{fontenc}

\newcommand{\vect}[1]{\boldsymbol{#1}}

\graphicspath{{./figures/}}

\begin{document}

\title{Valley-Free Silicon Fins Caused by Shear Strain}

	\author{Christoph Adelsberger}
	\author{Stefano Bosco}
	\author{Jelena Klinovaja}
	\author{Daniel Loss}
	\affiliation{Department of Physics, University of Basel, Klingelbergstrasse 82, CH-4056 Basel, Switzerland}

	\date{\today}
	
\begin{abstract}
Electron spins confined in silicon quantum dots are promising candidates for large-scale quantum computers. 
However, the degeneracy of the conduction band of bulk silicon introduces additional levels dangerously close to the window of computational energies, where the quantum information can leak. The energy of the valley states---typically $\SI{0.1}{\milli\electronvolt}$---depends on hardly controllable atomistic disorder and still constitutes a fundamental limit to the scalability of these architectures. 
In this work, we introduce designs of complementary metal-oxide-semiconductor (CMOS)-compatible silicon fin field-effect transistors that enhance the energy gap to noncomputational states by more than one order of magnitude. Our devices comprise realistic silicon-germanium nanostructures with a large shear strain, where troublesome valley degrees of freedom are completely removed. The energy of noncomputational states is therefore not affected by unavoidable atomistic disorder and can further be tuned \textit{in situ} by applied electric fields.  Our design ideas are directly applicable to a variety of setups and will offer a blueprint toward silicon-based large-scale quantum processors.
\end{abstract}
	
	\maketitle
	
\paragraph{Introduction.}
Spins in silicon and germanium quantum dots (QDs) are workhorses of modern semiconductor-based quantum technology~\cite{Geyer2024,Camenzind2022,Hendrickx2021,Piot2022,Jirovec2021,Jirovec2022,Madzik2022,Petit2022,Philips2022,Unseld2023,Denisov2022,Takeda2020}. 
The most advanced platforms to date utilize planar heterostructures comprising Si and SiGe alloys, where quantum information is carried by single electrons confined in the Si layer~\cite{Unseld2023,Denisov2022,Takeda2020,Philips2022,Xue2022}.
In these systems, long spin coherence is enabled by the weak spin-orbit interaction of the conduction band and by isotopically purifying Si~\cite{Itoh2014}. Electron spin resonance was harnessed to selectively control individual qubits~\cite{Veldhorst2014,Takeda2016,Yoneda2017} whereas tunable exchange interactions mediate fast, high-fidelity two-qubit gates~\cite{Veldhorst2015,Zajac2016,Watson2018,Huang2019,Xue2019,Sigillito2019,Petit2022}.
The versatility of these architectures permitted remote coupling of distant qubits via microwave cavities~\cite{Mi2018,Samkharadze2018,Bonsen2023} and spin shuttling~\cite{Li2018,Mills2019,Noiri2022,Seidler2022}, as well as entanglement of three spin states~\cite{Takeda2021}. Readout and two-qubit gate fidelities exceeding the error correction threshold~\cite{Mills2022a,Noiri2022a,Xue2022,Tanttu2023} and the recent demonstration of six-qubit quantum processors~\cite{Philips2022} constitute promising steps toward large-scale quantum processors.

Further progress in electron spin qubits in silicon is currently hindered by the valley degeneracy of the silicon conduction band. 
In planar Si/SiGe heterostructures tensile in-plane strain partially lifts the six-fold degeneracy of bulk silicon, pushing four valleys to higher energy; the ground state remains twofold degenerate~\cite{Ando1982,Zwanenburg2013,Ruskov2018}. The residual valleys introduce troublesome additional levels in the vicinity of the computational energies where the quantum information is processed. These states open the system to decoherence and relaxation channels and constitute a critical source of leakage~\cite{Borselli2011,Shi2011,Zajac2015,Hollmann2020,Chen2021,Scarlino2017,Mi2018a,Mi2017}.
The valley degeneracy can be lifted by strong electric fields, but the induced energy gap is relatively small $\SIrange{10}{100}{\micro\electronvolt}$ and dangerously close to the typical qubit energies $\sim \SI{10}{\micro\electronvolt}$. Because it strongly depends on atomistic details of the Si/SiGe interface, controlling this gap reliably and reproducibly  is challenging~\cite{Friesen2007,Chutia2008,Saraiva2009,Hosseinkhani2020,Hosseinkhani2021,Lima2023,Wuetz2022}. Moreover, in hot qubits, valley states can be thermally excited, hindering the scalability of quantum processors~\cite{Yang2020}. 
Larger valley splittings are reached by periodically altering the concentration of Ge in the well~\cite{McJunkin2022,Woods2023}. In metal-oxide-semiconductor (MOS) structures, splittings $\sim\SI{0.5}{\milli\electronvolt}$ are reached by tightly confined electrons at the interfaces between Si and SiO$_2$, but these values largely depend on interface disorder~\cite{Yang2013,Yang2020,Saraiva2021,Cifuentes2023}.

\begin{figure}[]
	\includegraphics[width=\columnwidth]{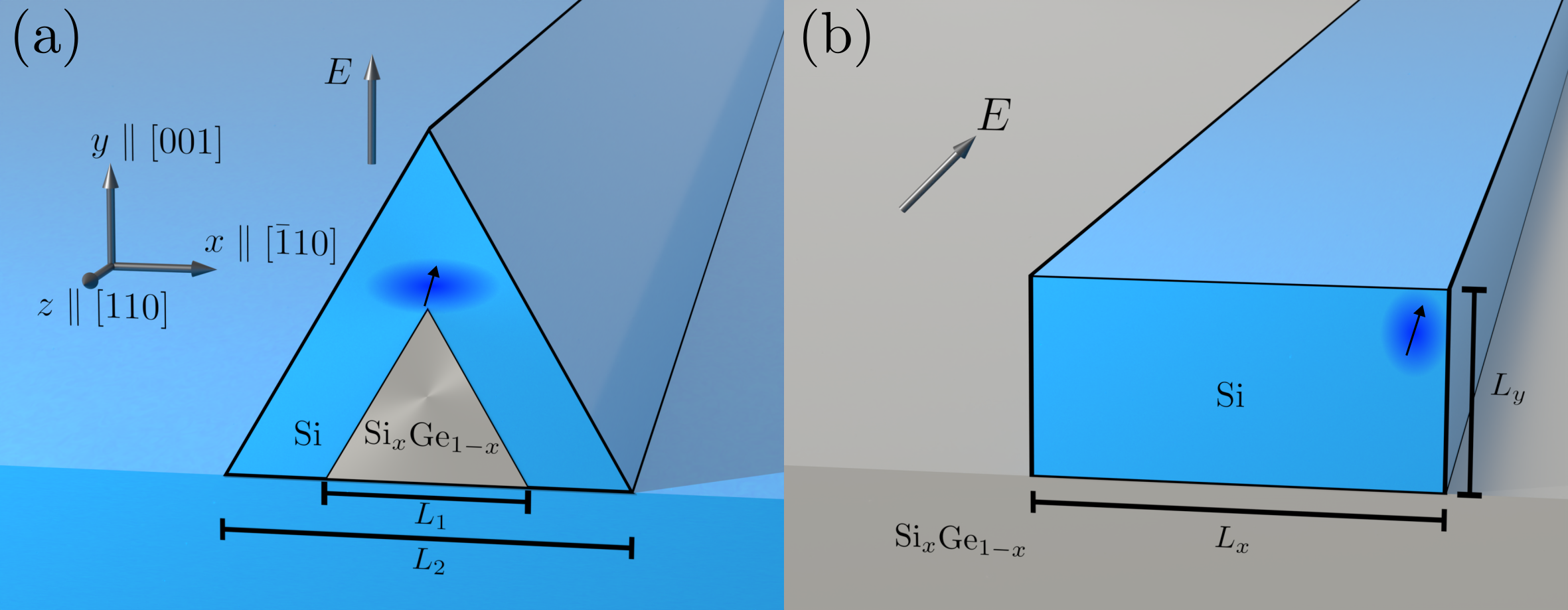}
	\caption{Design of valley-free fins in Si/Si$_x$Ge$_{1-x}$ heterostructures. (a) Equilateral triangular fin with inner and outer side lengths $L_1$ and $L_2$, respectively. The wave function is localized in the Si shell by uniaxial strain and the electric field $\vect{E}$. The fin is assumed to be grown on a Si substrate, but the results are similar for a Ge substrate. (b) Rectangular Si slab with side lengths $L_x$ and $L_y$ on Si$_x$Ge$_{1-x}$ substrate. The electrons are confined at the corners by $\vect{E}$ in the $x$-$y$ plane. The blue dots show the position of the QD hosting the spin qubit. We assume infinitely long systems in the $z$ direction. The coordinate system in (a) shows the crystallographic growth directions for both geometries.  \label{fig:schematic}}
\end{figure}

In this work, we propose alternative Si/SiGe nanostructures that completely lift the valley degeneracy and thus provide ideal platforms for future spin-based quantum processors.
In our designs the electrons are confined in quasi-one-dimensional (1D) Si fins compatible with CMOS technology~\cite{Choudhary2023},
see Fig.~\ref{fig:schematic}, where, in contrast to planar heterostructures, the SiGe induces a large \textit{shear} strain. The growth of similar fin structures as in Fig.~\ref{fig:schematic}(a)  has been demonstrated in Ref.~\cite{Ramanandan2024}. 
By detailed simulations based on continuum elasticity theory and microscopic $\vect{k}\cdot \vect{p}$ theory, we show that our engineered strain profile enables a nondegenerate ground state split from the excited states by energies $\sim \SIrange{1}{10}{\milli\electronvolt}$, 2 to 3 orders of magnitude larger than in current devices. Importantly, this energy gap remains large for realistic values of applied electric fields and is independent of atomistic disorder at the interfaces, rendering our design  robust in a wide variety of different fins. 
	
\paragraph{Theoretical model.}

The conduction band of bulk Si has six degenerate minima in the first Brillouin zone (BZ), which are located at distance $\pm k_0/2\pi = \pm 0.15/a_\text{Si}$ from the $X$ points (Si lattice constant $a_\text{Si}  = \SI{5.43}{\angstrom}$~\cite{Reeber1996}).
Its low-energy electronic states are described by the microscopic two-band $\vect{k}\cdot \vect{p}$ Hamiltonian~\cite{Hensel1965,Stanojevic2010}:
\begin{align}
	H = \frac{\hbar^2}{2 m_t} (k_{t_1}^2+k_{t_2}^2) + \frac{\hbar^2}{2 m_l} k_l^2 +\Xi_u \varepsilon_{ll} + e \vect{E}\cdot\vect{r} + V(\vect{r}) \nonumber\\
	+ \frac{\hbar^2}{m_l} k_0 k_l
\tau_x -\left(\frac{\hbar^2}{M}k_{t_1}k_{t_2} - 2\Xi_{u'} \varepsilon_{t_1t_2} \right) \tau_z\ , \label{eq:SiHam}
\end{align}
where $l$ is the longitudinal direction, and $t_1$ and $t_2$ are the transversal directions, which are aligned with the main crystallographic axes $[100]$, $[010]$, and $[001]$. The canonical momentum operators are $\hbar k_j = -i\hbar \partial_j$ ($j=l, t_1, t_2$).

This Hamiltonian is based on a small-momentum expansion of the band structure around the $X$ points and the Pauli matrices $\tau_i$ ($i=x,y,z$) refer to the two bands crossing there. Because there are three inequivalent $X$ points, the six valleys are described by three independent copies of Eq.~\eqref{eq:SiHam}.
The spin degree of freedom is neglected in $H$.
The transversal and longitudinal masses are $m_t = 0.19\, m_e$ and $m_l = 0.91\, m_e$, respectively, with the free electron mass $m_e$; $M\approx\left(m_t^{-1}-m_e^{-1}\right)^{-1}$ is the band-coupling mass~\cite{Hensel1965,Sverdlov2008}. 

The lattice constant of Ge $a_\text{Ge}=\SI{5.66}{\angstrom}$~\cite{Reeber1996} is larger than $a_\text{Si}$.  Thus, the Si is strained in Si/SiGe heterostructures. 
The uniaxial strain $\varepsilon_{ll}$ and the shear strain  $\varepsilon_{t_1t_2}$ modify the electron energy depending on the deformation potentials $\Xi_u = \SI{9}{\electronvolt}$~\cite{Fischetti1996,Walle1986,Tserbak1993,Friedel1989,Balslev1966,Rieger1993,Goroff1963} and $\Xi_{u'} = \SI{7}{\electronvolt}$~\cite{Hensel1965,Laude1971,Li2021}, respectively.
In the nanostructures sketched in Fig.~\ref{fig:schematic}, we simulate the strain tensor elements $\underline{\vect{\varepsilon}}$ by finite-element methods (FEM) based on continuum elasticity theory~\cite{Bosco2022,Kloeffel2014,Kosevich1986,Niquet2012}. A short review of linear elasticity theory and details on strain simulations is provided in the Supplemental Material (SM)~\cite{smStrain}. We assume that the lattice constant of an alloy of Si$_x$Ge$_{1-x}$ changes linearly from $a_\text{Si} $ to $a_\text{Ge}$, and thus we use the relation  
\begin{equation}
	\underline{\vect{\varepsilon}}_{\mathrm{Si/Si}_x\mathrm{Ge}_{1-x}}=(1-x)\underline{\vect{\varepsilon}}_\mathrm{Si/Ge} \ , \label{eq:linStrain}
\end{equation}
interpolating linearly from the minimal strain at $1-x=0$ to the maximal strain at $1-x=1$, which is a good approximation for the actual relation~\cite{Bosco2021,Terrazos2021,Rieger1993}.

Equation~\eqref{eq:SiHam} includes the homogeneous electric field $\vect{E}$ resulting in the electrostatic potential $-e\vect{E}\cdot \vect{r}$, with the electron charge $e>0$, $\vect{r}=(x,y,z)$, and the confinement potential $V(\vect{r})$. We model the sharp interface between Si and a Si$_x$Ge$_{1-x}$ alloy by using the steplike potential function 
 \begin{equation}
	V(\vect{r}) = \begin{cases}
		0 &\text{ for } \vect{r}\in R_\text{Si},\\
		(1-x)\,\SI{500}{\milli\electronvolt} & \text{ for } \vect{r} \in R_\text{Si$_x$Ge$_{1-x}$},
	\end{cases} \ ,
\end{equation}
where $R_\text{Si}$ $(R_\text{Si$_x$Ge$_{1-x}$})$ indicates the region in the cross section occupied by Si (Si$_x$Ge$_{1-x}$ ). In analogy to Eq.~\eqref{eq:linStrain}, we also assume that $V(\vect{r})$ decreases linearly from the maximal value of $\SI{500}{\milli\electronvolt} $ (band gap difference between Si and Ge) as $x$ increases. 

Because Si has an anisotropic dispersion relation, we emphasize electrons lying in the three different pairs of valley states generally experience different confinement potentials. To account for this effect, we fix the $z$ direction to be aligned to the fin and the $y$ direction to be perpendicular to the substrate, see Fig.~\ref{fig:schematic}. We also restrict ourselves to the analysis of Si $[001]$, with fins that are aligned to the $[110]$ crystallographic axis, i.e. $y\parallel [001]$ and $z\parallel[110]$. This is the standard orientation of current devices~\cite{Geyer2024,Camenzind2022,Maurand2016,Piot2022}.

\paragraph{Shear-strain-induced lifting of the valley degeneracy.}

The Hamiltonian in Eq.~\eqref{eq:SiHam} allows us to accurately analyze the physics of conduction-band electrons in the fins shown in Fig.~\ref{fig:schematic}.
We discretize $H$ for different cross sections with lattice spacings $a_x$ and $a_y$ and we analyze the dispersion relation of the lowest energy states. Importantly, we include the inhomogeneous strain tensor $\underline{\vect{\varepsilon}}$ simulated by FEM with COMSOL Multiphysics\,\textsuperscript{\tiny\textregistered}~\cite{COMSOL}; see SM~\cite{smStrain}.

\begin{figure}[h!]
	\includegraphics[width=0.9\columnwidth]{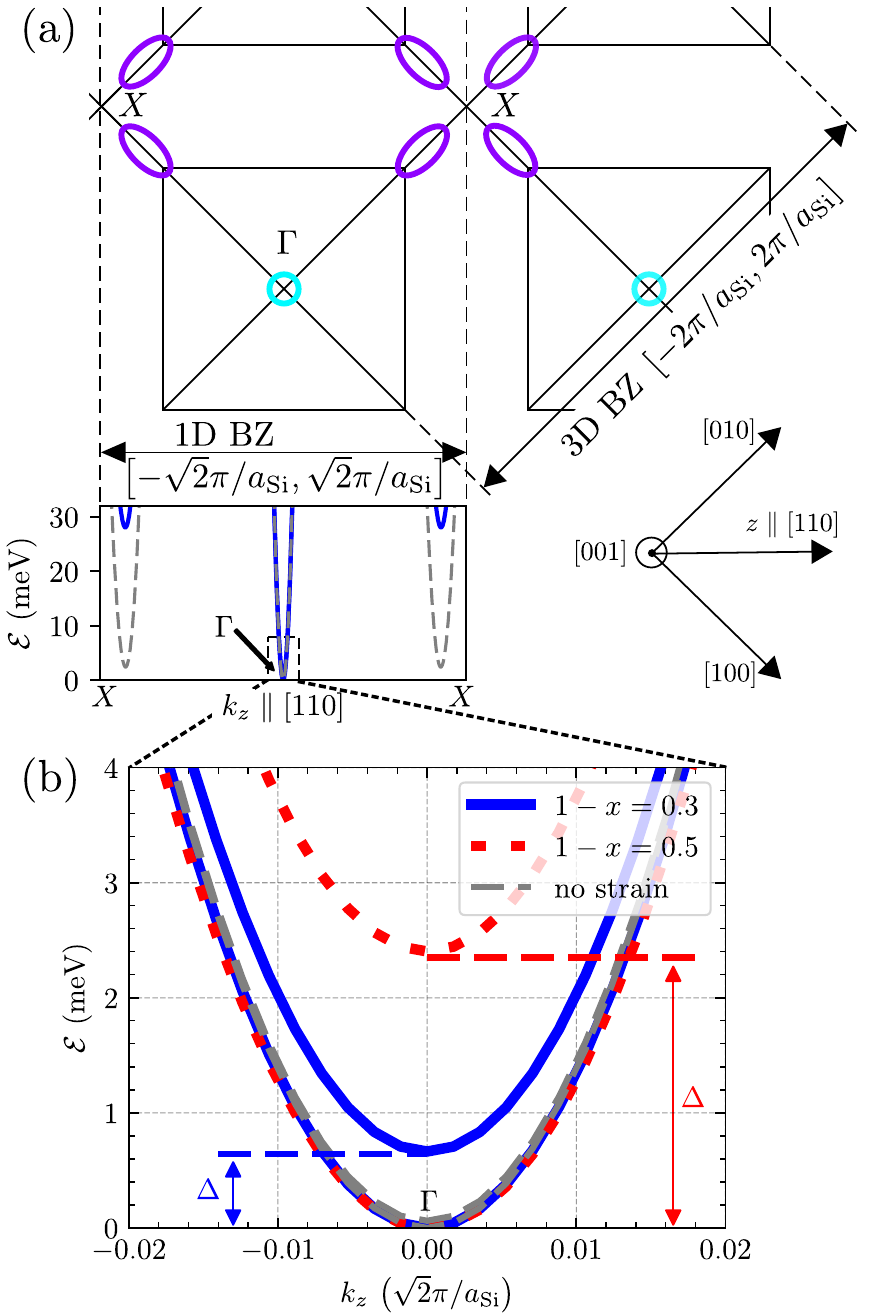}
	\caption{Band dispersion of electrons confined in strained Si fins. (a) Projection of the six valleys of the 3D BZ of bulk Si onto the 1D BZ with momentum $k_z\parallel[110]$. The purple ellipses indicate the four valleys in the $xz$ plane, while the turquoise circles indicate the two valleys near the $X$ points in the out-of-plane direction $y$. At the bottom, we show the dispersion relation $\mathcal{E} (k_z)$ of the lowest pair of sub-bands at the 1D $\Gamma$ point and close to the $X$ point of the triangular Si fin sketched in Fig.~\ref{fig:schematic}(a). Without strain (dashed gray lines) all the states are close in energy. 
		Including  moderate strain induced by a Si$_x$Ge$_{1-x}$ alloy with $x=0.7$ (solid blue lines), we observe that uniaxial strain $\varepsilon_{ll}$ pushes the states with minimum at finite $k_z$ several tens of meV away and shear strain $\varepsilon_{t_1t_2}$ induces a gap in the remaining two valleys in the $y$ direction (turquoise circles). 
		(b) Zoom into dispersion relation around the 1D $\Gamma$ point. When $\varepsilon_{t_1t_2}=0$, the two valleys are quasidegenerate. The degeneracy is lifted by shear strain induced by Si$_x$Ge$_{1-x}$. The resulting energy gap $\Delta\sim \SI{1}{\milli\electronvolt}$ increases with decreasing concentration of Si $x$, as shown by the blue and red lines obtained for $x= 0.7$ and $x=0.5$, respectively. The dispersion relation is qualitatively similar for the strained fin in Fig.~\ref{fig:schematic}(b). We used $L_1= \SI{9.5}{\nano\meter}$, $L_2 = \SI{19}{\nano\meter}$, $E_y=\SI{1}{\volt\per\micro\meter}$ (pointing along $[001]$), $a_x = \SI{0.32}{\nano\meter}$, and $a_y = \SI{0.28}{\nano\meter}$. \label{fig:dispRel}}
\end{figure}

The effect of strain in our fins is illustrated in Fig.~\ref{fig:dispRel}(a), where we show the projection of the three-dimensional (3D) bulk valleys in Si onto the 1D BZ along $z\parallel[110]$.
Along $z$, the four bulk Si valleys belonging to the $xz$ plane (purple ellipses) are projected close to the $X$ points and the two valleys along the  $y\parallel [001]$ axis  (turquoise circles) onto the $\Gamma$ point ($k_z=0$). 
When $\underline{\vect{\varepsilon}}=0$ (dashed gray lines) all valleys are close in energy up to a small contribution caused by the anisotropic confinement. In analogy to planar heterostructures~\cite{Stanojevic2010, Hong2008, Fischetti1996,Niquet2012} at finite values $\underline{\vect{\varepsilon}}$ (blue solid lines) uniaxial strain splits away the states with minimum close to the $X$ points by several tens of meV.

\begin{figure*}[ht]
	\includegraphics[width=\textwidth]{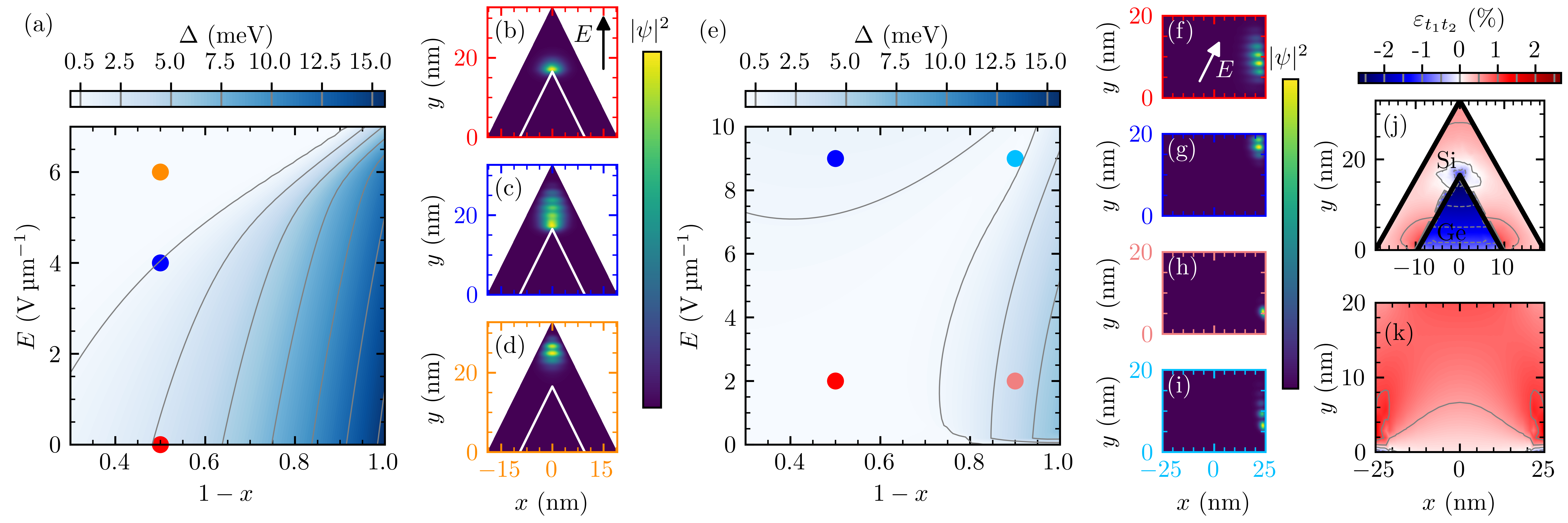}
	\caption{Valley splitting $\Delta$ in Si/Si$_x$Ge$_{1-x}$ fins:
		(a) $\Delta$ against Ge concentration $1-x$ and electric field $E$ in the triangular fin sketched in Fig.~\ref{fig:schematic}(a). For a wide range of experimentally relevant parameters $\Delta\gtrsim \SI{0.5}{\milli\electronvolt}$, substantially larger than in planar heterostructures; $\Delta$ is maximized when shear strain increases, i.e. at $1-x=1$, and at small values of $E$. The isocontours are defined in the colorbar.
		(b-d) Probability densities $\abs{\psi}^2$ of electron wave function at $1-x=0.5$ for (b) $E=0$, (c) $E=\SI{4}{\volt\per\micro\meter}$, and (d) $E=\SI{6}{\volt\per\micro\meter}$, marked in (a).
		The inhogomeneous uniaxial strain localizes the electron at the top of the fin. At $E=0$ the electron lies at the Si/SiGe interface, where shear strain is maximal, and resulting in the largest  $\Delta$. Increasing the electric field the electron is pushed toward the tip of the Si shell, where  shear strain is weaker and $\Delta$ decreases. We used $L_1= \SI{9.5}{\nano\meter}$ and $L_2 = \SI{19}{\nano\meter}$, as in Fig.~\ref{fig:dispRel}.
	The same quantities (e) $\Delta$, and (f-i) $\abs{\psi}^2$ in the rectangular fin sketched in Fig.~\ref{fig:schematic}(b).
		Note the different directions of $E$ in the two setups indicated by the arrows in (b) and (f). 
		Here shear strain is larger at the left and right sides of the cross section. At $1-x<0.75$ and for small $E$ the electron is localized at the bottom of the cross section (not shown). (f) At $1-x=0.5$ and $E=\SI{2}{\volt\per\micro\meter}$, the electron is weakly localized at the right of the fin, resulting in $\Delta<\SI{0.5}{\milli\electronvolt}$. (g) At $1-x=0.5$ and $E=\SI{9}{\volt\per\micro\meter}$, the electron is pushed toward the upper-right corner of the fin and the valley splitting increases to $\Delta>\SI{0.5}{\milli\electronvolt}$. (h) At $1-x=0.9$ and $E=\SI{2}{\volt\per\micro\meter}$, the larger concentration of Ge enhances locally the strain at the bottom corner of the rectangle.
		  (i) At $1-x=0.9$ and $E=\SI{9}{\volt\per\micro\meter}$, the large values of $E$ in a strongly strained device cause the wave function to spread out across the side. Then the situation is similar to (f) and $\Delta<\SI{0.5}{\milli\electronvolt}$. 
		We used $L_x= \SI{50}{\nano\meter}$, $L_y = \SI{20}{\nano\meter}$, $a_x = \SI{0.28}{\nano\meter}$, and $a_y = \SI{0.34}{\nano\meter}$. (j,k) Shear strain $\varepsilon_{t_1t_2}$ simulated with FEM for the two devices for pure Ge instead of SiGe. (j) $\varepsilon_{t_1t_2}$ is large above the tip of the inner Ge fin and becomes first weaker toward the tip of the Si shell and then increases with opposite sign. (k) Large $\varepsilon_{t_1t_2}$ is found close to the left and right side of the Si slab. In case of $1-x=0.3$ the range of the color bar for (j,k) is $[-0.82\,\%, 0.82\,\%]$. In the SM~\cite{smStrain}, we report cuts through (a) and (e) at certain values of $1-x$ and misaligned electric fields as well as a larger version of (j) and (k) and the strain simulation parameters used for our simulations.\label{fig:NW_WF}}
	\end{figure*}

Shear strain results in the splitting of the remaining two valleys. A closeup into the dispersion relation in the vicinity of $\Gamma$, highlighting the shear-strain-induced valley splitting $\Delta$, is shown in Fig.~\ref{fig:dispRel}(b). We focus here on the triangular fin sketched in Fig.~\ref{fig:schematic}(a); however, the results discussed are  valid also for rectangular fins. In the absence of strain the shear strain element $\varepsilon_{t_1t_2}=0$ and the lowest two energy states (dashed gray line) are quasidegenerate~\cite{Huang2014,Yang2013}, see Eq.~\eqref{eq:SiHam}. The Si$_x$Ge$_{1-x}$ induces a finite $\varepsilon_{t_1t_2}$ in the Si shell which lifts the valley degeneracy. Details on the strain profile in our fins are provided in Fig.~\ref{fig:NW_WF} and the SM~\cite{smStrain}. Considering a moderate Ge concentration of  $1-x=0.3$, we estimate $\Delta =\SI{ 0.65}{\milli\electronvolt}$ for $E_y=\SI{1}{\volt\per\micro\meter}$ pointing along $[001]$, significantly larger than what is obtained in planar heterostructures~\cite{Borselli2011,Shi2011,Zajac2015,Hollmann2020,Chen2021,Scarlino2017,Mi2018a,Mi2017}. By increasing the Ge amount to $1-x = 0.5$, $\varepsilon_{t_1t_2}$ increases [see Eq.~\eqref{eq:linStrain}] and consequently a larger value of $\Delta = \SI{2.35}{\milli\electronvolt}$ is reached. The split states from the same sub-band at $k_z=0$ are strongly hybridized.

Because the electron is localized at the top of the fin, the substrate does not affect the values of $\Delta$. 
We emphasize in striking contrast to valley splittings arising  in planar heterostructures, our $\Delta$ arises from shear strain and therefore is reproducible and robust against atomistic disorder at the Si/SiGe interfaces~\cite{Zwanenburg2013,Boykin2004} and modifications of the cross section (see SM~\cite{smStrain}).

\paragraph{Electric-field-dependence of valley splitting.} 

In planar Si/SiGe structures, the valley splitting $\Delta$ strongly depends on the applied electric field $E$. We show that in our fins, $\Delta$ can also be tuned \textit{in situ} by $E$; however at large enough concentrations of Ge in the Si$_x$Ge$_{1-x}$ alloy, $\Delta$  remains large.

The dependence of $\Delta$ on $E$ and $1-x$ is analyzed in Fig.~\ref{fig:NW_WF}(a,e).
In the triangular fin sketched in  Fig.~\ref{fig:schematic}(a), a positive electric field tends to decrease $\Delta$. 
This trend can be understood by observing that $E$ shifts the wave function toward the upper tip of the Si shell [see Fig.~\ref{fig:NW_WF}(b-d)], where shear strain first decreases and then slightly increases with opposite sign, see Fig.~\ref{fig:NW_WF}(j). A detailed explanation of FEM simulations is provided in the SM~\cite{smStrain}. As the concentration of Ge increases, strain also increases, resulting in $\Delta\gtrsim\SI{15}{\milli\electronvolt}$ for a wide range of $E$.
Because $\Delta$ depends on shear strain, our results are robust against variations in the shape of the cross section, which we verify in the SM~\cite{smStrain}.

Large values of $\Delta$ also emerge for a wide range of parameters in the rectangular Si fins on a SiGe substrate sketched  in  Fig.~\ref{fig:schematic}(b). Similar Si nanostructures are current state-of-the-art for spin qubits~\cite{Maurand2016,Piot2022,Yu2023,GonzalezZalba2021} and can be adapted to our proposal by replacing the oxide substrate by SiGe.
In this device, we observe in Fig.~\ref{fig:NW_WF}(e-i) a nontrivial interplay of $1-x$ and $E$, which we relate to the position and spacial spread of the wave function in the cross section. In this case, we study the effect of an electric field $E$ pushing the electron wave function toward the upper-right corner of the fin; due to symmetry, the results are equivalent if the electric field pushes the electron toward the upper-left corner.
In particular, $\varepsilon_{t_1t_2}$ is maximal at the left and right bottom sides of the fin, see Fig.~\ref{fig:NW_WF}(k), and, thus, $\Delta$ is large when the electron is localized close to these areas. We should emphasize, however, that  when it is too close to the interface the electron risks leaking into the substrate. In addition, the QD is also easier to  control electrostatically when it is localized at an upper corner. 

For low concentrations of Ge ($1-x\lesssim 0.7$) electrons are localized at the upper corner of the cross section by a strong field $E$, and thus $\Delta$ increases with increasing $E$. For weak $E$ the wave function is spread over the right side and $\Delta$ is small. At larger values of $1-x\gtrsim 0.7$, the inhomogeneous uniaxial strain localizes the electrons close to the edges already at weak $E$, thus resulting in large values of $\Delta$. In this case, $\Delta$ is only weakly dependent on $E$, and it decreases with increasing $E$ because the electrons are pushed away from the substrate. Details on the electric field dependence are provided in the SM~\cite{smStrain}.

The large valley splitting due to shear strain has important consequences for spin qubits realized in gate-defined QDs in Si fins. The spin qubit lifetime in planar Si/SiGe heterostructures is strongly limited at spin-valley relaxation hot spots where the qubit Zeeman and valley splittings become comparable~\cite{Hosseinkhani2020,Hosseinkhani2021,Hollmann2020,Yang2013,Borjans2019}. These hot spots are naturally avoided in our devices because of the large difference between the typically small qubit Zeeman splitting of $\sim \SI{10}{\micro\electronvolt}$ and the valley splitting of $\sim \SIrange{1}{10}{\milli\electronvolt}$ we predict.

Finally, we note that in quantum dot devices, an additional shear strain contribution arises because of the oxide and gate stacking. The shear strain values are strongly device dependent but simulations in Si/Ge heterostructures estimate these terms to be approximately one order of magnitude smaller than the values predicted here~\cite{Asaad2020,AbadilloUriel2023}. As a result, we expect that gate-induced strain will only weakly renormalize the large valley splitting in our devices.

\paragraph{Conclusion.}

In this work we show that shear strain substantially enhances the valley splitting in Si/SiGe heterostructures. In realistic Si fins we predict valley splittings $\sim \SIrange{1}{10}{\milli\electronvolt}$, orders of magnitude larger than in current devices. 
We show that the amplitude of the gap can be engineered by varying the composition of the Si$_x$Ge$_{1-x}$ alloy and is controllable \textit{in situ} by electric fields. Importantly, due to the large valley splitting spin-valley relaxation hot spots are avoided naturally in our proposed Si fins.
Our designs are robust against variations of the fin shape and, in contrast to planar systems, not affected by atomistic disorder.
By removing a critical issue of current electron spin qubits in Si, our devices will push these architectures toward new coherence standards and pave the way toward large-scale semiconductor-based quantum processors.

\acknowledgements{
	We thank Dominik Zumb\"uhl for giving access to the license for COMSOL Multiphysics\,\textsuperscript{\tiny\textregistered} and Andreas V. Kuhlmann for useful comments. 
This work was supported as a part of NCCR SPIN, a National Centre of Competence (or Excellence) in Research, funded by the Swiss National Science Foundation (grant number 51NF40-180604).
}


\bibliographystyle{apsrev4-1}
\bibliography{LiteratureCQW}

\begin{thebibliography}{13}%
\makeatletter
\providecommand \@ifxundefined [1]{%
 \@ifx{#1\undefined}
}%
\providecommand \@ifnum [1]{%
 \ifnum #1\expandafter \@firstoftwo
 \else \expandafter \@secondoftwo
 \fi
}%
\providecommand \@ifx [1]{%
 \ifx #1\expandafter \@firstoftwo
 \else \expandafter \@secondoftwo
 \fi
}%
\providecommand \natexlab [1]{#1}%
\providecommand \enquote  [1]{``#1''}%
\providecommand \bibnamefont  [1]{#1}%
\providecommand \bibfnamefont [1]{#1}%
\providecommand \citenamefont [1]{#1}%
\providecommand \href@noop [0]{\@secondoftwo}%
\providecommand \href [0]{\begingroup \@sanitize@url \@href}%
\providecommand \@href[1]{\@@startlink{#1}\@@href}%
\providecommand \@@href[1]{\endgroup#1\@@endlink}%
\providecommand \@sanitize@url [0]{\catcode `\\12\catcode `\$12\catcode
  `\&12\catcode `\#12\catcode `\^12\catcode `\_12\catcode `\%12\relax}%
\providecommand \@@startlink[1]{}%
\providecommand \@@endlink[0]{}%
\providecommand \url  [0]{\begingroup\@sanitize@url \@url }%
\providecommand \@url [1]{\endgroup\@href {#1}{\urlprefix }}%
\providecommand \urlprefix  [0]{URL }%
\providecommand \Eprint [0]{\href }%
\providecommand \doibase [0]{http://dx.doi.org/}%
\providecommand \selectlanguage [0]{\@gobble}%
\providecommand \bibinfo  [0]{\@secondoftwo}%
\providecommand \bibfield  [0]{\@secondoftwo}%
\providecommand \translation [1]{[#1]}%
\providecommand \BibitemOpen [0]{}%
\providecommand \bibitemStop [0]{}%
\providecommand \bibitemNoStop [0]{.\EOS\space}%
\providecommand \EOS [0]{\spacefactor3000\relax}%
\providecommand \BibitemShut  [1]{\csname bibitem#1\endcsname}%
\let\auto@bib@innerbib\@empty
\bibitem [{\citenamefont {Kosevich}\ \emph {et~al.}(1986)\citenamefont
  {Kosevich}, \citenamefont {Lifshitz}, \citenamefont {Landau},\ and\
  \citenamefont {Pitaevskii}}]{Kosevich1986sm}%
  \BibitemOpen
  \bibfield  {author} {\bibinfo {author} {\bibfnamefont {A.~M.}\ \bibnamefont
  {Kosevich}}, \bibinfo {author} {\bibfnamefont {E.~M.}\ \bibnamefont
  {Lifshitz}}, \bibinfo {author} {\bibfnamefont {L.~D.}\ \bibnamefont
  {Landau}}, \ and\ \bibinfo {author} {\bibfnamefont {L.~P.}\ \bibnamefont
  {Pitaevskii}},\ }\href@noop {} {\emph {\bibinfo {title} {Theory of
  Elasticity}}},\ Vol.~\bibinfo {volume} {7}\ (\bibinfo  {publisher}
  {Butterworth-Heinemann},\ \bibinfo {year} {1986})\BibitemShut {NoStop}%
\bibitem [{\citenamefont {Mengistu}\ and\ \citenamefont
  {Garc{\'{\i}}a-Crist{\'{o}}bal}(2016)}]{Mengistu2016sm}%
  \BibitemOpen
  \bibfield  {author} {\bibinfo {author} {\bibfnamefont {H.~T.}\ \bibnamefont
  {Mengistu}}\ and\ \bibinfo {author} {\bibfnamefont {A.}~\bibnamefont
  {Garc{\'{\i}}a-Crist{\'{o}}bal}},\ }\href {\doibase
  10.1016/j.ijsolstr.2016.08.022} {\bibfield  {journal} {\bibinfo  {journal}
  {Int. J. Solids Struct.}\ }\textbf {\bibinfo {volume} {100-101}},\ \bibinfo
  {pages} {257} (\bibinfo {year} {2016})}\BibitemShut {NoStop}%
\bibitem [{\citenamefont {Kloeffel}\ \emph {et~al.}(2014)\citenamefont
  {Kloeffel}, \citenamefont {Trif},\ and\ \citenamefont
  {Loss}}]{Kloeffel2014sm}%
  \BibitemOpen
  \bibfield  {author} {\bibinfo {author} {\bibfnamefont {C.}~\bibnamefont
  {Kloeffel}}, \bibinfo {author} {\bibfnamefont {M.}~\bibnamefont {Trif}}, \
  and\ \bibinfo {author} {\bibfnamefont {D.}~\bibnamefont {Loss}},\ }\href
  {\doibase 10.1103/physrevb.90.115419} {\bibfield  {journal} {\bibinfo
  {journal} {Phys. Rev. B}\ }\textbf {\bibinfo {volume} {90}},\ \bibinfo
  {pages} {115419} (\bibinfo {year} {2014})}\BibitemShut {NoStop}%
\bibitem [{\citenamefont {McSkimin}\ and\ \citenamefont
  {Andreatch}(1964)}]{McSkimin1964sm}%
  \BibitemOpen
  \bibfield  {author} {\bibinfo {author} {\bibfnamefont {H.~J.}\ \bibnamefont
  {McSkimin}}\ and\ \bibinfo {author} {\bibfnamefont {P.}~\bibnamefont
  {Andreatch}},\ }\href {\doibase 10.1063/1.1702809} {\bibfield  {journal}
  {\bibinfo  {journal} {J. Appl. Phys.}\ }\textbf {\bibinfo {volume} {35}},\
  \bibinfo {pages} {2161} (\bibinfo {year} {1964})}\BibitemShut {NoStop}%
\bibitem [{\citenamefont {McSkimin}\ and\ \citenamefont
  {Andreatch}(1963)}]{McSkimin1963sm}%
  \BibitemOpen
  \bibfield  {author} {\bibinfo {author} {\bibfnamefont {H.~J.}\ \bibnamefont
  {McSkimin}}\ and\ \bibinfo {author} {\bibfnamefont {P.}~\bibnamefont
  {Andreatch}},\ }\href {\doibase 10.1063/1.1729323} {\bibfield  {journal}
  {\bibinfo  {journal} {J. Appl. Phys.}\ }\textbf {\bibinfo {volume} {34}},\
  \bibinfo {pages} {651} (\bibinfo {year} {1963})}\BibitemShut {NoStop}%
\bibitem [{COM()}]{COMSOLsm}%
  \BibitemOpen
  \href@noop {} {\enquote {\bibinfo {title} {{COMSOL
  Multiphysics{\textregistered} v. 6.1. \url{www.comsol.com}. COMSOL AB,
  Stockholm, Sweden.}}}\ }\BibitemShut {NoStop}%
\bibitem [{\citenamefont {Friesen}\ \emph {et~al.}(2007)\citenamefont
  {Friesen}, \citenamefont {Chutia}, \citenamefont {Tahan},\ and\ \citenamefont
  {Coppersmith}}]{Friesen2007sm}%
  \BibitemOpen
  \bibfield  {author} {\bibinfo {author} {\bibfnamefont {M.}~\bibnamefont
  {Friesen}}, \bibinfo {author} {\bibfnamefont {S.}~\bibnamefont {Chutia}},
  \bibinfo {author} {\bibfnamefont {C.}~\bibnamefont {Tahan}}, \ and\ \bibinfo
  {author} {\bibfnamefont {S.~N.}\ \bibnamefont {Coppersmith}},\ }\href
  {\doibase 10.1103/physrevb.75.115318} {\bibfield  {journal} {\bibinfo
  {journal} {Phys. Rev. B}\ }\textbf {\bibinfo {volume} {75}},\ \bibinfo
  {pages} {115318} (\bibinfo {year} {2007})}\BibitemShut {NoStop}%
\bibitem [{\citenamefont {Chutia}\ \emph {et~al.}(2008)\citenamefont {Chutia},
  \citenamefont {Coppersmith},\ and\ \citenamefont {Friesen}}]{Chutia2008sm}%
  \BibitemOpen
  \bibfield  {author} {\bibinfo {author} {\bibfnamefont {S.}~\bibnamefont
  {Chutia}}, \bibinfo {author} {\bibfnamefont {S.~N.}\ \bibnamefont
  {Coppersmith}}, \ and\ \bibinfo {author} {\bibfnamefont {M.}~\bibnamefont
  {Friesen}},\ }\href {\doibase 10.1103/physrevb.77.193311} {\bibfield
  {journal} {\bibinfo  {journal} {Phys. Rev. B}\ }\textbf {\bibinfo {volume}
  {77}},\ \bibinfo {pages} {193311} (\bibinfo {year} {2008})}\BibitemShut
  {NoStop}%
\bibitem [{\citenamefont {Saraiva}\ \emph {et~al.}(2009)\citenamefont
  {Saraiva}, \citenamefont {Calder{\'{o}}n}, \citenamefont {Hu}, \citenamefont
  {{Das Sarma}},\ and\ \citenamefont {Koiller}}]{Saraiva2009sm}%
  \BibitemOpen
  \bibfield  {author} {\bibinfo {author} {\bibfnamefont {A.~L.}\ \bibnamefont
  {Saraiva}}, \bibinfo {author} {\bibfnamefont {M.~J.}\ \bibnamefont
  {Calder{\'{o}}n}}, \bibinfo {author} {\bibfnamefont {X.}~\bibnamefont {Hu}},
  \bibinfo {author} {\bibfnamefont {S.}~\bibnamefont {{Das Sarma}}}, \ and\
  \bibinfo {author} {\bibfnamefont {B.}~\bibnamefont {Koiller}},\ }\href
  {\doibase 10.1103/physrevb.80.081305} {\bibfield  {journal} {\bibinfo
  {journal} {Phys. Rev. B}\ }\textbf {\bibinfo {volume} {80}},\ \bibinfo
  {pages} {081305(R)} (\bibinfo {year} {2009})}\BibitemShut {NoStop}%
\bibitem [{\citenamefont {Hosseinkhani}\ and\ \citenamefont
  {Burkard}(2020)}]{Hosseinkhani2020sm}%
  \BibitemOpen
  \bibfield  {author} {\bibinfo {author} {\bibfnamefont {A.}~\bibnamefont
  {Hosseinkhani}}\ and\ \bibinfo {author} {\bibfnamefont {G.}~\bibnamefont
  {Burkard}},\ }\href {\doibase 10.1103/physrevresearch.2.043180} {\bibfield
  {journal} {\bibinfo  {journal} {Phys. Rev. Res.}\ }\textbf {\bibinfo {volume}
  {2}},\ \bibinfo {pages} {043180} (\bibinfo {year} {2020})}\BibitemShut
  {NoStop}%
\bibitem [{\citenamefont {Hosseinkhani}\ and\ \citenamefont
  {Burkard}(2021)}]{Hosseinkhani2021sm}%
  \BibitemOpen
  \bibfield  {author} {\bibinfo {author} {\bibfnamefont {A.}~\bibnamefont
  {Hosseinkhani}}\ and\ \bibinfo {author} {\bibfnamefont {G.}~\bibnamefont
  {Burkard}},\ }\href {\doibase 10.1103/physrevb.104.085309} {\bibfield
  {journal} {\bibinfo  {journal} {Phys. Rev. B}\ }\textbf {\bibinfo {volume}
  {104}},\ \bibinfo {pages} {085309} (\bibinfo {year} {2021})}\BibitemShut
  {NoStop}%
\bibitem [{\citenamefont {Lima}\ and\ \citenamefont
  {Burkard}(2023)}]{Lima2023sm}%
  \BibitemOpen
  \bibfield  {author} {\bibinfo {author} {\bibfnamefont {J.~R.~F.}\
  \bibnamefont {Lima}}\ and\ \bibinfo {author} {\bibfnamefont {G.}~\bibnamefont
  {Burkard}},\ }\href {\doibase 10.1088/2633-4356/acd743} {\bibfield  {journal}
  {\bibinfo  {journal} {Mater. Quantum. Technol.}\ }\textbf {\bibinfo {volume}
  {3}},\ \bibinfo {pages} {025004} (\bibinfo {year} {2023})}\BibitemShut
  {NoStop}%
\bibitem [{\citenamefont {Wuetz}\ \emph {et~al.}(2022)\citenamefont {Wuetz},
  \citenamefont {Losert}, \citenamefont {Koelling}, \citenamefont {Stehouwer},
  \citenamefont {Zwerver}, \citenamefont {Philips}, \citenamefont
  {M{\k{a}}dzik}, \citenamefont {Xue}, \citenamefont {Zheng}, \citenamefont
  {Lodari}, \citenamefont {Amitonov}, \citenamefont {Samkharadze},
  \citenamefont {Sammak}, \citenamefont {Vandersypen}, \citenamefont {Rahman},
  \citenamefont {Coppersmith}, \citenamefont {Moutanabbir}, \citenamefont
  {Friesen},\ and\ \citenamefont {Scappucci}}]{Wuetz2022sm}%
  \BibitemOpen
  \bibfield  {author} {\bibinfo {author} {\bibfnamefont {B.~P.}\ \bibnamefont
  {Wuetz}}, \bibinfo {author} {\bibfnamefont {M.~P.}\ \bibnamefont {Losert}},
  \bibinfo {author} {\bibfnamefont {S.}~\bibnamefont {Koelling}}, \bibinfo
  {author} {\bibfnamefont {L.~E.~A.}\ \bibnamefont {Stehouwer}}, \bibinfo
  {author} {\bibfnamefont {A.-M.~J.}\ \bibnamefont {Zwerver}}, \bibinfo
  {author} {\bibfnamefont {S.~G.~J.}\ \bibnamefont {Philips}}, \bibinfo
  {author} {\bibfnamefont {M.~T.}\ \bibnamefont {M{\k{a}}dzik}}, \bibinfo
  {author} {\bibfnamefont {X.}~\bibnamefont {Xue}}, \bibinfo {author}
  {\bibfnamefont {G.}~\bibnamefont {Zheng}}, \bibinfo {author} {\bibfnamefont
  {M.}~\bibnamefont {Lodari}}, \bibinfo {author} {\bibfnamefont {S.~V.}\
  \bibnamefont {Amitonov}}, \bibinfo {author} {\bibfnamefont {N.}~\bibnamefont
  {Samkharadze}}, \bibinfo {author} {\bibfnamefont {A.}~\bibnamefont {Sammak}},
  \bibinfo {author} {\bibfnamefont {L.~M.~K.}\ \bibnamefont {Vandersypen}},
  \bibinfo {author} {\bibfnamefont {R.}~\bibnamefont {Rahman}}, \bibinfo
  {author} {\bibfnamefont {S.~N.}\ \bibnamefont {Coppersmith}}, \bibinfo
  {author} {\bibfnamefont {O.}~\bibnamefont {Moutanabbir}}, \bibinfo {author}
  {\bibfnamefont {M.}~\bibnamefont {Friesen}}, \ and\ \bibinfo {author}
  {\bibfnamefont {G.}~\bibnamefont {Scappucci}},\ }\href {\doibase
  10.1038/s41467-022-35458-0} {\bibfield  {journal} {\bibinfo  {journal} {Nat.
  Commun.}\ }\textbf {\bibinfo {volume} {13}},\ \bibinfo {pages} {7730}
  (\bibinfo {year} {2022})}\BibitemShut {NoStop}%
\end{thebibliography}%


\begin{thebibliography}{93}%
\makeatletter
\providecommand \@ifxundefined [1]{%
 \@ifx{#1\undefined}
}%
\providecommand \@ifnum [1]{%
 \ifnum #1\expandafter \@firstoftwo
 \else \expandafter \@secondoftwo
 \fi
}%
\providecommand \@ifx [1]{%
 \ifx #1\expandafter \@firstoftwo
 \else \expandafter \@secondoftwo
 \fi
}%
\providecommand \natexlab [1]{#1}%
\providecommand \enquote  [1]{``#1''}%
\providecommand \bibnamefont  [1]{#1}%
\providecommand \bibfnamefont [1]{#1}%
\providecommand \citenamefont [1]{#1}%
\providecommand \href@noop [0]{\@secondoftwo}%
\providecommand \href [0]{\begingroup \@sanitize@url \@href}%
\providecommand \@href[1]{\@@startlink{#1}\@@href}%
\providecommand \@@href[1]{\endgroup#1\@@endlink}%
\providecommand \@sanitize@url [0]{\catcode `\\12\catcode `\$12\catcode
  `\&12\catcode `\#12\catcode `\^12\catcode `\_12\catcode `\%12\relax}%
\providecommand \@@startlink[1]{}%
\providecommand \@@endlink[0]{}%
\providecommand \url  [0]{\begingroup\@sanitize@url \@url }%
\providecommand \@url [1]{\endgroup\@href {#1}{\urlprefix }}%
\providecommand \urlprefix  [0]{URL }%
\providecommand \Eprint [0]{\href }%
\providecommand \doibase [0]{http://dx.doi.org/}%
\providecommand \selectlanguage [0]{\@gobble}%
\providecommand \bibinfo  [0]{\@secondoftwo}%
\providecommand \bibfield  [0]{\@secondoftwo}%
\providecommand \translation [1]{[#1]}%
\providecommand \BibitemOpen [0]{}%
\providecommand \bibitemStop [0]{}%
\providecommand \bibitemNoStop [0]{.\EOS\space}%
\providecommand \EOS [0]{\spacefactor3000\relax}%
\providecommand \BibitemShut  [1]{\csname bibitem#1\endcsname}%
\let\auto@bib@innerbib\@empty
\bibitem [{\citenamefont {Geyer}\ \emph {et~al.}(2024)\citenamefont {Geyer},
  \citenamefont {Hetényi}, \citenamefont {Bosco}, \citenamefont {Camenzind},
  \citenamefont {Eggli}, \citenamefont {Fuhrer}, \citenamefont {Loss},
  \citenamefont {Warburton}, \citenamefont {Zumbühl},\ and\ \citenamefont
  {Kuhlmann}}]{Geyer2024}%
  \BibitemOpen
  \bibfield  {author} {\bibinfo {author} {\bibfnamefont {S.}~\bibnamefont
  {Geyer}}, \bibinfo {author} {\bibfnamefont {B.}~\bibnamefont {Hetényi}},
  \bibinfo {author} {\bibfnamefont {S.}~\bibnamefont {Bosco}}, \bibinfo
  {author} {\bibfnamefont {L.~C.}\ \bibnamefont {Camenzind}}, \bibinfo {author}
  {\bibfnamefont {R.~S.}\ \bibnamefont {Eggli}}, \bibinfo {author}
  {\bibfnamefont {A.}~\bibnamefont {Fuhrer}}, \bibinfo {author} {\bibfnamefont
  {D.}~\bibnamefont {Loss}}, \bibinfo {author} {\bibfnamefont {R.~J.}\
  \bibnamefont {Warburton}}, \bibinfo {author} {\bibfnamefont {D.~M.}\
  \bibnamefont {Zumbühl}}, \ and\ \bibinfo {author} {\bibfnamefont {A.~V.}\
  \bibnamefont {Kuhlmann}},\ }\href {\doibase 10.1038/s41567-024-02481-5}
  {\bibfield  {journal} {\bibinfo  {journal} {Nat. Phys.}\ } (\bibinfo {year}
  {2024}),\ 10.1038/s41567-024-02481-5}\BibitemShut {NoStop}%
\bibitem [{\citenamefont {Camenzind}\ \emph {et~al.}(2022)\citenamefont
  {Camenzind}, \citenamefont {Geyer}, \citenamefont {Fuhrer}, \citenamefont
  {Warburton}, \citenamefont {Zumbühl},\ and\ \citenamefont
  {Kuhlmann}}]{Camenzind2022}%
  \BibitemOpen
  \bibfield  {author} {\bibinfo {author} {\bibfnamefont {L.~C.}\ \bibnamefont
  {Camenzind}}, \bibinfo {author} {\bibfnamefont {S.}~\bibnamefont {Geyer}},
  \bibinfo {author} {\bibfnamefont {A.}~\bibnamefont {Fuhrer}}, \bibinfo
  {author} {\bibfnamefont {R.~J.}\ \bibnamefont {Warburton}}, \bibinfo {author}
  {\bibfnamefont {D.~M.}\ \bibnamefont {Zumbühl}}, \ and\ \bibinfo {author}
  {\bibfnamefont {A.~V.}\ \bibnamefont {Kuhlmann}},\ }\href {\doibase
  10.1038/s41928-022-00722-0} {\bibfield  {journal} {\bibinfo  {journal} {Nat.
  Electron.}\ }\textbf {\bibinfo {volume} {5}},\ \bibinfo {pages} {178}
  (\bibinfo {year} {2022})}\BibitemShut {NoStop}%
\bibitem [{\citenamefont {Hendrickx}\ \emph {et~al.}(2021)\citenamefont
  {Hendrickx}, \citenamefont {Lawrie}, \citenamefont {Russ}, \citenamefont {van
  Riggelen}, \citenamefont {de~Snoo}, \citenamefont {Schouten}, \citenamefont
  {Sammak}, \citenamefont {Scappucci},\ and\ \citenamefont
  {Veldhorst}}]{Hendrickx2021}%
  \BibitemOpen
  \bibfield  {author} {\bibinfo {author} {\bibfnamefont {N.~W.}\ \bibnamefont
  {Hendrickx}}, \bibinfo {author} {\bibfnamefont {W.~I.~L.}\ \bibnamefont
  {Lawrie}}, \bibinfo {author} {\bibfnamefont {M.}~\bibnamefont {Russ}},
  \bibinfo {author} {\bibfnamefont {F.}~\bibnamefont {van Riggelen}}, \bibinfo
  {author} {\bibfnamefont {S.~L.}\ \bibnamefont {de~Snoo}}, \bibinfo {author}
  {\bibfnamefont {R.~N.}\ \bibnamefont {Schouten}}, \bibinfo {author}
  {\bibfnamefont {A.}~\bibnamefont {Sammak}}, \bibinfo {author} {\bibfnamefont
  {G.}~\bibnamefont {Scappucci}}, \ and\ \bibinfo {author} {\bibfnamefont
  {M.}~\bibnamefont {Veldhorst}},\ }\href {\doibase 10.1038/s41586-021-03332-6}
  {\bibfield  {journal} {\bibinfo  {journal} {Nature}\ }\textbf {\bibinfo
  {volume} {591}},\ \bibinfo {pages} {580} (\bibinfo {year}
  {2021})}\BibitemShut {NoStop}%
\bibitem [{\citenamefont {Piot}\ \emph {et~al.}(2022)\citenamefont {Piot},
  \citenamefont {Brun}, \citenamefont {Schmitt}, \citenamefont {Zihlmann},
  \citenamefont {Michal}, \citenamefont {Apra}, \citenamefont {Abadillo-Uriel},
  \citenamefont {Jehl}, \citenamefont {Bertrand}, \citenamefont {Niebojewski},
  \citenamefont {Hutin}, \citenamefont {Vinet}, \citenamefont {Urdampilleta},
  \citenamefont {Meunier}, \citenamefont {Niquet}, \citenamefont {Maurand},\
  and\ \citenamefont {Franceschi}}]{Piot2022}%
  \BibitemOpen
  \bibfield  {author} {\bibinfo {author} {\bibfnamefont {N.}~\bibnamefont
  {Piot}}, \bibinfo {author} {\bibfnamefont {B.}~\bibnamefont {Brun}}, \bibinfo
  {author} {\bibfnamefont {V.}~\bibnamefont {Schmitt}}, \bibinfo {author}
  {\bibfnamefont {S.}~\bibnamefont {Zihlmann}}, \bibinfo {author}
  {\bibfnamefont {V.~P.}\ \bibnamefont {Michal}}, \bibinfo {author}
  {\bibfnamefont {A.}~\bibnamefont {Apra}}, \bibinfo {author} {\bibfnamefont
  {J.~C.}\ \bibnamefont {Abadillo-Uriel}}, \bibinfo {author} {\bibfnamefont
  {X.}~\bibnamefont {Jehl}}, \bibinfo {author} {\bibfnamefont {B.}~\bibnamefont
  {Bertrand}}, \bibinfo {author} {\bibfnamefont {H.}~\bibnamefont
  {Niebojewski}}, \bibinfo {author} {\bibfnamefont {L.}~\bibnamefont {Hutin}},
  \bibinfo {author} {\bibfnamefont {M.}~\bibnamefont {Vinet}}, \bibinfo
  {author} {\bibfnamefont {M.}~\bibnamefont {Urdampilleta}}, \bibinfo {author}
  {\bibfnamefont {T.}~\bibnamefont {Meunier}}, \bibinfo {author} {\bibfnamefont
  {Y.-M.}\ \bibnamefont {Niquet}}, \bibinfo {author} {\bibfnamefont
  {R.}~\bibnamefont {Maurand}}, \ and\ \bibinfo {author} {\bibfnamefont
  {S.~D.}\ \bibnamefont {Franceschi}},\ }\href {\doibase
  10.1038/s41565-022-01196-z} {\bibfield  {journal} {\bibinfo  {journal} {Nat.
  Nanotechnol.}\ }\textbf {\bibinfo {volume} {17}},\ \bibinfo {pages} {1072}
  (\bibinfo {year} {2022})}\BibitemShut {NoStop}%
\bibitem [{\citenamefont {Jirovec}\ \emph {et~al.}(2021)\citenamefont
  {Jirovec}, \citenamefont {Hofmann}, \citenamefont {Ballabio}, \citenamefont
  {Mutter}, \citenamefont {Tavani}, \citenamefont {Botifoll}, \citenamefont
  {Crippa}, \citenamefont {Kukucka}, \citenamefont {Sagi}, \citenamefont
  {Martins}, \citenamefont {Saez-Mollejo}, \citenamefont {Prieto},
  \citenamefont {Borovkov}, \citenamefont {Arbiol}, \citenamefont {Chrastina},
  \citenamefont {Isella},\ and\ \citenamefont {Katsaros}}]{Jirovec2021}%
  \BibitemOpen
  \bibfield  {author} {\bibinfo {author} {\bibfnamefont {D.}~\bibnamefont
  {Jirovec}}, \bibinfo {author} {\bibfnamefont {A.}~\bibnamefont {Hofmann}},
  \bibinfo {author} {\bibfnamefont {A.}~\bibnamefont {Ballabio}}, \bibinfo
  {author} {\bibfnamefont {P.~M.}\ \bibnamefont {Mutter}}, \bibinfo {author}
  {\bibfnamefont {G.}~\bibnamefont {Tavani}}, \bibinfo {author} {\bibfnamefont
  {M.}~\bibnamefont {Botifoll}}, \bibinfo {author} {\bibfnamefont
  {A.}~\bibnamefont {Crippa}}, \bibinfo {author} {\bibfnamefont
  {J.}~\bibnamefont {Kukucka}}, \bibinfo {author} {\bibfnamefont
  {O.}~\bibnamefont {Sagi}}, \bibinfo {author} {\bibfnamefont {F.}~\bibnamefont
  {Martins}}, \bibinfo {author} {\bibfnamefont {J.}~\bibnamefont
  {Saez-Mollejo}}, \bibinfo {author} {\bibfnamefont {I.}~\bibnamefont
  {Prieto}}, \bibinfo {author} {\bibfnamefont {M.}~\bibnamefont {Borovkov}},
  \bibinfo {author} {\bibfnamefont {J.}~\bibnamefont {Arbiol}}, \bibinfo
  {author} {\bibfnamefont {D.}~\bibnamefont {Chrastina}}, \bibinfo {author}
  {\bibfnamefont {G.}~\bibnamefont {Isella}}, \ and\ \bibinfo {author}
  {\bibfnamefont {G.}~\bibnamefont {Katsaros}},\ }\href {\doibase
  10.1038/s41563-021-01022-2} {\bibfield  {journal} {\bibinfo  {journal} {Nat.
  Mater.}\ }\textbf {\bibinfo {volume} {20}},\ \bibinfo {pages} {1106}
  (\bibinfo {year} {2021})}\BibitemShut {NoStop}%
\bibitem [{\citenamefont {Jirovec}\ \emph {et~al.}(2022)\citenamefont
  {Jirovec}, \citenamefont {Mutter}, \citenamefont {Hofmann}, \citenamefont
  {Crippa}, \citenamefont {Rychetsky}, \citenamefont {Craig}, \citenamefont
  {Kukucka}, \citenamefont {Martins}, \citenamefont {Ballabio}, \citenamefont
  {Ares}, \citenamefont {Chrastina}, \citenamefont {Isella}, \citenamefont
  {Burkard},\ and\ \citenamefont {Katsaros}}]{Jirovec2022}%
  \BibitemOpen
  \bibfield  {author} {\bibinfo {author} {\bibfnamefont {D.}~\bibnamefont
  {Jirovec}}, \bibinfo {author} {\bibfnamefont {P.~M.}\ \bibnamefont {Mutter}},
  \bibinfo {author} {\bibfnamefont {A.}~\bibnamefont {Hofmann}}, \bibinfo
  {author} {\bibfnamefont {A.}~\bibnamefont {Crippa}}, \bibinfo {author}
  {\bibfnamefont {M.}~\bibnamefont {Rychetsky}}, \bibinfo {author}
  {\bibfnamefont {D.~L.}\ \bibnamefont {Craig}}, \bibinfo {author}
  {\bibfnamefont {J.}~\bibnamefont {Kukucka}}, \bibinfo {author} {\bibfnamefont
  {F.}~\bibnamefont {Martins}}, \bibinfo {author} {\bibfnamefont
  {A.}~\bibnamefont {Ballabio}}, \bibinfo {author} {\bibfnamefont
  {N.}~\bibnamefont {Ares}}, \bibinfo {author} {\bibfnamefont {D.}~\bibnamefont
  {Chrastina}}, \bibinfo {author} {\bibfnamefont {G.}~\bibnamefont {Isella}},
  \bibinfo {author} {\bibfnamefont {G.}~\bibnamefont {Burkard}}, \ and\
  \bibinfo {author} {\bibfnamefont {G.}~\bibnamefont {Katsaros}},\ }\href
  {\doibase 10.1103/physrevlett.128.126803} {\bibfield  {journal} {\bibinfo
  {journal} {Phys. Rev. Lett.}\ }\textbf {\bibinfo {volume} {128}},\ \bibinfo
  {pages} {126803} (\bibinfo {year} {2022})}\BibitemShut {NoStop}%
\bibitem [{\citenamefont {M{\k{a}}dzik}\ \emph {et~al.}(2022)\citenamefont
  {M{\k{a}}dzik}, \citenamefont {Asaad}, \citenamefont {Youssry}, \citenamefont
  {Joecker}, \citenamefont {Rudinger}, \citenamefont {Nielsen}, \citenamefont
  {Young}, \citenamefont {Proctor}, \citenamefont {Baczewski}, \citenamefont
  {Laucht}, \citenamefont {Schmitt}, \citenamefont {Hudson}, \citenamefont
  {Itoh}, \citenamefont {Jakob}, \citenamefont {Johnson}, \citenamefont
  {Jamieson}, \citenamefont {Dzurak}, \citenamefont {Ferrie}, \citenamefont
  {Blume-Kohout},\ and\ \citenamefont {Morello}}]{Madzik2022}%
  \BibitemOpen
  \bibfield  {author} {\bibinfo {author} {\bibfnamefont {M.~T.}\ \bibnamefont
  {M{\k{a}}dzik}}, \bibinfo {author} {\bibfnamefont {S.}~\bibnamefont {Asaad}},
  \bibinfo {author} {\bibfnamefont {A.}~\bibnamefont {Youssry}}, \bibinfo
  {author} {\bibfnamefont {B.}~\bibnamefont {Joecker}}, \bibinfo {author}
  {\bibfnamefont {K.~M.}\ \bibnamefont {Rudinger}}, \bibinfo {author}
  {\bibfnamefont {E.}~\bibnamefont {Nielsen}}, \bibinfo {author} {\bibfnamefont
  {K.~C.}\ \bibnamefont {Young}}, \bibinfo {author} {\bibfnamefont {T.~J.}\
  \bibnamefont {Proctor}}, \bibinfo {author} {\bibfnamefont {A.~D.}\
  \bibnamefont {Baczewski}}, \bibinfo {author} {\bibfnamefont {A.}~\bibnamefont
  {Laucht}}, \bibinfo {author} {\bibfnamefont {V.}~\bibnamefont {Schmitt}},
  \bibinfo {author} {\bibfnamefont {F.~E.}\ \bibnamefont {Hudson}}, \bibinfo
  {author} {\bibfnamefont {K.~M.}\ \bibnamefont {Itoh}}, \bibinfo {author}
  {\bibfnamefont {A.~M.}\ \bibnamefont {Jakob}}, \bibinfo {author}
  {\bibfnamefont {B.~C.}\ \bibnamefont {Johnson}}, \bibinfo {author}
  {\bibfnamefont {D.~N.}\ \bibnamefont {Jamieson}}, \bibinfo {author}
  {\bibfnamefont {A.~S.}\ \bibnamefont {Dzurak}}, \bibinfo {author}
  {\bibfnamefont {C.}~\bibnamefont {Ferrie}}, \bibinfo {author} {\bibfnamefont
  {R.}~\bibnamefont {Blume-Kohout}}, \ and\ \bibinfo {author} {\bibfnamefont
  {A.}~\bibnamefont {Morello}},\ }\href {\doibase 10.1038/s41586-021-04292-7}
  {\bibfield  {journal} {\bibinfo  {journal} {Nature}\ }\textbf {\bibinfo
  {volume} {601}},\ \bibinfo {pages} {348} (\bibinfo {year}
  {2022})}\BibitemShut {NoStop}%
\bibitem [{\citenamefont {Petit}\ \emph {et~al.}(2022)\citenamefont {Petit},
  \citenamefont {Russ}, \citenamefont {Eenink}, \citenamefont {Lawrie},
  \citenamefont {Clarke}, \citenamefont {Vandersypen},\ and\ \citenamefont
  {Veldhorst}}]{Petit2022}%
  \BibitemOpen
  \bibfield  {author} {\bibinfo {author} {\bibfnamefont {L.}~\bibnamefont
  {Petit}}, \bibinfo {author} {\bibfnamefont {M.}~\bibnamefont {Russ}},
  \bibinfo {author} {\bibfnamefont {G.~H. G.~J.}\ \bibnamefont {Eenink}},
  \bibinfo {author} {\bibfnamefont {W.~I.~L.}\ \bibnamefont {Lawrie}}, \bibinfo
  {author} {\bibfnamefont {J.~S.}\ \bibnamefont {Clarke}}, \bibinfo {author}
  {\bibfnamefont {L.~M.~K.}\ \bibnamefont {Vandersypen}}, \ and\ \bibinfo
  {author} {\bibfnamefont {M.}~\bibnamefont {Veldhorst}},\ }\href {\doibase
  10.1038/s43246-022-00304-9} {\bibfield  {journal} {\bibinfo  {journal}
  {Commun. Mater.}\ }\textbf {\bibinfo {volume} {3}},\ \bibinfo {pages} {82}
  (\bibinfo {year} {2022})}\BibitemShut {NoStop}%
\bibitem [{\citenamefont {Philips}\ \emph {et~al.}(2022)\citenamefont
  {Philips}, \citenamefont {M\k{a}dzik}, \citenamefont {Amitonov},
  \citenamefont {de~Snoo}, \citenamefont {Russ}, \citenamefont {Kalhor},
  \citenamefont {Volk}, \citenamefont {Lawrie}, \citenamefont {Brousse},
  \citenamefont {Tryputen}, \citenamefont {Wuetz}, \citenamefont {Sammak},
  \citenamefont {Veldhorst}, \citenamefont {Scappucci},\ and\ \citenamefont
  {Vandersypen}}]{Philips2022}%
  \BibitemOpen
  \bibfield  {author} {\bibinfo {author} {\bibfnamefont {S.~G.~J.}\
  \bibnamefont {Philips}}, \bibinfo {author} {\bibfnamefont {M.~T.}\
  \bibnamefont {M\k{a}dzik}}, \bibinfo {author} {\bibfnamefont {S.~V.}\
  \bibnamefont {Amitonov}}, \bibinfo {author} {\bibfnamefont {S.~L.}\
  \bibnamefont {de~Snoo}}, \bibinfo {author} {\bibfnamefont {M.}~\bibnamefont
  {Russ}}, \bibinfo {author} {\bibfnamefont {N.}~\bibnamefont {Kalhor}},
  \bibinfo {author} {\bibfnamefont {C.}~\bibnamefont {Volk}}, \bibinfo {author}
  {\bibfnamefont {W.~I.~L.}\ \bibnamefont {Lawrie}}, \bibinfo {author}
  {\bibfnamefont {D.}~\bibnamefont {Brousse}}, \bibinfo {author} {\bibfnamefont
  {L.}~\bibnamefont {Tryputen}}, \bibinfo {author} {\bibfnamefont {B.~P.}\
  \bibnamefont {Wuetz}}, \bibinfo {author} {\bibfnamefont {A.}~\bibnamefont
  {Sammak}}, \bibinfo {author} {\bibfnamefont {M.}~\bibnamefont {Veldhorst}},
  \bibinfo {author} {\bibfnamefont {G.}~\bibnamefont {Scappucci}}, \ and\
  \bibinfo {author} {\bibfnamefont {L.~M.~K.}\ \bibnamefont {Vandersypen}},\
  }\href {\doibase 10.1038/s41586-022-05117-x} {\bibfield  {journal} {\bibinfo
  {journal} {Nature}\ }\textbf {\bibinfo {volume} {609}},\ \bibinfo {pages}
  {919} (\bibinfo {year} {2022})}\BibitemShut {NoStop}%
\bibitem [{\citenamefont {Unseld}\ \emph {et~al.}(2023)\citenamefont {Unseld},
  \citenamefont {Meyer}, \citenamefont {Mądzik}, \citenamefont {Borsoi},
  \citenamefont {de~Snoo}, \citenamefont {Amitonov}, \citenamefont {Sammak},
  \citenamefont {Scappucci}, \citenamefont {Veldhorst},\ and\ \citenamefont
  {Vandersypen}}]{Unseld2023}%
  \BibitemOpen
  \bibfield  {author} {\bibinfo {author} {\bibfnamefont {F.~K.}\ \bibnamefont
  {Unseld}}, \bibinfo {author} {\bibfnamefont {M.}~\bibnamefont {Meyer}},
  \bibinfo {author} {\bibfnamefont {M.~T.}\ \bibnamefont {Mądzik}}, \bibinfo
  {author} {\bibfnamefont {F.}~\bibnamefont {Borsoi}}, \bibinfo {author}
  {\bibfnamefont {S.~L.}\ \bibnamefont {de~Snoo}}, \bibinfo {author}
  {\bibfnamefont {S.~V.}\ \bibnamefont {Amitonov}}, \bibinfo {author}
  {\bibfnamefont {A.}~\bibnamefont {Sammak}}, \bibinfo {author} {\bibfnamefont
  {G.}~\bibnamefont {Scappucci}}, \bibinfo {author} {\bibfnamefont
  {M.}~\bibnamefont {Veldhorst}}, \ and\ \bibinfo {author} {\bibfnamefont
  {L.~M.~K.}\ \bibnamefont {Vandersypen}},\ }\href {\doibase 10.1063/5.0160847}
  {\bibfield  {journal} {\bibinfo  {journal} {Appl. Phys. Lett.}\ }\textbf
  {\bibinfo {volume} {123}},\ \bibinfo {pages} {084002} (\bibinfo {year}
  {2023})}\BibitemShut {NoStop}%
\bibitem [{\citenamefont {Denisov}\ \emph {et~al.}(2022)\citenamefont
  {Denisov}, \citenamefont {Oh}, \citenamefont {Fuchs}, \citenamefont {Mills},
  \citenamefont {Chen}, \citenamefont {Anderson}, \citenamefont {Gyure},
  \citenamefont {Barnard},\ and\ \citenamefont {Petta}}]{Denisov2022}%
  \BibitemOpen
  \bibfield  {author} {\bibinfo {author} {\bibfnamefont {A.~O.}\ \bibnamefont
  {Denisov}}, \bibinfo {author} {\bibfnamefont {S.~W.}\ \bibnamefont {Oh}},
  \bibinfo {author} {\bibfnamefont {G.}~\bibnamefont {Fuchs}}, \bibinfo
  {author} {\bibfnamefont {A.~R.}\ \bibnamefont {Mills}}, \bibinfo {author}
  {\bibfnamefont {P.}~\bibnamefont {Chen}}, \bibinfo {author} {\bibfnamefont
  {C.~R.}\ \bibnamefont {Anderson}}, \bibinfo {author} {\bibfnamefont {M.~F.}\
  \bibnamefont {Gyure}}, \bibinfo {author} {\bibfnamefont {A.~W.}\ \bibnamefont
  {Barnard}}, \ and\ \bibinfo {author} {\bibfnamefont {J.~R.}\ \bibnamefont
  {Petta}},\ }\href {\doibase 10.1021/acs.nanolett.2c01098} {\bibfield
  {journal} {\bibinfo  {journal} {Nano Lett.}\ }\textbf {\bibinfo {volume}
  {22}},\ \bibinfo {pages} {4807} (\bibinfo {year} {2022})}\BibitemShut
  {NoStop}%
\bibitem [{\citenamefont {Takeda}\ \emph {et~al.}(2020)\citenamefont {Takeda},
  \citenamefont {Noiri}, \citenamefont {Yoneda}, \citenamefont {Nakajima},\
  and\ \citenamefont {Tarucha}}]{Takeda2020}%
  \BibitemOpen
  \bibfield  {author} {\bibinfo {author} {\bibfnamefont {K.}~\bibnamefont
  {Takeda}}, \bibinfo {author} {\bibfnamefont {A.}~\bibnamefont {Noiri}},
  \bibinfo {author} {\bibfnamefont {J.}~\bibnamefont {Yoneda}}, \bibinfo
  {author} {\bibfnamefont {T.}~\bibnamefont {Nakajima}}, \ and\ \bibinfo
  {author} {\bibfnamefont {S.}~\bibnamefont {Tarucha}},\ }\href {\doibase
  10.1103/physrevlett.124.117701} {\bibfield  {journal} {\bibinfo  {journal}
  {Phys. Rev. Lett.}\ }\textbf {\bibinfo {volume} {124}},\ \bibinfo {pages}
  {117701} (\bibinfo {year} {2020})}\BibitemShut {NoStop}%
\bibitem [{\citenamefont {Xue}\ \emph {et~al.}(2022)\citenamefont {Xue},
  \citenamefont {Russ}, \citenamefont {Samkharadze}, \citenamefont {Undseth},
  \citenamefont {Sammak}, \citenamefont {Scappucci},\ and\ \citenamefont
  {Vandersypen}}]{Xue2022}%
  \BibitemOpen
  \bibfield  {author} {\bibinfo {author} {\bibfnamefont {X.}~\bibnamefont
  {Xue}}, \bibinfo {author} {\bibfnamefont {M.}~\bibnamefont {Russ}}, \bibinfo
  {author} {\bibfnamefont {N.}~\bibnamefont {Samkharadze}}, \bibinfo {author}
  {\bibfnamefont {B.}~\bibnamefont {Undseth}}, \bibinfo {author} {\bibfnamefont
  {A.}~\bibnamefont {Sammak}}, \bibinfo {author} {\bibfnamefont
  {G.}~\bibnamefont {Scappucci}}, \ and\ \bibinfo {author} {\bibfnamefont
  {L.~M.~K.}\ \bibnamefont {Vandersypen}},\ }\href {\doibase
  10.1038/s41586-021-04273-w} {\bibfield  {journal} {\bibinfo  {journal}
  {Nature}\ }\textbf {\bibinfo {volume} {601}},\ \bibinfo {pages} {343}
  (\bibinfo {year} {2022})}\BibitemShut {NoStop}%
\bibitem [{\citenamefont {Itoh}\ and\ \citenamefont
  {Watanabe}(2014)}]{Itoh2014}%
  \BibitemOpen
  \bibfield  {author} {\bibinfo {author} {\bibfnamefont {K.~M.}\ \bibnamefont
  {Itoh}}\ and\ \bibinfo {author} {\bibfnamefont {H.}~\bibnamefont
  {Watanabe}},\ }\href {\doibase 10.1557/mrc.2014.32} {\bibfield  {journal}
  {\bibinfo  {journal} {{MRS} Commun.}\ }\textbf {\bibinfo {volume} {4}},\
  \bibinfo {pages} {143} (\bibinfo {year} {2014})}\BibitemShut {NoStop}%
\bibitem [{\citenamefont {Veldhorst}\ \emph {et~al.}(2014)\citenamefont
  {Veldhorst}, \citenamefont {Hwang}, \citenamefont {Yang}, \citenamefont
  {Leenstra}, \citenamefont {de~Ronde}, \citenamefont {Dehollain},
  \citenamefont {Muhonen}, \citenamefont {Hudson}, \citenamefont {Itoh},
  \citenamefont {Morello},\ and\ \citenamefont {Dzurak}}]{Veldhorst2014}%
  \BibitemOpen
  \bibfield  {author} {\bibinfo {author} {\bibfnamefont {M.}~\bibnamefont
  {Veldhorst}}, \bibinfo {author} {\bibfnamefont {J.~C.~C.}\ \bibnamefont
  {Hwang}}, \bibinfo {author} {\bibfnamefont {C.~H.}\ \bibnamefont {Yang}},
  \bibinfo {author} {\bibfnamefont {A.~W.}\ \bibnamefont {Leenstra}}, \bibinfo
  {author} {\bibfnamefont {B.}~\bibnamefont {de~Ronde}}, \bibinfo {author}
  {\bibfnamefont {J.~P.}\ \bibnamefont {Dehollain}}, \bibinfo {author}
  {\bibfnamefont {J.~T.}\ \bibnamefont {Muhonen}}, \bibinfo {author}
  {\bibfnamefont {F.~E.}\ \bibnamefont {Hudson}}, \bibinfo {author}
  {\bibfnamefont {K.~M.}\ \bibnamefont {Itoh}}, \bibinfo {author}
  {\bibfnamefont {A.}~\bibnamefont {Morello}}, \ and\ \bibinfo {author}
  {\bibfnamefont {A.~S.}\ \bibnamefont {Dzurak}},\ }\href {\doibase
  10.1038/nnano.2014.216} {\bibfield  {journal} {\bibinfo  {journal} {Nat.
  Nanotechnol.}\ }\textbf {\bibinfo {volume} {9}},\ \bibinfo {pages} {981}
  (\bibinfo {year} {2014})}\BibitemShut {NoStop}%
\bibitem [{\citenamefont {Takeda}\ \emph {et~al.}(2016)\citenamefont {Takeda},
  \citenamefont {Kamioka}, \citenamefont {Otsuka}, \citenamefont {Yoneda},
  \citenamefont {Nakajima}, \citenamefont {Delbecq}, \citenamefont {Amaha},
  \citenamefont {Allison}, \citenamefont {Kodera}, \citenamefont {Oda},\ and\
  \citenamefont {Tarucha}}]{Takeda2016}%
  \BibitemOpen
  \bibfield  {author} {\bibinfo {author} {\bibfnamefont {K.}~\bibnamefont
  {Takeda}}, \bibinfo {author} {\bibfnamefont {J.}~\bibnamefont {Kamioka}},
  \bibinfo {author} {\bibfnamefont {T.}~\bibnamefont {Otsuka}}, \bibinfo
  {author} {\bibfnamefont {J.}~\bibnamefont {Yoneda}}, \bibinfo {author}
  {\bibfnamefont {T.}~\bibnamefont {Nakajima}}, \bibinfo {author}
  {\bibfnamefont {M.~R.}\ \bibnamefont {Delbecq}}, \bibinfo {author}
  {\bibfnamefont {S.}~\bibnamefont {Amaha}}, \bibinfo {author} {\bibfnamefont
  {G.}~\bibnamefont {Allison}}, \bibinfo {author} {\bibfnamefont
  {T.}~\bibnamefont {Kodera}}, \bibinfo {author} {\bibfnamefont
  {S.}~\bibnamefont {Oda}}, \ and\ \bibinfo {author} {\bibfnamefont
  {S.}~\bibnamefont {Tarucha}},\ }\href {\doibase 10.1126/sciadv.1600694}
  {\bibfield  {journal} {\bibinfo  {journal} {Sci. Adv.}\ }\textbf {\bibinfo
  {volume} {2}},\ \bibinfo {pages} {e1600694} (\bibinfo {year}
  {2016})}\BibitemShut {NoStop}%
\bibitem [{\citenamefont {Yoneda}\ \emph {et~al.}(2018)\citenamefont {Yoneda},
  \citenamefont {Takeda}, \citenamefont {Otsuka}, \citenamefont {Nakajima},
  \citenamefont {Delbecq}, \citenamefont {Allison}, \citenamefont {Honda},
  \citenamefont {Kodera}, \citenamefont {Oda}, \citenamefont {Hoshi},
  \citenamefont {Usami}, \citenamefont {Itoh},\ and\ \citenamefont
  {Tarucha}}]{Yoneda2017}%
  \BibitemOpen
  \bibfield  {author} {\bibinfo {author} {\bibfnamefont {J.}~\bibnamefont
  {Yoneda}}, \bibinfo {author} {\bibfnamefont {K.}~\bibnamefont {Takeda}},
  \bibinfo {author} {\bibfnamefont {T.}~\bibnamefont {Otsuka}}, \bibinfo
  {author} {\bibfnamefont {T.}~\bibnamefont {Nakajima}}, \bibinfo {author}
  {\bibfnamefont {M.~R.}\ \bibnamefont {Delbecq}}, \bibinfo {author}
  {\bibfnamefont {G.}~\bibnamefont {Allison}}, \bibinfo {author} {\bibfnamefont
  {T.}~\bibnamefont {Honda}}, \bibinfo {author} {\bibfnamefont
  {T.}~\bibnamefont {Kodera}}, \bibinfo {author} {\bibfnamefont
  {S.}~\bibnamefont {Oda}}, \bibinfo {author} {\bibfnamefont {Y.}~\bibnamefont
  {Hoshi}}, \bibinfo {author} {\bibfnamefont {N.}~\bibnamefont {Usami}},
  \bibinfo {author} {\bibfnamefont {K.~M.}\ \bibnamefont {Itoh}}, \ and\
  \bibinfo {author} {\bibfnamefont {S.}~\bibnamefont {Tarucha}},\ }\href
  {\doibase 10.1038/s41565-017-0014-x} {\bibfield  {journal} {\bibinfo
  {journal} {Nat. Nanotechnol.}\ }\textbf {\bibinfo {volume} {13}},\ \bibinfo
  {pages} {102} (\bibinfo {year} {2018})}\BibitemShut {NoStop}%
\bibitem [{\citenamefont {Veldhorst}\ \emph {et~al.}(2015)\citenamefont
  {Veldhorst}, \citenamefont {Yang}, \citenamefont {Hwang}, \citenamefont
  {Huang}, \citenamefont {Dehollain}, \citenamefont {Muhonen}, \citenamefont
  {Simmons}, \citenamefont {Laucht}, \citenamefont {Hudson}, \citenamefont
  {Itoh}, \citenamefont {Morello},\ and\ \citenamefont
  {Dzurak}}]{Veldhorst2015}%
  \BibitemOpen
  \bibfield  {author} {\bibinfo {author} {\bibfnamefont {M.}~\bibnamefont
  {Veldhorst}}, \bibinfo {author} {\bibfnamefont {C.~H.}\ \bibnamefont {Yang}},
  \bibinfo {author} {\bibfnamefont {J.~C.~C.}\ \bibnamefont {Hwang}}, \bibinfo
  {author} {\bibfnamefont {W.}~\bibnamefont {Huang}}, \bibinfo {author}
  {\bibfnamefont {J.~P.}\ \bibnamefont {Dehollain}}, \bibinfo {author}
  {\bibfnamefont {J.~T.}\ \bibnamefont {Muhonen}}, \bibinfo {author}
  {\bibfnamefont {S.}~\bibnamefont {Simmons}}, \bibinfo {author} {\bibfnamefont
  {A.}~\bibnamefont {Laucht}}, \bibinfo {author} {\bibfnamefont {F.~E.}\
  \bibnamefont {Hudson}}, \bibinfo {author} {\bibfnamefont {K.~M.}\
  \bibnamefont {Itoh}}, \bibinfo {author} {\bibfnamefont {A.}~\bibnamefont
  {Morello}}, \ and\ \bibinfo {author} {\bibfnamefont {A.~S.}\ \bibnamefont
  {Dzurak}},\ }\href {\doibase 10.1038/nature15263} {\bibfield  {journal}
  {\bibinfo  {journal} {Nature}\ }\textbf {\bibinfo {volume} {526}},\ \bibinfo
  {pages} {410} (\bibinfo {year} {2015})}\BibitemShut {NoStop}%
\bibitem [{\citenamefont {Zajac}\ \emph {et~al.}(2016)\citenamefont {Zajac},
  \citenamefont {Hazard}, \citenamefont {Mi}, \citenamefont {Nielsen},\ and\
  \citenamefont {Petta}}]{Zajac2016}%
  \BibitemOpen
  \bibfield  {author} {\bibinfo {author} {\bibfnamefont {D.~M.}\ \bibnamefont
  {Zajac}}, \bibinfo {author} {\bibfnamefont {T.~M.}\ \bibnamefont {Hazard}},
  \bibinfo {author} {\bibfnamefont {X.}~\bibnamefont {Mi}}, \bibinfo {author}
  {\bibfnamefont {E.}~\bibnamefont {Nielsen}}, \ and\ \bibinfo {author}
  {\bibfnamefont {J.~R.}\ \bibnamefont {Petta}},\ }\href {\doibase
  10.1103/physrevapplied.6.054013} {\bibfield  {journal} {\bibinfo  {journal}
  {Phys. Rev. Appl.}\ }\textbf {\bibinfo {volume} {6}},\ \bibinfo {pages}
  {054013} (\bibinfo {year} {2016})}\BibitemShut {NoStop}%
\bibitem [{\citenamefont {Watson}\ \emph {et~al.}(2018)\citenamefont {Watson},
  \citenamefont {Philips}, \citenamefont {Kawakami}, \citenamefont {Ward},
  \citenamefont {Scarlino}, \citenamefont {Veldhorst}, \citenamefont {Savage},
  \citenamefont {Lagally}, \citenamefont {Friesen}, \citenamefont
  {Coppersmith}, \citenamefont {Eriksson},\ and\ \citenamefont
  {Vandersypen}}]{Watson2018}%
  \BibitemOpen
  \bibfield  {author} {\bibinfo {author} {\bibfnamefont {T.~F.}\ \bibnamefont
  {Watson}}, \bibinfo {author} {\bibfnamefont {S.~G.~J.}\ \bibnamefont
  {Philips}}, \bibinfo {author} {\bibfnamefont {E.}~\bibnamefont {Kawakami}},
  \bibinfo {author} {\bibfnamefont {D.~R.}\ \bibnamefont {Ward}}, \bibinfo
  {author} {\bibfnamefont {P.}~\bibnamefont {Scarlino}}, \bibinfo {author}
  {\bibfnamefont {M.}~\bibnamefont {Veldhorst}}, \bibinfo {author}
  {\bibfnamefont {D.~E.}\ \bibnamefont {Savage}}, \bibinfo {author}
  {\bibfnamefont {M.~G.}\ \bibnamefont {Lagally}}, \bibinfo {author}
  {\bibfnamefont {M.}~\bibnamefont {Friesen}}, \bibinfo {author} {\bibfnamefont
  {S.~N.}\ \bibnamefont {Coppersmith}}, \bibinfo {author} {\bibfnamefont
  {M.~A.}\ \bibnamefont {Eriksson}}, \ and\ \bibinfo {author} {\bibfnamefont
  {L.~M.~K.}\ \bibnamefont {Vandersypen}},\ }\href {\doibase
  10.1038/nature25766} {\bibfield  {journal} {\bibinfo  {journal} {Nature}\
  }\textbf {\bibinfo {volume} {555}},\ \bibinfo {pages} {633} (\bibinfo {year}
  {2018})}\BibitemShut {NoStop}%
\bibitem [{\citenamefont {Huang}\ \emph {et~al.}(2019)\citenamefont {Huang},
  \citenamefont {Yang}, \citenamefont {Chan}, \citenamefont {Tanttu},
  \citenamefont {Hensen}, \citenamefont {Leon}, \citenamefont {Fogarty},
  \citenamefont {Hwang}, \citenamefont {Hudson}, \citenamefont {Itoh},
  \citenamefont {Morello}, \citenamefont {Laucht},\ and\ \citenamefont
  {Dzurak}}]{Huang2019}%
  \BibitemOpen
  \bibfield  {author} {\bibinfo {author} {\bibfnamefont {W.}~\bibnamefont
  {Huang}}, \bibinfo {author} {\bibfnamefont {C.~H.}\ \bibnamefont {Yang}},
  \bibinfo {author} {\bibfnamefont {K.~W.}\ \bibnamefont {Chan}}, \bibinfo
  {author} {\bibfnamefont {T.}~\bibnamefont {Tanttu}}, \bibinfo {author}
  {\bibfnamefont {B.}~\bibnamefont {Hensen}}, \bibinfo {author} {\bibfnamefont
  {R.~C.~C.}\ \bibnamefont {Leon}}, \bibinfo {author} {\bibfnamefont {M.~A.}\
  \bibnamefont {Fogarty}}, \bibinfo {author} {\bibfnamefont {J.~C.~C.}\
  \bibnamefont {Hwang}}, \bibinfo {author} {\bibfnamefont {F.~E.}\ \bibnamefont
  {Hudson}}, \bibinfo {author} {\bibfnamefont {K.~M.}\ \bibnamefont {Itoh}},
  \bibinfo {author} {\bibfnamefont {A.}~\bibnamefont {Morello}}, \bibinfo
  {author} {\bibfnamefont {A.}~\bibnamefont {Laucht}}, \ and\ \bibinfo {author}
  {\bibfnamefont {A.~S.}\ \bibnamefont {Dzurak}},\ }\href {\doibase
  10.1038/s41586-019-1197-0} {\bibfield  {journal} {\bibinfo  {journal}
  {Nature}\ }\textbf {\bibinfo {volume} {569}},\ \bibinfo {pages} {532}
  (\bibinfo {year} {2019})}\BibitemShut {NoStop}%
\bibitem [{\citenamefont {Xue}\ \emph {et~al.}(2019)\citenamefont {Xue},
  \citenamefont {Watson}, \citenamefont {Helsen}, \citenamefont {Ward},
  \citenamefont {Savage}, \citenamefont {Lagally}, \citenamefont {Coppersmith},
  \citenamefont {Eriksson}, \citenamefont {Wehner},\ and\ \citenamefont
  {Vandersypen}}]{Xue2019}%
  \BibitemOpen
  \bibfield  {author} {\bibinfo {author} {\bibfnamefont {X.}~\bibnamefont
  {Xue}}, \bibinfo {author} {\bibfnamefont {T.~F.}\ \bibnamefont {Watson}},
  \bibinfo {author} {\bibfnamefont {J.}~\bibnamefont {Helsen}}, \bibinfo
  {author} {\bibfnamefont {D.~R.}\ \bibnamefont {Ward}}, \bibinfo {author}
  {\bibfnamefont {D.~E.}\ \bibnamefont {Savage}}, \bibinfo {author}
  {\bibfnamefont {M.~G.}\ \bibnamefont {Lagally}}, \bibinfo {author}
  {\bibfnamefont {S.~N.}\ \bibnamefont {Coppersmith}}, \bibinfo {author}
  {\bibfnamefont {M.~A.}\ \bibnamefont {Eriksson}}, \bibinfo {author}
  {\bibfnamefont {S.}~\bibnamefont {Wehner}}, \ and\ \bibinfo {author}
  {\bibfnamefont {L.~M.~K.}\ \bibnamefont {Vandersypen}},\ }\href {\doibase
  10.1103/physrevx.9.021011} {\bibfield  {journal} {\bibinfo  {journal} {Phys.
  Rev. X}\ }\textbf {\bibinfo {volume} {9}},\ \bibinfo {pages} {021011}
  (\bibinfo {year} {2019})}\BibitemShut {NoStop}%
\bibitem [{\citenamefont {Sigillito}\ \emph {et~al.}(2019)\citenamefont
  {Sigillito}, \citenamefont {Loy}, \citenamefont {Zajac}, \citenamefont
  {Gullans}, \citenamefont {Edge},\ and\ \citenamefont
  {Petta}}]{Sigillito2019}%
  \BibitemOpen
  \bibfield  {author} {\bibinfo {author} {\bibfnamefont {A.~J.}\ \bibnamefont
  {Sigillito}}, \bibinfo {author} {\bibfnamefont {J.~C.}\ \bibnamefont {Loy}},
  \bibinfo {author} {\bibfnamefont {D.~M.}\ \bibnamefont {Zajac}}, \bibinfo
  {author} {\bibfnamefont {M.~J.}\ \bibnamefont {Gullans}}, \bibinfo {author}
  {\bibfnamefont {L.~F.}\ \bibnamefont {Edge}}, \ and\ \bibinfo {author}
  {\bibfnamefont {J.~R.}\ \bibnamefont {Petta}},\ }\href {\doibase
  10.1103/physrevapplied.11.061006} {\bibfield  {journal} {\bibinfo  {journal}
  {Phys. Rev. Appl.}\ }\textbf {\bibinfo {volume} {11}},\ \bibinfo {pages}
  {061006(R)} (\bibinfo {year} {2019})}\BibitemShut {NoStop}%
\bibitem [{\citenamefont {Mi}\ \emph {et~al.}(2018{\natexlab{a}})\citenamefont
  {Mi}, \citenamefont {Benito}, \citenamefont {Putz}, \citenamefont {Zajac},
  \citenamefont {Taylor}, \citenamefont {Burkard},\ and\ \citenamefont
  {Petta}}]{Mi2018}%
  \BibitemOpen
  \bibfield  {author} {\bibinfo {author} {\bibfnamefont {X.}~\bibnamefont
  {Mi}}, \bibinfo {author} {\bibfnamefont {M.}~\bibnamefont {Benito}}, \bibinfo
  {author} {\bibfnamefont {S.}~\bibnamefont {Putz}}, \bibinfo {author}
  {\bibfnamefont {D.~M.}\ \bibnamefont {Zajac}}, \bibinfo {author}
  {\bibfnamefont {J.~M.}\ \bibnamefont {Taylor}}, \bibinfo {author}
  {\bibfnamefont {G.}~\bibnamefont {Burkard}}, \ and\ \bibinfo {author}
  {\bibfnamefont {J.~R.}\ \bibnamefont {Petta}},\ }\href {\doibase
  10.1038/nature25769} {\bibfield  {journal} {\bibinfo  {journal} {Nature}\
  }\textbf {\bibinfo {volume} {555}},\ \bibinfo {pages} {599} (\bibinfo {year}
  {2018}{\natexlab{a}})}\BibitemShut {NoStop}%
\bibitem [{\citenamefont {Samkharadze}\ \emph {et~al.}(2018)\citenamefont
  {Samkharadze}, \citenamefont {Zheng}, \citenamefont {Kalhor}, \citenamefont
  {Brousse}, \citenamefont {Sammak}, \citenamefont {Mendes}, \citenamefont
  {Blais}, \citenamefont {Scappucci},\ and\ \citenamefont
  {Vandersypen}}]{Samkharadze2018}%
  \BibitemOpen
  \bibfield  {author} {\bibinfo {author} {\bibfnamefont {N.}~\bibnamefont
  {Samkharadze}}, \bibinfo {author} {\bibfnamefont {G.}~\bibnamefont {Zheng}},
  \bibinfo {author} {\bibfnamefont {N.}~\bibnamefont {Kalhor}}, \bibinfo
  {author} {\bibfnamefont {D.}~\bibnamefont {Brousse}}, \bibinfo {author}
  {\bibfnamefont {A.}~\bibnamefont {Sammak}}, \bibinfo {author} {\bibfnamefont
  {U.~C.}\ \bibnamefont {Mendes}}, \bibinfo {author} {\bibfnamefont
  {A.}~\bibnamefont {Blais}}, \bibinfo {author} {\bibfnamefont
  {G.}~\bibnamefont {Scappucci}}, \ and\ \bibinfo {author} {\bibfnamefont
  {L.~M.~K.}\ \bibnamefont {Vandersypen}},\ }\href {\doibase
  10.1126/science.aar4054} {\bibfield  {journal} {\bibinfo  {journal}
  {Science}\ }\textbf {\bibinfo {volume} {359}},\ \bibinfo {pages} {1123}
  (\bibinfo {year} {2018})}\BibitemShut {NoStop}%
\bibitem [{\citenamefont {Bonsen}\ \emph {et~al.}(2023)\citenamefont {Bonsen},
  \citenamefont {Harvey-Collard}, \citenamefont {Russ}, \citenamefont
  {Dijkema}, \citenamefont {Sammak}, \citenamefont {Scappucci},\ and\
  \citenamefont {Vandersypen}}]{Bonsen2023}%
  \BibitemOpen
  \bibfield  {author} {\bibinfo {author} {\bibfnamefont {T.}~\bibnamefont
  {Bonsen}}, \bibinfo {author} {\bibfnamefont {P.}~\bibnamefont
  {Harvey-Collard}}, \bibinfo {author} {\bibfnamefont {M.}~\bibnamefont
  {Russ}}, \bibinfo {author} {\bibfnamefont {J.}~\bibnamefont {Dijkema}},
  \bibinfo {author} {\bibfnamefont {A.}~\bibnamefont {Sammak}}, \bibinfo
  {author} {\bibfnamefont {G.}~\bibnamefont {Scappucci}}, \ and\ \bibinfo
  {author} {\bibfnamefont {L.~M.~K.}\ \bibnamefont {Vandersypen}},\ }\href
  {\doibase 10.1103/physrevlett.130.137001} {\bibfield  {journal} {\bibinfo
  {journal} {Phys. Rev. Lett.}\ }\textbf {\bibinfo {volume} {130}},\ \bibinfo
  {pages} {137001} (\bibinfo {year} {2023})}\BibitemShut {NoStop}%
\bibitem [{\citenamefont {Li}\ \emph {et~al.}(2018)\citenamefont {Li},
  \citenamefont {Petit}, \citenamefont {Franke}, \citenamefont {Dehollain},
  \citenamefont {Helsen}, \citenamefont {Steudtner}, \citenamefont {Thomas},
  \citenamefont {Yoscovits}, \citenamefont {Singh}, \citenamefont {Wehner},
  \citenamefont {Vandersypen}, \citenamefont {Clarke},\ and\ \citenamefont
  {Veldhorst}}]{Li2018}%
  \BibitemOpen
  \bibfield  {author} {\bibinfo {author} {\bibfnamefont {R.}~\bibnamefont
  {Li}}, \bibinfo {author} {\bibfnamefont {L.}~\bibnamefont {Petit}}, \bibinfo
  {author} {\bibfnamefont {D.~P.}\ \bibnamefont {Franke}}, \bibinfo {author}
  {\bibfnamefont {J.~P.}\ \bibnamefont {Dehollain}}, \bibinfo {author}
  {\bibfnamefont {J.}~\bibnamefont {Helsen}}, \bibinfo {author} {\bibfnamefont
  {M.}~\bibnamefont {Steudtner}}, \bibinfo {author} {\bibfnamefont {N.~K.}\
  \bibnamefont {Thomas}}, \bibinfo {author} {\bibfnamefont {Z.~R.}\
  \bibnamefont {Yoscovits}}, \bibinfo {author} {\bibfnamefont {K.~J.}\
  \bibnamefont {Singh}}, \bibinfo {author} {\bibfnamefont {S.}~\bibnamefont
  {Wehner}}, \bibinfo {author} {\bibfnamefont {L.~M.~K.}\ \bibnamefont
  {Vandersypen}}, \bibinfo {author} {\bibfnamefont {J.~S.}\ \bibnamefont
  {Clarke}}, \ and\ \bibinfo {author} {\bibfnamefont {M.}~\bibnamefont
  {Veldhorst}},\ }\href {\doibase 10.1126/sciadv.aar3960} {\bibfield  {journal}
  {\bibinfo  {journal} {Sci. Adv.}\ }\textbf {\bibinfo {volume} {4}},\ \bibinfo
  {pages} {eaar3960} (\bibinfo {year} {2018})}\BibitemShut {NoStop}%
\bibitem [{\citenamefont {Mills}\ \emph {et~al.}(2019)\citenamefont {Mills},
  \citenamefont {Zajac}, \citenamefont {Gullans}, \citenamefont {Schupp},
  \citenamefont {Hazard},\ and\ \citenamefont {Petta}}]{Mills2019}%
  \BibitemOpen
  \bibfield  {author} {\bibinfo {author} {\bibfnamefont {A.~R.}\ \bibnamefont
  {Mills}}, \bibinfo {author} {\bibfnamefont {D.~M.}\ \bibnamefont {Zajac}},
  \bibinfo {author} {\bibfnamefont {M.~J.}\ \bibnamefont {Gullans}}, \bibinfo
  {author} {\bibfnamefont {F.~J.}\ \bibnamefont {Schupp}}, \bibinfo {author}
  {\bibfnamefont {T.~M.}\ \bibnamefont {Hazard}}, \ and\ \bibinfo {author}
  {\bibfnamefont {J.~R.}\ \bibnamefont {Petta}},\ }\href {\doibase
  10.1038/s41467-019-08970-z} {\bibfield  {journal} {\bibinfo  {journal} {Nat.
  Commun.}\ }\textbf {\bibinfo {volume} {10}},\ \bibinfo {pages} {1063}
  (\bibinfo {year} {2019})}\BibitemShut {NoStop}%
\bibitem [{\citenamefont {Noiri}\ \emph
  {et~al.}(2022{\natexlab{a}})\citenamefont {Noiri}, \citenamefont {Takeda},
  \citenamefont {Nakajima}, \citenamefont {Kobayashi}, \citenamefont {Sammak},
  \citenamefont {Scappucci},\ and\ \citenamefont {Tarucha}}]{Noiri2022}%
  \BibitemOpen
  \bibfield  {author} {\bibinfo {author} {\bibfnamefont {A.}~\bibnamefont
  {Noiri}}, \bibinfo {author} {\bibfnamefont {K.}~\bibnamefont {Takeda}},
  \bibinfo {author} {\bibfnamefont {T.}~\bibnamefont {Nakajima}}, \bibinfo
  {author} {\bibfnamefont {T.}~\bibnamefont {Kobayashi}}, \bibinfo {author}
  {\bibfnamefont {A.}~\bibnamefont {Sammak}}, \bibinfo {author} {\bibfnamefont
  {G.}~\bibnamefont {Scappucci}}, \ and\ \bibinfo {author} {\bibfnamefont
  {S.}~\bibnamefont {Tarucha}},\ }\href {\doibase 10.1038/s41467-022-33453-z}
  {\bibfield  {journal} {\bibinfo  {journal} {Nat. Commun.}\ }\textbf {\bibinfo
  {volume} {13}},\ \bibinfo {pages} {5740} (\bibinfo {year}
  {2022}{\natexlab{a}})}\BibitemShut {NoStop}%
\bibitem [{\citenamefont {Seidler}\ \emph {et~al.}(2022)\citenamefont
  {Seidler}, \citenamefont {Struck}, \citenamefont {Xue}, \citenamefont
  {Focke}, \citenamefont {Trellenkamp}, \citenamefont {Bluhm},\ and\
  \citenamefont {Schreiber}}]{Seidler2022}%
  \BibitemOpen
  \bibfield  {author} {\bibinfo {author} {\bibfnamefont {I.}~\bibnamefont
  {Seidler}}, \bibinfo {author} {\bibfnamefont {T.}~\bibnamefont {Struck}},
  \bibinfo {author} {\bibfnamefont {R.}~\bibnamefont {Xue}}, \bibinfo {author}
  {\bibfnamefont {N.}~\bibnamefont {Focke}}, \bibinfo {author} {\bibfnamefont
  {S.}~\bibnamefont {Trellenkamp}}, \bibinfo {author} {\bibfnamefont
  {H.}~\bibnamefont {Bluhm}}, \ and\ \bibinfo {author} {\bibfnamefont {L.~R.}\
  \bibnamefont {Schreiber}},\ }\href {\doibase 10.1038/s41534-022-00615-2}
  {\bibfield  {journal} {\bibinfo  {journal} {npj Quantum Inf.}\ }\textbf
  {\bibinfo {volume} {8}},\ \bibinfo {pages} {100} (\bibinfo {year}
  {2022})}\BibitemShut {NoStop}%
\bibitem [{\citenamefont {Takeda}\ \emph {et~al.}(2021)\citenamefont {Takeda},
  \citenamefont {Noiri}, \citenamefont {Nakajima}, \citenamefont {Yoneda},
  \citenamefont {Kobayashi},\ and\ \citenamefont {Tarucha}}]{Takeda2021}%
  \BibitemOpen
  \bibfield  {author} {\bibinfo {author} {\bibfnamefont {K.}~\bibnamefont
  {Takeda}}, \bibinfo {author} {\bibfnamefont {A.}~\bibnamefont {Noiri}},
  \bibinfo {author} {\bibfnamefont {T.}~\bibnamefont {Nakajima}}, \bibinfo
  {author} {\bibfnamefont {J.}~\bibnamefont {Yoneda}}, \bibinfo {author}
  {\bibfnamefont {T.}~\bibnamefont {Kobayashi}}, \ and\ \bibinfo {author}
  {\bibfnamefont {S.}~\bibnamefont {Tarucha}},\ }\href {\doibase
  10.1038/s41565-021-00925-0} {\bibfield  {journal} {\bibinfo  {journal} {Nat.
  Nanotechnol.}\ }\textbf {\bibinfo {volume} {16}},\ \bibinfo {pages} {965}
  (\bibinfo {year} {2021})}\BibitemShut {NoStop}%
\bibitem [{\citenamefont {Mills}\ \emph {et~al.}(2022)\citenamefont {Mills},
  \citenamefont {Guinn}, \citenamefont {Feldman}, \citenamefont {Sigillito},
  \citenamefont {Gullans}, \citenamefont {Rakher}, \citenamefont {Kerckhoff},
  \citenamefont {Jackson},\ and\ \citenamefont {Petta}}]{Mills2022a}%
  \BibitemOpen
  \bibfield  {author} {\bibinfo {author} {\bibfnamefont {A.~R.}\ \bibnamefont
  {Mills}}, \bibinfo {author} {\bibfnamefont {C.~R.}\ \bibnamefont {Guinn}},
  \bibinfo {author} {\bibfnamefont {M.~M.}\ \bibnamefont {Feldman}}, \bibinfo
  {author} {\bibfnamefont {A.~J.}\ \bibnamefont {Sigillito}}, \bibinfo {author}
  {\bibfnamefont {M.~J.}\ \bibnamefont {Gullans}}, \bibinfo {author}
  {\bibfnamefont {M.~T.}\ \bibnamefont {Rakher}}, \bibinfo {author}
  {\bibfnamefont {J.}~\bibnamefont {Kerckhoff}}, \bibinfo {author}
  {\bibfnamefont {C.~A.~C.}\ \bibnamefont {Jackson}}, \ and\ \bibinfo {author}
  {\bibfnamefont {J.~R.}\ \bibnamefont {Petta}},\ }\href {\doibase
  10.1103/physrevapplied.18.064028} {\bibfield  {journal} {\bibinfo  {journal}
  {Phys. Rev. Appl.}\ }\textbf {\bibinfo {volume} {18}},\ \bibinfo {pages}
  {064028} (\bibinfo {year} {2022})}\BibitemShut {NoStop}%
\bibitem [{\citenamefont {Noiri}\ \emph
  {et~al.}(2022{\natexlab{b}})\citenamefont {Noiri}, \citenamefont {Takeda},
  \citenamefont {Nakajima}, \citenamefont {Kobayashi}, \citenamefont {Sammak},
  \citenamefont {Scappucci},\ and\ \citenamefont {Tarucha}}]{Noiri2022a}%
  \BibitemOpen
  \bibfield  {author} {\bibinfo {author} {\bibfnamefont {A.}~\bibnamefont
  {Noiri}}, \bibinfo {author} {\bibfnamefont {K.}~\bibnamefont {Takeda}},
  \bibinfo {author} {\bibfnamefont {T.}~\bibnamefont {Nakajima}}, \bibinfo
  {author} {\bibfnamefont {T.}~\bibnamefont {Kobayashi}}, \bibinfo {author}
  {\bibfnamefont {A.}~\bibnamefont {Sammak}}, \bibinfo {author} {\bibfnamefont
  {G.}~\bibnamefont {Scappucci}}, \ and\ \bibinfo {author} {\bibfnamefont
  {S.}~\bibnamefont {Tarucha}},\ }\href {\doibase 10.1038/s41586-021-04182-y}
  {\bibfield  {journal} {\bibinfo  {journal} {Nature}\ }\textbf {\bibinfo
  {volume} {601}},\ \bibinfo {pages} {338} (\bibinfo {year}
  {2022}{\natexlab{b}})}\BibitemShut {NoStop}%
\bibitem [{\citenamefont {Tanttu}\ \emph {et~al.}(2023)\citenamefont {Tanttu},
  \citenamefont {Lim}, \citenamefont {Huang}, \citenamefont {Stuyck},
  \citenamefont {Gilbert}, \citenamefont {Su}, \citenamefont {Feng},
  \citenamefont {Cifuentes}, \citenamefont {Seedhouse}, \citenamefont
  {Seritan}, \citenamefont {Ostrove}, \citenamefont {Rudinger}, \citenamefont
  {Leon}, \citenamefont {Huang}, \citenamefont {Escott}, \citenamefont {Itoh},
  \citenamefont {Abrosimov}, \citenamefont {Pohl}, \citenamefont {Thewalt},
  \citenamefont {Hudson}, \citenamefont {Blume-Kohout}, \citenamefont
  {Bartlett}, \citenamefont {Morello}, \citenamefont {Laucht}, \citenamefont
  {Yang}, \citenamefont {Saraiva},\ and\ \citenamefont {Dzurak}}]{Tanttu2023}%
  \BibitemOpen
  \bibfield  {author} {\bibinfo {author} {\bibfnamefont {T.}~\bibnamefont
  {Tanttu}}, \bibinfo {author} {\bibfnamefont {W.~H.}\ \bibnamefont {Lim}},
  \bibinfo {author} {\bibfnamefont {J.~Y.}\ \bibnamefont {Huang}}, \bibinfo
  {author} {\bibfnamefont {N.~D.}\ \bibnamefont {Stuyck}}, \bibinfo {author}
  {\bibfnamefont {W.}~\bibnamefont {Gilbert}}, \bibinfo {author} {\bibfnamefont
  {R.~Y.}\ \bibnamefont {Su}}, \bibinfo {author} {\bibfnamefont
  {M.}~\bibnamefont {Feng}}, \bibinfo {author} {\bibfnamefont {J.~D.}\
  \bibnamefont {Cifuentes}}, \bibinfo {author} {\bibfnamefont {A.~E.}\
  \bibnamefont {Seedhouse}}, \bibinfo {author} {\bibfnamefont {S.~K.}\
  \bibnamefont {Seritan}}, \bibinfo {author} {\bibfnamefont {C.~I.}\
  \bibnamefont {Ostrove}}, \bibinfo {author} {\bibfnamefont {K.~M.}\
  \bibnamefont {Rudinger}}, \bibinfo {author} {\bibfnamefont {R.~C.~C.}\
  \bibnamefont {Leon}}, \bibinfo {author} {\bibfnamefont {W.}~\bibnamefont
  {Huang}}, \bibinfo {author} {\bibfnamefont {C.~C.}\ \bibnamefont {Escott}},
  \bibinfo {author} {\bibfnamefont {K.~M.}\ \bibnamefont {Itoh}}, \bibinfo
  {author} {\bibfnamefont {N.~V.}\ \bibnamefont {Abrosimov}}, \bibinfo {author}
  {\bibfnamefont {H.-J.}\ \bibnamefont {Pohl}}, \bibinfo {author}
  {\bibfnamefont {M.~L.~W.}\ \bibnamefont {Thewalt}}, \bibinfo {author}
  {\bibfnamefont {F.~E.}\ \bibnamefont {Hudson}}, \bibinfo {author}
  {\bibfnamefont {R.}~\bibnamefont {Blume-Kohout}}, \bibinfo {author}
  {\bibfnamefont {S.~D.}\ \bibnamefont {Bartlett}}, \bibinfo {author}
  {\bibfnamefont {A.}~\bibnamefont {Morello}}, \bibinfo {author} {\bibfnamefont
  {A.}~\bibnamefont {Laucht}}, \bibinfo {author} {\bibfnamefont {C.~H.}\
  \bibnamefont {Yang}}, \bibinfo {author} {\bibfnamefont {A.}~\bibnamefont
  {Saraiva}}, \ and\ \bibinfo {author} {\bibfnamefont {A.~S.}\ \bibnamefont
  {Dzurak}},\ }\href@noop {} {\  (\bibinfo {year} {2023})},\ \Eprint
  {http://arxiv.org/abs/2303.04090v2} {arXiv:2303.04090v2 [quant-ph]}
  \BibitemShut {NoStop}%
\bibitem [{\citenamefont {Ando}\ \emph {et~al.}(1982)\citenamefont {Ando},
  \citenamefont {Fowler},\ and\ \citenamefont {Stern}}]{Ando1982}%
  \BibitemOpen
  \bibfield  {author} {\bibinfo {author} {\bibfnamefont {T.}~\bibnamefont
  {Ando}}, \bibinfo {author} {\bibfnamefont {A.~B.}\ \bibnamefont {Fowler}}, \
  and\ \bibinfo {author} {\bibfnamefont {F.}~\bibnamefont {Stern}},\ }\href
  {\doibase 10.1103/revmodphys.54.437} {\bibfield  {journal} {\bibinfo
  {journal} {Rev. Mod. Phys.}\ }\textbf {\bibinfo {volume} {54}},\ \bibinfo
  {pages} {437} (\bibinfo {year} {1982})}\BibitemShut {NoStop}%
\bibitem [{\citenamefont {Zwanenburg}\ \emph {et~al.}(2013)\citenamefont
  {Zwanenburg}, \citenamefont {Dzurak}, \citenamefont {Morello}, \citenamefont
  {Simmons}, \citenamefont {Hollenberg}, \citenamefont {Klimeck}, \citenamefont
  {Rogge}, \citenamefont {Coppersmith},\ and\ \citenamefont
  {Eriksson}}]{Zwanenburg2013}%
  \BibitemOpen
  \bibfield  {author} {\bibinfo {author} {\bibfnamefont {F.~A.}\ \bibnamefont
  {Zwanenburg}}, \bibinfo {author} {\bibfnamefont {A.~S.}\ \bibnamefont
  {Dzurak}}, \bibinfo {author} {\bibfnamefont {A.}~\bibnamefont {Morello}},
  \bibinfo {author} {\bibfnamefont {M.~Y.}\ \bibnamefont {Simmons}}, \bibinfo
  {author} {\bibfnamefont {L.~C.~L.}\ \bibnamefont {Hollenberg}}, \bibinfo
  {author} {\bibfnamefont {G.}~\bibnamefont {Klimeck}}, \bibinfo {author}
  {\bibfnamefont {S.}~\bibnamefont {Rogge}}, \bibinfo {author} {\bibfnamefont
  {S.~N.}\ \bibnamefont {Coppersmith}}, \ and\ \bibinfo {author} {\bibfnamefont
  {M.~A.}\ \bibnamefont {Eriksson}},\ }\href {\doibase
  10.1103/revmodphys.85.961} {\bibfield  {journal} {\bibinfo  {journal} {Rev.
  Mod. Phys.}\ }\textbf {\bibinfo {volume} {85}},\ \bibinfo {pages} {961}
  (\bibinfo {year} {2013})}\BibitemShut {NoStop}%
\bibitem [{\citenamefont {Ruskov}\ \emph {et~al.}(2018)\citenamefont {Ruskov},
  \citenamefont {Veldhorst}, \citenamefont {Dzurak},\ and\ \citenamefont
  {Tahan}}]{Ruskov2018}%
  \BibitemOpen
  \bibfield  {author} {\bibinfo {author} {\bibfnamefont {R.}~\bibnamefont
  {Ruskov}}, \bibinfo {author} {\bibfnamefont {M.}~\bibnamefont {Veldhorst}},
  \bibinfo {author} {\bibfnamefont {A.~S.}\ \bibnamefont {Dzurak}}, \ and\
  \bibinfo {author} {\bibfnamefont {C.}~\bibnamefont {Tahan}},\ }\href
  {\doibase 10.1103/physrevb.98.245424} {\bibfield  {journal} {\bibinfo
  {journal} {Phys. Rev. B}\ }\textbf {\bibinfo {volume} {98}},\ \bibinfo
  {pages} {245424} (\bibinfo {year} {2018})}\BibitemShut {NoStop}%
\bibitem [{\citenamefont {Borselli}\ \emph {et~al.}(2011)\citenamefont
  {Borselli}, \citenamefont {Ross}, \citenamefont {Kiselev}, \citenamefont
  {Croke}, \citenamefont {Holabird}, \citenamefont {Deelman}, \citenamefont
  {Warren}, \citenamefont {Alvarado-Rodriguez}, \citenamefont {Milosavljevic},
  \citenamefont {Ku}, \citenamefont {Wong}, \citenamefont {Schmitz},
  \citenamefont {Sokolich}, \citenamefont {Gyure},\ and\ \citenamefont
  {Hunter}}]{Borselli2011}%
  \BibitemOpen
  \bibfield  {author} {\bibinfo {author} {\bibfnamefont {M.~G.}\ \bibnamefont
  {Borselli}}, \bibinfo {author} {\bibfnamefont {R.~S.}\ \bibnamefont {Ross}},
  \bibinfo {author} {\bibfnamefont {A.~A.}\ \bibnamefont {Kiselev}}, \bibinfo
  {author} {\bibfnamefont {E.~T.}\ \bibnamefont {Croke}}, \bibinfo {author}
  {\bibfnamefont {K.~S.}\ \bibnamefont {Holabird}}, \bibinfo {author}
  {\bibfnamefont {P.~W.}\ \bibnamefont {Deelman}}, \bibinfo {author}
  {\bibfnamefont {L.~D.}\ \bibnamefont {Warren}}, \bibinfo {author}
  {\bibfnamefont {I.}~\bibnamefont {Alvarado-Rodriguez}}, \bibinfo {author}
  {\bibfnamefont {I.}~\bibnamefont {Milosavljevic}}, \bibinfo {author}
  {\bibfnamefont {F.~C.}\ \bibnamefont {Ku}}, \bibinfo {author} {\bibfnamefont
  {W.~S.}\ \bibnamefont {Wong}}, \bibinfo {author} {\bibfnamefont {A.~E.}\
  \bibnamefont {Schmitz}}, \bibinfo {author} {\bibfnamefont {M.}~\bibnamefont
  {Sokolich}}, \bibinfo {author} {\bibfnamefont {M.~F.}\ \bibnamefont {Gyure}},
  \ and\ \bibinfo {author} {\bibfnamefont {A.~T.}\ \bibnamefont {Hunter}},\
  }\href {\doibase 10.1063/1.3569717} {\bibfield  {journal} {\bibinfo
  {journal} {Appl. Phys. Lett.}\ }\textbf {\bibinfo {volume} {98}},\ \bibinfo
  {pages} {123118} (\bibinfo {year} {2011})}\BibitemShut {NoStop}%
\bibitem [{\citenamefont {Shi}\ \emph {et~al.}(2011)\citenamefont {Shi},
  \citenamefont {Simmons}, \citenamefont {Prance}, \citenamefont {Gamble},
  \citenamefont {Friesen}, \citenamefont {Savage}, \citenamefont {Lagally},
  \citenamefont {Coppersmith},\ and\ \citenamefont {Eriksson}}]{Shi2011}%
  \BibitemOpen
  \bibfield  {author} {\bibinfo {author} {\bibfnamefont {Z.}~\bibnamefont
  {Shi}}, \bibinfo {author} {\bibfnamefont {C.~B.}\ \bibnamefont {Simmons}},
  \bibinfo {author} {\bibfnamefont {J.~R.}\ \bibnamefont {Prance}}, \bibinfo
  {author} {\bibfnamefont {J.~K.}\ \bibnamefont {Gamble}}, \bibinfo {author}
  {\bibfnamefont {M.}~\bibnamefont {Friesen}}, \bibinfo {author} {\bibfnamefont
  {D.~E.}\ \bibnamefont {Savage}}, \bibinfo {author} {\bibfnamefont {M.~G.}\
  \bibnamefont {Lagally}}, \bibinfo {author} {\bibfnamefont {S.~N.}\
  \bibnamefont {Coppersmith}}, \ and\ \bibinfo {author} {\bibfnamefont {M.~A.}\
  \bibnamefont {Eriksson}},\ }\href {\doibase 10.1063/1.3666232} {\bibfield
  {journal} {\bibinfo  {journal} {Appl. Phys. Lett.}\ }\textbf {\bibinfo
  {volume} {99}},\ \bibinfo {pages} {233108} (\bibinfo {year}
  {2011})}\BibitemShut {NoStop}%
\bibitem [{\citenamefont {Zajac}\ \emph {et~al.}(2015)\citenamefont {Zajac},
  \citenamefont {Hazard}, \citenamefont {Mi}, \citenamefont {Wang},\ and\
  \citenamefont {Petta}}]{Zajac2015}%
  \BibitemOpen
  \bibfield  {author} {\bibinfo {author} {\bibfnamefont {D.~M.}\ \bibnamefont
  {Zajac}}, \bibinfo {author} {\bibfnamefont {T.~M.}\ \bibnamefont {Hazard}},
  \bibinfo {author} {\bibfnamefont {X.}~\bibnamefont {Mi}}, \bibinfo {author}
  {\bibfnamefont {K.}~\bibnamefont {Wang}}, \ and\ \bibinfo {author}
  {\bibfnamefont {J.~R.}\ \bibnamefont {Petta}},\ }\href {\doibase
  10.1063/1.4922249} {\bibfield  {journal} {\bibinfo  {journal} {Appl. Phys.
  Lett.}\ }\textbf {\bibinfo {volume} {106}},\ \bibinfo {pages} {223507}
  (\bibinfo {year} {2015})}\BibitemShut {NoStop}%
\bibitem [{\citenamefont {Hollmann}\ \emph {et~al.}(2020)\citenamefont
  {Hollmann}, \citenamefont {Struck}, \citenamefont {Langrock}, \citenamefont
  {Schmidbauer}, \citenamefont {Schauer}, \citenamefont {Leonhardt},
  \citenamefont {Sawano}, \citenamefont {Riemann}, \citenamefont {Abrosimov},
  \citenamefont {Bougeard},\ and\ \citenamefont {Schreiber}}]{Hollmann2020}%
  \BibitemOpen
  \bibfield  {author} {\bibinfo {author} {\bibfnamefont {A.}~\bibnamefont
  {Hollmann}}, \bibinfo {author} {\bibfnamefont {T.}~\bibnamefont {Struck}},
  \bibinfo {author} {\bibfnamefont {V.}~\bibnamefont {Langrock}}, \bibinfo
  {author} {\bibfnamefont {A.}~\bibnamefont {Schmidbauer}}, \bibinfo {author}
  {\bibfnamefont {F.}~\bibnamefont {Schauer}}, \bibinfo {author} {\bibfnamefont
  {T.}~\bibnamefont {Leonhardt}}, \bibinfo {author} {\bibfnamefont
  {K.}~\bibnamefont {Sawano}}, \bibinfo {author} {\bibfnamefont
  {H.}~\bibnamefont {Riemann}}, \bibinfo {author} {\bibfnamefont {N.~V.}\
  \bibnamefont {Abrosimov}}, \bibinfo {author} {\bibfnamefont {D.}~\bibnamefont
  {Bougeard}}, \ and\ \bibinfo {author} {\bibfnamefont {L.~R.}\ \bibnamefont
  {Schreiber}},\ }\href {\doibase 10.1103/physrevapplied.13.034068} {\bibfield
  {journal} {\bibinfo  {journal} {Phys. Rev. Appl.}\ }\textbf {\bibinfo
  {volume} {13}},\ \bibinfo {pages} {034068} (\bibinfo {year}
  {2020})}\BibitemShut {NoStop}%
\bibitem [{\citenamefont {Chen}\ \emph {et~al.}(2021)\citenamefont {Chen},
  \citenamefont {Raach}, \citenamefont {Pan}, \citenamefont {Kiselev},
  \citenamefont {Acuna}, \citenamefont {Blumoff}, \citenamefont {Brecht},
  \citenamefont {Choi}, \citenamefont {Ha}, \citenamefont {Hulbert},
  \citenamefont {Jura}, \citenamefont {Keating}, \citenamefont {Noah},
  \citenamefont {Sun}, \citenamefont {Thomas}, \citenamefont {Borselli},
  \citenamefont {Jackson}, \citenamefont {Rakher},\ and\ \citenamefont
  {Ross}}]{Chen2021}%
  \BibitemOpen
  \bibfield  {author} {\bibinfo {author} {\bibfnamefont {E.~H.}\ \bibnamefont
  {Chen}}, \bibinfo {author} {\bibfnamefont {K.}~\bibnamefont {Raach}},
  \bibinfo {author} {\bibfnamefont {A.}~\bibnamefont {Pan}}, \bibinfo {author}
  {\bibfnamefont {A.~A.}\ \bibnamefont {Kiselev}}, \bibinfo {author}
  {\bibfnamefont {E.}~\bibnamefont {Acuna}}, \bibinfo {author} {\bibfnamefont
  {J.~Z.}\ \bibnamefont {Blumoff}}, \bibinfo {author} {\bibfnamefont
  {T.}~\bibnamefont {Brecht}}, \bibinfo {author} {\bibfnamefont {M.~D.}\
  \bibnamefont {Choi}}, \bibinfo {author} {\bibfnamefont {W.}~\bibnamefont
  {Ha}}, \bibinfo {author} {\bibfnamefont {D.~R.}\ \bibnamefont {Hulbert}},
  \bibinfo {author} {\bibfnamefont {M.~P.}\ \bibnamefont {Jura}}, \bibinfo
  {author} {\bibfnamefont {T.~E.}\ \bibnamefont {Keating}}, \bibinfo {author}
  {\bibfnamefont {R.}~\bibnamefont {Noah}}, \bibinfo {author} {\bibfnamefont
  {B.}~\bibnamefont {Sun}}, \bibinfo {author} {\bibfnamefont {B.~J.}\
  \bibnamefont {Thomas}}, \bibinfo {author} {\bibfnamefont {M.~G.}\
  \bibnamefont {Borselli}}, \bibinfo {author} {\bibfnamefont {C.~A.~C.}\
  \bibnamefont {Jackson}}, \bibinfo {author} {\bibfnamefont {M.~T.}\
  \bibnamefont {Rakher}}, \ and\ \bibinfo {author} {\bibfnamefont {R.~S.}\
  \bibnamefont {Ross}},\ }\href {\doibase 10.1103/physrevapplied.15.044033}
  {\bibfield  {journal} {\bibinfo  {journal} {Phys. Rev. Appl.}\ }\textbf
  {\bibinfo {volume} {15}},\ \bibinfo {pages} {044033} (\bibinfo {year}
  {2021})}\BibitemShut {NoStop}%
\bibitem [{\citenamefont {Scarlino}\ \emph {et~al.}(2017)\citenamefont
  {Scarlino}, \citenamefont {Kawakami}, \citenamefont {Jullien}, \citenamefont
  {Ward}, \citenamefont {Savage}, \citenamefont {Lagally}, \citenamefont
  {Friesen}, \citenamefont {Coppersmith}, \citenamefont {Eriksson},\ and\
  \citenamefont {Vandersypen}}]{Scarlino2017}%
  \BibitemOpen
  \bibfield  {author} {\bibinfo {author} {\bibfnamefont {P.}~\bibnamefont
  {Scarlino}}, \bibinfo {author} {\bibfnamefont {E.}~\bibnamefont {Kawakami}},
  \bibinfo {author} {\bibfnamefont {T.}~\bibnamefont {Jullien}}, \bibinfo
  {author} {\bibfnamefont {D.~R.}\ \bibnamefont {Ward}}, \bibinfo {author}
  {\bibfnamefont {D.~E.}\ \bibnamefont {Savage}}, \bibinfo {author}
  {\bibfnamefont {M.~G.}\ \bibnamefont {Lagally}}, \bibinfo {author}
  {\bibfnamefont {M.}~\bibnamefont {Friesen}}, \bibinfo {author} {\bibfnamefont
  {S.~N.}\ \bibnamefont {Coppersmith}}, \bibinfo {author} {\bibfnamefont
  {M.~A.}\ \bibnamefont {Eriksson}}, \ and\ \bibinfo {author} {\bibfnamefont
  {L.~M.~K.}\ \bibnamefont {Vandersypen}},\ }\href {\doibase
  10.1103/physrevb.95.165429} {\bibfield  {journal} {\bibinfo  {journal} {Phys.
  Rev. B}\ }\textbf {\bibinfo {volume} {95}},\ \bibinfo {pages} {165429}
  (\bibinfo {year} {2017})}\BibitemShut {NoStop}%
\bibitem [{\citenamefont {Mi}\ \emph {et~al.}(2018{\natexlab{b}})\citenamefont
  {Mi}, \citenamefont {Kohler},\ and\ \citenamefont {Petta}}]{Mi2018a}%
  \BibitemOpen
  \bibfield  {author} {\bibinfo {author} {\bibfnamefont {X.}~\bibnamefont
  {Mi}}, \bibinfo {author} {\bibfnamefont {S.}~\bibnamefont {Kohler}}, \ and\
  \bibinfo {author} {\bibfnamefont {J.~R.}\ \bibnamefont {Petta}},\ }\href
  {\doibase 10.1103/physrevb.98.161404} {\bibfield  {journal} {\bibinfo
  {journal} {Phys. Rev. B}\ }\textbf {\bibinfo {volume} {98}},\ \bibinfo
  {pages} {161404(R)} (\bibinfo {year} {2018}{\natexlab{b}})}\BibitemShut
  {NoStop}%
\bibitem [{\citenamefont {Mi}\ \emph {et~al.}(2017)\citenamefont {Mi},
  \citenamefont {P{\'{e}}terfalvi}, \citenamefont {Burkard},\ and\
  \citenamefont {Petta}}]{Mi2017}%
  \BibitemOpen
  \bibfield  {author} {\bibinfo {author} {\bibfnamefont {X.}~\bibnamefont
  {Mi}}, \bibinfo {author} {\bibfnamefont {C.~G.}\ \bibnamefont
  {P{\'{e}}terfalvi}}, \bibinfo {author} {\bibfnamefont {G.}~\bibnamefont
  {Burkard}}, \ and\ \bibinfo {author} {\bibfnamefont {J.~R.}\ \bibnamefont
  {Petta}},\ }\href {\doibase 10.1103/physrevlett.119.176803} {\bibfield
  {journal} {\bibinfo  {journal} {Phys. Rev. Lett.}\ }\textbf {\bibinfo
  {volume} {119}},\ \bibinfo {pages} {176803} (\bibinfo {year}
  {2017})}\BibitemShut {NoStop}%
\bibitem [{\citenamefont {Friesen}\ \emph {et~al.}(2007)\citenamefont
  {Friesen}, \citenamefont {Chutia}, \citenamefont {Tahan},\ and\ \citenamefont
  {Coppersmith}}]{Friesen2007}%
  \BibitemOpen
  \bibfield  {author} {\bibinfo {author} {\bibfnamefont {M.}~\bibnamefont
  {Friesen}}, \bibinfo {author} {\bibfnamefont {S.}~\bibnamefont {Chutia}},
  \bibinfo {author} {\bibfnamefont {C.}~\bibnamefont {Tahan}}, \ and\ \bibinfo
  {author} {\bibfnamefont {S.~N.}\ \bibnamefont {Coppersmith}},\ }\href
  {\doibase 10.1103/physrevb.75.115318} {\bibfield  {journal} {\bibinfo
  {journal} {Phys. Rev. B}\ }\textbf {\bibinfo {volume} {75}},\ \bibinfo
  {pages} {115318} (\bibinfo {year} {2007})}\BibitemShut {NoStop}%
\bibitem [{\citenamefont {Chutia}\ \emph {et~al.}(2008)\citenamefont {Chutia},
  \citenamefont {Coppersmith},\ and\ \citenamefont {Friesen}}]{Chutia2008}%
  \BibitemOpen
  \bibfield  {author} {\bibinfo {author} {\bibfnamefont {S.}~\bibnamefont
  {Chutia}}, \bibinfo {author} {\bibfnamefont {S.~N.}\ \bibnamefont
  {Coppersmith}}, \ and\ \bibinfo {author} {\bibfnamefont {M.}~\bibnamefont
  {Friesen}},\ }\href {\doibase 10.1103/physrevb.77.193311} {\bibfield
  {journal} {\bibinfo  {journal} {Phys. Rev. B}\ }\textbf {\bibinfo {volume}
  {77}},\ \bibinfo {pages} {193311} (\bibinfo {year} {2008})}\BibitemShut
  {NoStop}%
\bibitem [{\citenamefont {Saraiva}\ \emph {et~al.}(2009)\citenamefont
  {Saraiva}, \citenamefont {Calder{\'{o}}n}, \citenamefont {Hu}, \citenamefont
  {{Das Sarma}},\ and\ \citenamefont {Koiller}}]{Saraiva2009}%
  \BibitemOpen
  \bibfield  {author} {\bibinfo {author} {\bibfnamefont {A.~L.}\ \bibnamefont
  {Saraiva}}, \bibinfo {author} {\bibfnamefont {M.~J.}\ \bibnamefont
  {Calder{\'{o}}n}}, \bibinfo {author} {\bibfnamefont {X.}~\bibnamefont {Hu}},
  \bibinfo {author} {\bibfnamefont {S.}~\bibnamefont {{Das Sarma}}}, \ and\
  \bibinfo {author} {\bibfnamefont {B.}~\bibnamefont {Koiller}},\ }\href
  {\doibase 10.1103/physrevb.80.081305} {\bibfield  {journal} {\bibinfo
  {journal} {Phys. Rev. B}\ }\textbf {\bibinfo {volume} {80}},\ \bibinfo
  {pages} {081305(R)} (\bibinfo {year} {2009})}\BibitemShut {NoStop}%
\bibitem [{\citenamefont {Hosseinkhani}\ and\ \citenamefont
  {Burkard}(2020)}]{Hosseinkhani2020}%
  \BibitemOpen
  \bibfield  {author} {\bibinfo {author} {\bibfnamefont {A.}~\bibnamefont
  {Hosseinkhani}}\ and\ \bibinfo {author} {\bibfnamefont {G.}~\bibnamefont
  {Burkard}},\ }\href {\doibase 10.1103/physrevresearch.2.043180} {\bibfield
  {journal} {\bibinfo  {journal} {Phys. Rev. Res.}\ }\textbf {\bibinfo {volume}
  {2}},\ \bibinfo {pages} {043180} (\bibinfo {year} {2020})}\BibitemShut
  {NoStop}%
\bibitem [{\citenamefont {Hosseinkhani}\ and\ \citenamefont
  {Burkard}(2021)}]{Hosseinkhani2021}%
  \BibitemOpen
  \bibfield  {author} {\bibinfo {author} {\bibfnamefont {A.}~\bibnamefont
  {Hosseinkhani}}\ and\ \bibinfo {author} {\bibfnamefont {G.}~\bibnamefont
  {Burkard}},\ }\href {\doibase 10.1103/physrevb.104.085309} {\bibfield
  {journal} {\bibinfo  {journal} {Phys. Rev. B}\ }\textbf {\bibinfo {volume}
  {104}},\ \bibinfo {pages} {085309} (\bibinfo {year} {2021})}\BibitemShut
  {NoStop}%
\bibitem [{\citenamefont {Lima}\ and\ \citenamefont
  {Burkard}(2023)}]{Lima2023}%
  \BibitemOpen
  \bibfield  {author} {\bibinfo {author} {\bibfnamefont {J.~R.~F.}\
  \bibnamefont {Lima}}\ and\ \bibinfo {author} {\bibfnamefont {G.}~\bibnamefont
  {Burkard}},\ }\href {\doibase 10.1088/2633-4356/acd743} {\bibfield  {journal}
  {\bibinfo  {journal} {Mater. Quantum. Technol.}\ }\textbf {\bibinfo {volume}
  {3}},\ \bibinfo {pages} {025004} (\bibinfo {year} {2023})}\BibitemShut
  {NoStop}%
\bibitem [{\citenamefont {Wuetz}\ \emph {et~al.}(2022)\citenamefont {Wuetz},
  \citenamefont {Losert}, \citenamefont {Koelling}, \citenamefont {Stehouwer},
  \citenamefont {Zwerver}, \citenamefont {Philips}, \citenamefont
  {M{\k{a}}dzik}, \citenamefont {Xue}, \citenamefont {Zheng}, \citenamefont
  {Lodari}, \citenamefont {Amitonov}, \citenamefont {Samkharadze},
  \citenamefont {Sammak}, \citenamefont {Vandersypen}, \citenamefont {Rahman},
  \citenamefont {Coppersmith}, \citenamefont {Moutanabbir}, \citenamefont
  {Friesen},\ and\ \citenamefont {Scappucci}}]{Wuetz2022}%
  \BibitemOpen
  \bibfield  {author} {\bibinfo {author} {\bibfnamefont {B.~P.}\ \bibnamefont
  {Wuetz}}, \bibinfo {author} {\bibfnamefont {M.~P.}\ \bibnamefont {Losert}},
  \bibinfo {author} {\bibfnamefont {S.}~\bibnamefont {Koelling}}, \bibinfo
  {author} {\bibfnamefont {L.~E.~A.}\ \bibnamefont {Stehouwer}}, \bibinfo
  {author} {\bibfnamefont {A.-M.~J.}\ \bibnamefont {Zwerver}}, \bibinfo
  {author} {\bibfnamefont {S.~G.~J.}\ \bibnamefont {Philips}}, \bibinfo
  {author} {\bibfnamefont {M.~T.}\ \bibnamefont {M{\k{a}}dzik}}, \bibinfo
  {author} {\bibfnamefont {X.}~\bibnamefont {Xue}}, \bibinfo {author}
  {\bibfnamefont {G.}~\bibnamefont {Zheng}}, \bibinfo {author} {\bibfnamefont
  {M.}~\bibnamefont {Lodari}}, \bibinfo {author} {\bibfnamefont {S.~V.}\
  \bibnamefont {Amitonov}}, \bibinfo {author} {\bibfnamefont {N.}~\bibnamefont
  {Samkharadze}}, \bibinfo {author} {\bibfnamefont {A.}~\bibnamefont {Sammak}},
  \bibinfo {author} {\bibfnamefont {L.~M.~K.}\ \bibnamefont {Vandersypen}},
  \bibinfo {author} {\bibfnamefont {R.}~\bibnamefont {Rahman}}, \bibinfo
  {author} {\bibfnamefont {S.~N.}\ \bibnamefont {Coppersmith}}, \bibinfo
  {author} {\bibfnamefont {O.}~\bibnamefont {Moutanabbir}}, \bibinfo {author}
  {\bibfnamefont {M.}~\bibnamefont {Friesen}}, \ and\ \bibinfo {author}
  {\bibfnamefont {G.}~\bibnamefont {Scappucci}},\ }\href {\doibase
  10.1038/s41467-022-35458-0} {\bibfield  {journal} {\bibinfo  {journal} {Nat.
  Commun.}\ }\textbf {\bibinfo {volume} {13}},\ \bibinfo {pages} {7730}
  (\bibinfo {year} {2022})}\BibitemShut {NoStop}%
\bibitem [{\citenamefont {Yang}\ \emph {et~al.}(2020)\citenamefont {Yang},
  \citenamefont {Leon}, \citenamefont {Hwang}, \citenamefont {Saraiva},
  \citenamefont {Tanttu}, \citenamefont {Huang}, \citenamefont {Lemyre},
  \citenamefont {Chan}, \citenamefont {Tan}, \citenamefont {Hudson},
  \citenamefont {Itoh}, \citenamefont {Morello}, \citenamefont
  {Pioro-Ladri{\`{e}}re}, \citenamefont {Laucht},\ and\ \citenamefont
  {Dzurak}}]{Yang2020}%
  \BibitemOpen
  \bibfield  {author} {\bibinfo {author} {\bibfnamefont {C.~H.}\ \bibnamefont
  {Yang}}, \bibinfo {author} {\bibfnamefont {R.~C.~C.}\ \bibnamefont {Leon}},
  \bibinfo {author} {\bibfnamefont {J.~C.~C.}\ \bibnamefont {Hwang}}, \bibinfo
  {author} {\bibfnamefont {A.}~\bibnamefont {Saraiva}}, \bibinfo {author}
  {\bibfnamefont {T.}~\bibnamefont {Tanttu}}, \bibinfo {author} {\bibfnamefont
  {W.}~\bibnamefont {Huang}}, \bibinfo {author} {\bibfnamefont {J.~C.}\
  \bibnamefont {Lemyre}}, \bibinfo {author} {\bibfnamefont {K.~W.}\
  \bibnamefont {Chan}}, \bibinfo {author} {\bibfnamefont {K.~Y.}\ \bibnamefont
  {Tan}}, \bibinfo {author} {\bibfnamefont {F.~E.}\ \bibnamefont {Hudson}},
  \bibinfo {author} {\bibfnamefont {K.~M.}\ \bibnamefont {Itoh}}, \bibinfo
  {author} {\bibfnamefont {A.}~\bibnamefont {Morello}}, \bibinfo {author}
  {\bibfnamefont {M.}~\bibnamefont {Pioro-Ladri{\`{e}}re}}, \bibinfo {author}
  {\bibfnamefont {A.}~\bibnamefont {Laucht}}, \ and\ \bibinfo {author}
  {\bibfnamefont {A.~S.}\ \bibnamefont {Dzurak}},\ }\href {\doibase
  10.1038/s41586-020-2171-6} {\bibfield  {journal} {\bibinfo  {journal}
  {Nature}\ }\textbf {\bibinfo {volume} {580}},\ \bibinfo {pages} {350}
  (\bibinfo {year} {2020})}\BibitemShut {NoStop}%
\bibitem [{\citenamefont {McJunkin}\ \emph {et~al.}(2022)\citenamefont
  {McJunkin}, \citenamefont {Harpt}, \citenamefont {Feng}, \citenamefont
  {Losert}, \citenamefont {Rahman}, \citenamefont {Dodson}, \citenamefont
  {Wolfe}, \citenamefont {Savage}, \citenamefont {Lagally}, \citenamefont
  {Coppersmith}, \citenamefont {Friesen}, \citenamefont {Joynt},\ and\
  \citenamefont {Eriksson}}]{McJunkin2022}%
  \BibitemOpen
  \bibfield  {author} {\bibinfo {author} {\bibfnamefont {T.}~\bibnamefont
  {McJunkin}}, \bibinfo {author} {\bibfnamefont {B.}~\bibnamefont {Harpt}},
  \bibinfo {author} {\bibfnamefont {Y.}~\bibnamefont {Feng}}, \bibinfo {author}
  {\bibfnamefont {M.~P.}\ \bibnamefont {Losert}}, \bibinfo {author}
  {\bibfnamefont {R.}~\bibnamefont {Rahman}}, \bibinfo {author} {\bibfnamefont
  {J.~P.}\ \bibnamefont {Dodson}}, \bibinfo {author} {\bibfnamefont {M.~A.}\
  \bibnamefont {Wolfe}}, \bibinfo {author} {\bibfnamefont {D.~E.}\ \bibnamefont
  {Savage}}, \bibinfo {author} {\bibfnamefont {M.~G.}\ \bibnamefont {Lagally}},
  \bibinfo {author} {\bibfnamefont {S.~N.}\ \bibnamefont {Coppersmith}},
  \bibinfo {author} {\bibfnamefont {M.}~\bibnamefont {Friesen}}, \bibinfo
  {author} {\bibfnamefont {R.}~\bibnamefont {Joynt}}, \ and\ \bibinfo {author}
  {\bibfnamefont {M.~A.}\ \bibnamefont {Eriksson}},\ }\href {\doibase
  10.1038/s41467-022-35510-z} {\bibfield  {journal} {\bibinfo  {journal} {Nat.
  Commun.}\ }\textbf {\bibinfo {volume} {13}},\ \bibinfo {pages} {7777}
  (\bibinfo {year} {2022})}\BibitemShut {NoStop}%
\bibitem [{\citenamefont {Woods}\ \emph {et~al.}(2023)\citenamefont {Woods},
  \citenamefont {Eriksson}, \citenamefont {Joynt},\ and\ \citenamefont
  {Friesen}}]{Woods2023}%
  \BibitemOpen
  \bibfield  {author} {\bibinfo {author} {\bibfnamefont {B.~D.}\ \bibnamefont
  {Woods}}, \bibinfo {author} {\bibfnamefont {M.~A.}\ \bibnamefont {Eriksson}},
  \bibinfo {author} {\bibfnamefont {R.}~\bibnamefont {Joynt}}, \ and\ \bibinfo
  {author} {\bibfnamefont {M.}~\bibnamefont {Friesen}},\ }\href {\doibase
  10.1103/physrevb.107.035418} {\bibfield  {journal} {\bibinfo  {journal}
  {Phys. Rev. B}\ }\textbf {\bibinfo {volume} {107}},\ \bibinfo {pages}
  {035418} (\bibinfo {year} {2023})}\BibitemShut {NoStop}%
\bibitem [{\citenamefont {Yang}\ \emph {et~al.}(2013)\citenamefont {Yang},
  \citenamefont {Rossi}, \citenamefont {Ruskov}, \citenamefont {Lai},
  \citenamefont {Mohiyaddin}, \citenamefont {Lee}, \citenamefont {Tahan},
  \citenamefont {Klimeck}, \citenamefont {Morello},\ and\ \citenamefont
  {Dzurak}}]{Yang2013}%
  \BibitemOpen
  \bibfield  {author} {\bibinfo {author} {\bibfnamefont {C.~H.}\ \bibnamefont
  {Yang}}, \bibinfo {author} {\bibfnamefont {A.}~\bibnamefont {Rossi}},
  \bibinfo {author} {\bibfnamefont {R.}~\bibnamefont {Ruskov}}, \bibinfo
  {author} {\bibfnamefont {N.~S.}\ \bibnamefont {Lai}}, \bibinfo {author}
  {\bibfnamefont {F.~A.}\ \bibnamefont {Mohiyaddin}}, \bibinfo {author}
  {\bibfnamefont {S.}~\bibnamefont {Lee}}, \bibinfo {author} {\bibfnamefont
  {C.}~\bibnamefont {Tahan}}, \bibinfo {author} {\bibfnamefont
  {G.}~\bibnamefont {Klimeck}}, \bibinfo {author} {\bibfnamefont
  {A.}~\bibnamefont {Morello}}, \ and\ \bibinfo {author} {\bibfnamefont
  {A.~S.}\ \bibnamefont {Dzurak}},\ }\href {\doibase 10.1038/ncomms3069}
  {\bibfield  {journal} {\bibinfo  {journal} {Nat. Commun.}\ }\textbf {\bibinfo
  {volume} {4}},\ \bibinfo {pages} {2069} (\bibinfo {year} {2013})}\BibitemShut
  {NoStop}%
\bibitem [{\citenamefont {Saraiva}\ \emph {et~al.}(2021)\citenamefont
  {Saraiva}, \citenamefont {Lim}, \citenamefont {Yang}, \citenamefont {Escott},
  \citenamefont {Laucht},\ and\ \citenamefont {Dzurak}}]{Saraiva2021}%
  \BibitemOpen
  \bibfield  {author} {\bibinfo {author} {\bibfnamefont {A.}~\bibnamefont
  {Saraiva}}, \bibinfo {author} {\bibfnamefont {W.~H.}\ \bibnamefont {Lim}},
  \bibinfo {author} {\bibfnamefont {C.~H.}\ \bibnamefont {Yang}}, \bibinfo
  {author} {\bibfnamefont {C.~C.}\ \bibnamefont {Escott}}, \bibinfo {author}
  {\bibfnamefont {A.}~\bibnamefont {Laucht}}, \ and\ \bibinfo {author}
  {\bibfnamefont {A.~S.}\ \bibnamefont {Dzurak}},\ }\href {\doibase
  10.1002/adfm.202105488} {\bibfield  {journal} {\bibinfo  {journal} {Adv.
  Funct. Mater.}\ }\textbf {\bibinfo {volume} {32}},\ \bibinfo {pages}
  {2105488} (\bibinfo {year} {2021})}\BibitemShut {NoStop}%
\bibitem [{\citenamefont {Cifuentes}\ \emph {et~al.}(2023)\citenamefont
  {Cifuentes}, \citenamefont {Tanttu}, \citenamefont {Gilbert}, \citenamefont
  {Huang}, \citenamefont {Vahapoglu}, \citenamefont {Leon}, \citenamefont
  {Serrano}, \citenamefont {Otter}, \citenamefont {Dunmore}, \citenamefont
  {Mai}, \citenamefont {Schlattner}, \citenamefont {Feng}, \citenamefont
  {Itoh}, \citenamefont {Abrosimov}, \citenamefont {Pohl}, \citenamefont
  {Thewalt}, \citenamefont {Laucht}, \citenamefont {Yang}, \citenamefont
  {Escott}, \citenamefont {Lim}, \citenamefont {Hudson}, \citenamefont
  {Rahman}, \citenamefont {Saraiva},\ and\ \citenamefont
  {Dzurak}}]{Cifuentes2023}%
  \BibitemOpen
  \bibfield  {author} {\bibinfo {author} {\bibfnamefont {J.~D.}\ \bibnamefont
  {Cifuentes}}, \bibinfo {author} {\bibfnamefont {T.}~\bibnamefont {Tanttu}},
  \bibinfo {author} {\bibfnamefont {W.}~\bibnamefont {Gilbert}}, \bibinfo
  {author} {\bibfnamefont {J.~Y.}\ \bibnamefont {Huang}}, \bibinfo {author}
  {\bibfnamefont {E.}~\bibnamefont {Vahapoglu}}, \bibinfo {author}
  {\bibfnamefont {R.~C.~C.}\ \bibnamefont {Leon}}, \bibinfo {author}
  {\bibfnamefont {S.}~\bibnamefont {Serrano}}, \bibinfo {author} {\bibfnamefont
  {D.}~\bibnamefont {Otter}}, \bibinfo {author} {\bibfnamefont
  {D.}~\bibnamefont {Dunmore}}, \bibinfo {author} {\bibfnamefont {P.~Y.}\
  \bibnamefont {Mai}}, \bibinfo {author} {\bibfnamefont {F.}~\bibnamefont
  {Schlattner}}, \bibinfo {author} {\bibfnamefont {M.}~\bibnamefont {Feng}},
  \bibinfo {author} {\bibfnamefont {K.}~\bibnamefont {Itoh}}, \bibinfo {author}
  {\bibfnamefont {N.}~\bibnamefont {Abrosimov}}, \bibinfo {author}
  {\bibfnamefont {H.-J.}\ \bibnamefont {Pohl}}, \bibinfo {author}
  {\bibfnamefont {M.}~\bibnamefont {Thewalt}}, \bibinfo {author} {\bibfnamefont
  {A.}~\bibnamefont {Laucht}}, \bibinfo {author} {\bibfnamefont {C.~H.}\
  \bibnamefont {Yang}}, \bibinfo {author} {\bibfnamefont {C.~C.}\ \bibnamefont
  {Escott}}, \bibinfo {author} {\bibfnamefont {W.~H.}\ \bibnamefont {Lim}},
  \bibinfo {author} {\bibfnamefont {F.~E.}\ \bibnamefont {Hudson}}, \bibinfo
  {author} {\bibfnamefont {R.}~\bibnamefont {Rahman}}, \bibinfo {author}
  {\bibfnamefont {A.}~\bibnamefont {Saraiva}}, \ and\ \bibinfo {author}
  {\bibfnamefont {A.~S.}\ \bibnamefont {Dzurak}},\ }\href@noop {} {\  (\bibinfo
  {year} {2023})},\ \Eprint {http://arxiv.org/abs/2303.14864} {arXiv:2303.14864
  [quant-ph]} \BibitemShut {NoStop}%
\bibitem [{\citenamefont {Choudhary}\ \emph {et~al.}(2023)\citenamefont
  {Choudhary}, \citenamefont {Yogesh}, \citenamefont {Schwarz}, \citenamefont
  {Funk}, \citenamefont {Ghosh}, \citenamefont {Sharma}, \citenamefont
  {Schulze},\ and\ \citenamefont {Gonsalves}}]{Choudhary2023}%
  \BibitemOpen
  \bibfield  {author} {\bibinfo {author} {\bibfnamefont {S.}~\bibnamefont
  {Choudhary}}, \bibinfo {author} {\bibfnamefont {M.}~\bibnamefont {Yogesh}},
  \bibinfo {author} {\bibfnamefont {D.}~\bibnamefont {Schwarz}}, \bibinfo
  {author} {\bibfnamefont {H.~S.}\ \bibnamefont {Funk}}, \bibinfo {author}
  {\bibfnamefont {S.}~\bibnamefont {Ghosh}}, \bibinfo {author} {\bibfnamefont
  {S.~K.}\ \bibnamefont {Sharma}}, \bibinfo {author} {\bibfnamefont
  {J.}~\bibnamefont {Schulze}}, \ and\ \bibinfo {author} {\bibfnamefont
  {K.~E.}\ \bibnamefont {Gonsalves}},\ }\href {\doibase 10.1116/6.0002767}
  {\bibfield  {journal} {\bibinfo  {journal} {J. Vac. Sci. Technol. B}\
  }\textbf {\bibinfo {volume} {41}},\ \bibinfo {pages} {052203} (\bibinfo
  {year} {2023})}\BibitemShut {NoStop}%
\bibitem [{\citenamefont {Ramanandan}\ \emph {et~al.}(2024)\citenamefont
  {Ramanandan}, \citenamefont {Re\~{n}\'{e}~Sapera}, \citenamefont {Morelle},
  \citenamefont {Mart\'{i}-S\'{a}nchez}, \citenamefont {Rudra}, \citenamefont
  {Arbiol}, \citenamefont {Dubrovskii},\ and\ \citenamefont {Fontcuberta~i
  Morral}}]{Ramanandan2024}%
  \BibitemOpen
  \bibfield  {author} {\bibinfo {author} {\bibfnamefont {S.~P.}\ \bibnamefont
  {Ramanandan}}, \bibinfo {author} {\bibfnamefont {J.}~\bibnamefont
  {Re\~{n}\'{e}~Sapera}}, \bibinfo {author} {\bibfnamefont {A.}~\bibnamefont
  {Morelle}}, \bibinfo {author} {\bibfnamefont {S.}~\bibnamefont
  {Mart\'{i}-S\'{a}nchez}}, \bibinfo {author} {\bibfnamefont {A.}~\bibnamefont
  {Rudra}}, \bibinfo {author} {\bibfnamefont {J.}~\bibnamefont {Arbiol}},
  \bibinfo {author} {\bibfnamefont {V.~G.}\ \bibnamefont {Dubrovskii}}, \ and\
  \bibinfo {author} {\bibfnamefont {A.}~\bibnamefont {Fontcuberta~i Morral}},\
  }\href {\doibase 10.1039/d3nh00573a} {\bibfield  {journal} {\bibinfo
  {journal} {Nanoscale Horiz.}\ }\textbf {\bibinfo {volume} {9}},\ \bibinfo
  {pages} {555} (\bibinfo {year} {2024})}\BibitemShut {NoStop}%
\bibitem [{\citenamefont {Reeber}\ and\ \citenamefont
  {Wang}(1996)}]{Reeber1996}%
  \BibitemOpen
  \bibfield  {author} {\bibinfo {author} {\bibfnamefont {R.~R.}\ \bibnamefont
  {Reeber}}\ and\ \bibinfo {author} {\bibfnamefont {K.}~\bibnamefont {Wang}},\
  }\href {\doibase 10.1016/s0254-0584(96)01808-1} {\bibfield  {journal}
  {\bibinfo  {journal} {Mater. Chem. Phys.}\ }\textbf {\bibinfo {volume}
  {46}},\ \bibinfo {pages} {259} (\bibinfo {year} {1996})}\BibitemShut
  {NoStop}%
\bibitem [{\citenamefont {Hensel}\ \emph {et~al.}(1965)\citenamefont {Hensel},
  \citenamefont {Hasegawa},\ and\ \citenamefont {Nakayama}}]{Hensel1965}%
  \BibitemOpen
  \bibfield  {author} {\bibinfo {author} {\bibfnamefont {J.~C.}\ \bibnamefont
  {Hensel}}, \bibinfo {author} {\bibfnamefont {H.}~\bibnamefont {Hasegawa}}, \
  and\ \bibinfo {author} {\bibfnamefont {M.}~\bibnamefont {Nakayama}},\ }\href
  {\doibase 10.1103/physrev.138.a225} {\bibfield  {journal} {\bibinfo
  {journal} {Phys. Rev.}\ }\textbf {\bibinfo {volume} {138}},\ \bibinfo {pages}
  {A225} (\bibinfo {year} {1965})}\BibitemShut {NoStop}%
\bibitem [{\citenamefont {Stanojevic}\ \emph {et~al.}(2010)\citenamefont
  {Stanojevic}, \citenamefont {Baumgartner}, \citenamefont {Sverdlov},\ and\
  \citenamefont {Kosina}}]{Stanojevic2010}%
  \BibitemOpen
  \bibfield  {author} {\bibinfo {author} {\bibfnamefont {Z.}~\bibnamefont
  {Stanojevic}}, \bibinfo {author} {\bibfnamefont {O.}~\bibnamefont
  {Baumgartner}}, \bibinfo {author} {\bibfnamefont {V.}~\bibnamefont
  {Sverdlov}}, \ and\ \bibinfo {author} {\bibfnamefont {H.}~\bibnamefont
  {Kosina}},\ }in\ \href {\doibase 10.1109/iwce.2010.5677927} {\emph {\bibinfo
  {booktitle} {2010 14th International Workshop on Computational
  Electronics}}}\ (\bibinfo  {publisher} {{IEEE}},\ \bibinfo {year}
  {2010})\BibitemShut {NoStop}%
\bibitem [{\citenamefont {Sverdlov}\ \emph {et~al.}(2008)\citenamefont
  {Sverdlov}, \citenamefont {Karlowatz}, \citenamefont {Dhar}, \citenamefont
  {Kosina},\ and\ \citenamefont {Selberherr}}]{Sverdlov2008}%
  \BibitemOpen
  \bibfield  {author} {\bibinfo {author} {\bibfnamefont {V.}~\bibnamefont
  {Sverdlov}}, \bibinfo {author} {\bibfnamefont {G.}~\bibnamefont {Karlowatz}},
  \bibinfo {author} {\bibfnamefont {S.}~\bibnamefont {Dhar}}, \bibinfo {author}
  {\bibfnamefont {H.}~\bibnamefont {Kosina}}, \ and\ \bibinfo {author}
  {\bibfnamefont {S.}~\bibnamefont {Selberherr}},\ }\href {\doibase
  10.1016/j.sse.2008.06.019} {\bibfield  {journal} {\bibinfo  {journal}
  {Solid-State Electron.}\ }\textbf {\bibinfo {volume} {52}},\ \bibinfo {pages}
  {1563} (\bibinfo {year} {2008})}\BibitemShut {NoStop}%
\bibitem [{\citenamefont {Fischetti}\ and\ \citenamefont
  {Laux}(1996)}]{Fischetti1996}%
  \BibitemOpen
  \bibfield  {author} {\bibinfo {author} {\bibfnamefont {M.~V.}\ \bibnamefont
  {Fischetti}}\ and\ \bibinfo {author} {\bibfnamefont {S.~E.}\ \bibnamefont
  {Laux}},\ }\href {\doibase 10.1063/1.363052} {\bibfield  {journal} {\bibinfo
  {journal} {J. Appl. Phys.}\ }\textbf {\bibinfo {volume} {80}},\ \bibinfo
  {pages} {2234} (\bibinfo {year} {1996})}\BibitemShut {NoStop}%
\bibitem [{\citenamefont {{Van de Walle}}\ and\ \citenamefont
  {Martin}(1986)}]{Walle1986}%
  \BibitemOpen
  \bibfield  {author} {\bibinfo {author} {\bibfnamefont {C.~G.}\ \bibnamefont
  {{Van de Walle}}}\ and\ \bibinfo {author} {\bibfnamefont {R.~M.}\
  \bibnamefont {Martin}},\ }\href {\doibase 10.1103/physrevb.34.5621}
  {\bibfield  {journal} {\bibinfo  {journal} {Phys. Rev. B}\ }\textbf {\bibinfo
  {volume} {34}},\ \bibinfo {pages} {5621} (\bibinfo {year}
  {1986})}\BibitemShut {NoStop}%
\bibitem [{\citenamefont {Tserbak}\ \emph {et~al.}(1993)\citenamefont
  {Tserbak}, \citenamefont {Polatoglou},\ and\ \citenamefont
  {Theodorou}}]{Tserbak1993}%
  \BibitemOpen
  \bibfield  {author} {\bibinfo {author} {\bibfnamefont {C.}~\bibnamefont
  {Tserbak}}, \bibinfo {author} {\bibfnamefont {H.~M.}\ \bibnamefont
  {Polatoglou}}, \ and\ \bibinfo {author} {\bibfnamefont {G.}~\bibnamefont
  {Theodorou}},\ }\href {\doibase 10.1103/physrevb.47.7104} {\bibfield
  {journal} {\bibinfo  {journal} {Phys. Rev. B}\ }\textbf {\bibinfo {volume}
  {47}},\ \bibinfo {pages} {7104} (\bibinfo {year} {1993})}\BibitemShut
  {NoStop}%
\bibitem [{\citenamefont {Friedel}\ \emph {et~al.}(1989)\citenamefont
  {Friedel}, \citenamefont {Hybertsen},\ and\ \citenamefont
  {Schl\"uter}}]{Friedel1989}%
  \BibitemOpen
  \bibfield  {author} {\bibinfo {author} {\bibfnamefont {P.}~\bibnamefont
  {Friedel}}, \bibinfo {author} {\bibfnamefont {M.~S.}\ \bibnamefont
  {Hybertsen}}, \ and\ \bibinfo {author} {\bibfnamefont {M.}~\bibnamefont
  {Schl\"uter}},\ }\href {\doibase 10.1103/physrevb.39.7974} {\bibfield
  {journal} {\bibinfo  {journal} {Phys. Rev. B}\ }\textbf {\bibinfo {volume}
  {39}},\ \bibinfo {pages} {7974} (\bibinfo {year} {1989})}\BibitemShut
  {NoStop}%
\bibitem [{\citenamefont {Balslev}(1966)}]{Balslev1966}%
  \BibitemOpen
  \bibfield  {author} {\bibinfo {author} {\bibfnamefont {I.}~\bibnamefont
  {Balslev}},\ }\href {\doibase 10.1103/physrev.143.636} {\bibfield  {journal}
  {\bibinfo  {journal} {Phys. Rev.}\ }\textbf {\bibinfo {volume} {143}},\
  \bibinfo {pages} {636} (\bibinfo {year} {1966})}\BibitemShut {NoStop}%
\bibitem [{\citenamefont {Rieger}\ and\ \citenamefont
  {Vogl}(1993)}]{Rieger1993}%
  \BibitemOpen
  \bibfield  {author} {\bibinfo {author} {\bibfnamefont {M.~M.}\ \bibnamefont
  {Rieger}}\ and\ \bibinfo {author} {\bibfnamefont {P.}~\bibnamefont {Vogl}},\
  }\href {\doibase 10.1103/physrevb.48.14276} {\bibfield  {journal} {\bibinfo
  {journal} {Phys. Rev. B}\ }\textbf {\bibinfo {volume} {48}},\ \bibinfo
  {pages} {14276} (\bibinfo {year} {1993})}\BibitemShut {NoStop}%
\bibitem [{\citenamefont {Goroff}\ and\ \citenamefont
  {Kleinman}(1963)}]{Goroff1963}%
  \BibitemOpen
  \bibfield  {author} {\bibinfo {author} {\bibfnamefont {I.}~\bibnamefont
  {Goroff}}\ and\ \bibinfo {author} {\bibfnamefont {L.}~\bibnamefont
  {Kleinman}},\ }\href {\doibase 10.1103/physrev.132.1080} {\bibfield
  {journal} {\bibinfo  {journal} {Phys. Rev.}\ }\textbf {\bibinfo {volume}
  {132}},\ \bibinfo {pages} {1080} (\bibinfo {year} {1963})}\BibitemShut
  {NoStop}%
\bibitem [{\citenamefont {Laude}\ \emph {et~al.}(1971)\citenamefont {Laude},
  \citenamefont {Pollak},\ and\ \citenamefont {Cardona}}]{Laude1971}%
  \BibitemOpen
  \bibfield  {author} {\bibinfo {author} {\bibfnamefont {L.~D.}\ \bibnamefont
  {Laude}}, \bibinfo {author} {\bibfnamefont {F.~H.}\ \bibnamefont {Pollak}}, \
  and\ \bibinfo {author} {\bibfnamefont {M.}~\bibnamefont {Cardona}},\ }\href
  {\doibase 10.1103/physrevb.3.2623} {\bibfield  {journal} {\bibinfo  {journal}
  {Phys. Rev. B}\ }\textbf {\bibinfo {volume} {3}},\ \bibinfo {pages} {2623}
  (\bibinfo {year} {1971})}\BibitemShut {NoStop}%
\bibitem [{\citenamefont {Li}\ \emph {et~al.}(2021)\citenamefont {Li},
  \citenamefont {Graziosi},\ and\ \citenamefont {Neophytou}}]{Li2021}%
  \BibitemOpen
  \bibfield  {author} {\bibinfo {author} {\bibfnamefont {Z.}~\bibnamefont
  {Li}}, \bibinfo {author} {\bibfnamefont {P.}~\bibnamefont {Graziosi}}, \ and\
  \bibinfo {author} {\bibfnamefont {N.}~\bibnamefont {Neophytou}},\ }\href
  {\doibase 10.1103/physrevb.104.195201} {\bibfield  {journal} {\bibinfo
  {journal} {Phys. Rev. B}\ }\textbf {\bibinfo {volume} {104}},\ \bibinfo
  {pages} {195201} (\bibinfo {year} {2021})}\BibitemShut {NoStop}%
\bibitem [{\citenamefont {Bosco}\ and\ \citenamefont {Loss}(2022)}]{Bosco2022}%
  \BibitemOpen
  \bibfield  {author} {\bibinfo {author} {\bibfnamefont {S.}~\bibnamefont
  {Bosco}}\ and\ \bibinfo {author} {\bibfnamefont {D.}~\bibnamefont {Loss}},\
  }\href {\doibase 10.1103/physrevapplied.18.044038} {\bibfield  {journal}
  {\bibinfo  {journal} {Phys. Rev. Appl.}\ }\textbf {\bibinfo {volume} {18}},\
  \bibinfo {pages} {044038} (\bibinfo {year} {2022})}\BibitemShut {NoStop}%
\bibitem [{\citenamefont {Kloeffel}\ \emph {et~al.}(2014)\citenamefont
  {Kloeffel}, \citenamefont {Trif},\ and\ \citenamefont {Loss}}]{Kloeffel2014}%
  \BibitemOpen
  \bibfield  {author} {\bibinfo {author} {\bibfnamefont {C.}~\bibnamefont
  {Kloeffel}}, \bibinfo {author} {\bibfnamefont {M.}~\bibnamefont {Trif}}, \
  and\ \bibinfo {author} {\bibfnamefont {D.}~\bibnamefont {Loss}},\ }\href
  {\doibase 10.1103/physrevb.90.115419} {\bibfield  {journal} {\bibinfo
  {journal} {Phys. Rev. B}\ }\textbf {\bibinfo {volume} {90}},\ \bibinfo
  {pages} {115419} (\bibinfo {year} {2014})}\BibitemShut {NoStop}%
\bibitem [{\citenamefont {Kosevich}\ \emph {et~al.}(1986)\citenamefont
  {Kosevich}, \citenamefont {Lifshitz}, \citenamefont {Landau},\ and\
  \citenamefont {Pitaevskii}}]{Kosevich1986}%
  \BibitemOpen
  \bibfield  {author} {\bibinfo {author} {\bibfnamefont {A.~M.}\ \bibnamefont
  {Kosevich}}, \bibinfo {author} {\bibfnamefont {E.~M.}\ \bibnamefont
  {Lifshitz}}, \bibinfo {author} {\bibfnamefont {L.~D.}\ \bibnamefont
  {Landau}}, \ and\ \bibinfo {author} {\bibfnamefont {L.~P.}\ \bibnamefont
  {Pitaevskii}},\ }\href@noop {} {\emph {\bibinfo {title} {Theory of
  Elasticity}}},\ Vol.~\bibinfo {volume} {7}\ (\bibinfo  {publisher}
  {Butterworth-Heinemann},\ \bibinfo {address} {Oxford},\ \bibinfo {year}
  {1986})\BibitemShut {NoStop}%
\bibitem [{\citenamefont {Niquet}\ \emph {et~al.}(2012)\citenamefont {Niquet},
  \citenamefont {Delerue},\ and\ \citenamefont {Krzeminski}}]{Niquet2012}%
  \BibitemOpen
  \bibfield  {author} {\bibinfo {author} {\bibfnamefont {Y.-M.}\ \bibnamefont
  {Niquet}}, \bibinfo {author} {\bibfnamefont {C.}~\bibnamefont {Delerue}}, \
  and\ \bibinfo {author} {\bibfnamefont {C.}~\bibnamefont {Krzeminski}},\
  }\href {\doibase 10.1021/nl3010995} {\bibfield  {journal} {\bibinfo
  {journal} {Nano Lett.}\ }\textbf {\bibinfo {volume} {12}},\ \bibinfo {pages}
  {3545} (\bibinfo {year} {2012})}\BibitemShut {NoStop}%
\bibitem [{smS()}]{smStrain}%
  \BibitemOpen
  \href@noop {} {\enquote {\bibinfo {title} {{See Supplemental Material, which
  includes Refs.~\cite{Mengistu2016,McSkimin1964,McSkimin1963}, for a
  discussion of the model for strain used for FEM calculations and results for
  the uniaxial strain components, atomic size steps at the interface between
  the {Ge fin and the Si} shell, more details on the electric field dependence
  of the valley splitting, a discussion of the robustness of the valley
  splitting, and a semicylindrical device where a sizalbe valley splitting is
  found in a broad parameter regime of electric field strengths and Ge
  concentrations as well}},}\ }\BibitemShut {NoStop}%
\bibitem [{\citenamefont {Bosco}\ \emph {et~al.}(2021)\citenamefont {Bosco},
  \citenamefont {Benito}, \citenamefont {Adelsberger},\ and\ \citenamefont
  {Loss}}]{Bosco2021}%
  \BibitemOpen
  \bibfield  {author} {\bibinfo {author} {\bibfnamefont {S.}~\bibnamefont
  {Bosco}}, \bibinfo {author} {\bibfnamefont {M.}~\bibnamefont {Benito}},
  \bibinfo {author} {\bibfnamefont {C.}~\bibnamefont {Adelsberger}}, \ and\
  \bibinfo {author} {\bibfnamefont {D.}~\bibnamefont {Loss}},\ }\href {\doibase
  https://doi.org/10.1103/PhysRevB.104.115425} {\bibfield  {journal} {\bibinfo
  {journal} {Phys. Rev. B}\ }\textbf {\bibinfo {volume} {104}},\ \bibinfo
  {pages} {115425} (\bibinfo {year} {2021})}\BibitemShut {NoStop}%
\bibitem [{\citenamefont {Terrazos}\ \emph {et~al.}(2021)\citenamefont
  {Terrazos}, \citenamefont {Marcellina}, \citenamefont {Wang}, \citenamefont
  {Coppersmith}, \citenamefont {Friesen}, \citenamefont {Hamilton},
  \citenamefont {Hu}, \citenamefont {Koiller}, \citenamefont {Saraiva},
  \citenamefont {Culcer},\ and\ \citenamefont {Capaz}}]{Terrazos2021}%
  \BibitemOpen
  \bibfield  {author} {\bibinfo {author} {\bibfnamefont {L.~A.}\ \bibnamefont
  {Terrazos}}, \bibinfo {author} {\bibfnamefont {E.}~\bibnamefont
  {Marcellina}}, \bibinfo {author} {\bibfnamefont {Z.}~\bibnamefont {Wang}},
  \bibinfo {author} {\bibfnamefont {S.~N.}\ \bibnamefont {Coppersmith}},
  \bibinfo {author} {\bibfnamefont {M.}~\bibnamefont {Friesen}}, \bibinfo
  {author} {\bibfnamefont {A.~R.}\ \bibnamefont {Hamilton}}, \bibinfo {author}
  {\bibfnamefont {X.}~\bibnamefont {Hu}}, \bibinfo {author} {\bibfnamefont
  {B.}~\bibnamefont {Koiller}}, \bibinfo {author} {\bibfnamefont {A.~L.}\
  \bibnamefont {Saraiva}}, \bibinfo {author} {\bibfnamefont {D.}~\bibnamefont
  {Culcer}}, \ and\ \bibinfo {author} {\bibfnamefont {R.~B.}\ \bibnamefont
  {Capaz}},\ }\href {\doibase 10.1103/physrevb.103.125201} {\bibfield
  {journal} {\bibinfo  {journal} {Phys. Rev. B}\ }\textbf {\bibinfo {volume}
  {103}},\ \bibinfo {pages} {125201} (\bibinfo {year} {2021})}\BibitemShut
  {NoStop}%
\bibitem [{\citenamefont {Maurand}\ \emph {et~al.}(2016)\citenamefont
  {Maurand}, \citenamefont {Jehl}, \citenamefont {Kotekar-Patil}, \citenamefont
  {Corna}, \citenamefont {Bohuslavskyi}, \citenamefont {Lavi{\'{e}}ville},
  \citenamefont {Hutin}, \citenamefont {Barraud}, \citenamefont {Vinet},
  \citenamefont {Sanquer},\ and\ \citenamefont {Franceschi}}]{Maurand2016}%
  \BibitemOpen
  \bibfield  {author} {\bibinfo {author} {\bibfnamefont {R.}~\bibnamefont
  {Maurand}}, \bibinfo {author} {\bibfnamefont {X.}~\bibnamefont {Jehl}},
  \bibinfo {author} {\bibfnamefont {D.}~\bibnamefont {Kotekar-Patil}}, \bibinfo
  {author} {\bibfnamefont {A.}~\bibnamefont {Corna}}, \bibinfo {author}
  {\bibfnamefont {H.}~\bibnamefont {Bohuslavskyi}}, \bibinfo {author}
  {\bibfnamefont {R.}~\bibnamefont {Lavi{\'{e}}ville}}, \bibinfo {author}
  {\bibfnamefont {L.}~\bibnamefont {Hutin}}, \bibinfo {author} {\bibfnamefont
  {S.}~\bibnamefont {Barraud}}, \bibinfo {author} {\bibfnamefont
  {M.}~\bibnamefont {Vinet}}, \bibinfo {author} {\bibfnamefont
  {M.}~\bibnamefont {Sanquer}}, \ and\ \bibinfo {author} {\bibfnamefont
  {S.~D.}\ \bibnamefont {Franceschi}},\ }\href {\doibase 10.1038/ncomms13575}
  {\bibfield  {journal} {\bibinfo  {journal} {Nat. Commun.}\ }\textbf {\bibinfo
  {volume} {7}},\ \bibinfo {pages} {13575} (\bibinfo {year}
  {2016})}\BibitemShut {NoStop}%
\bibitem [{COM()}]{COMSOL}%
  \BibitemOpen
  \href@noop {} {\enquote {\bibinfo {title} {{COMSOL
  Multiphysics{\textregistered} v. 6.1. \url{www.comsol.com}. COMSOL AB,
  Stockholm, Sweden.}}}\ }\BibitemShut {NoStop}%
\bibitem [{\citenamefont {Hong}\ \emph {et~al.}(2008)\citenamefont {Hong},
  \citenamefont {Kim}, \citenamefont {Lee},\ and\ \citenamefont
  {Shin}}]{Hong2008}%
  \BibitemOpen
  \bibfield  {author} {\bibinfo {author} {\bibfnamefont {K.-H.}\ \bibnamefont
  {Hong}}, \bibinfo {author} {\bibfnamefont {J.}~\bibnamefont {Kim}}, \bibinfo
  {author} {\bibfnamefont {S.-H.}\ \bibnamefont {Lee}}, \ and\ \bibinfo
  {author} {\bibfnamefont {J.~K.}\ \bibnamefont {Shin}},\ }\href {\doibase
  10.1021/nl0734140} {\bibfield  {journal} {\bibinfo  {journal} {Nano Lett.}\
  }\textbf {\bibinfo {volume} {8}},\ \bibinfo {pages} {1335} (\bibinfo {year}
  {2008})}\BibitemShut {NoStop}%
\bibitem [{\citenamefont {Huang}\ and\ \citenamefont {Hu}(2014)}]{Huang2014}%
  \BibitemOpen
  \bibfield  {author} {\bibinfo {author} {\bibfnamefont {P.}~\bibnamefont
  {Huang}}\ and\ \bibinfo {author} {\bibfnamefont {X.}~\bibnamefont {Hu}},\
  }\href {\doibase 10.1103/physrevb.90.235315} {\bibfield  {journal} {\bibinfo
  {journal} {Phys. Rev. B}\ }\textbf {\bibinfo {volume} {90}},\ \bibinfo
  {pages} {235315} (\bibinfo {year} {2014})}\BibitemShut {NoStop}%
\bibitem [{\citenamefont {Boykin}\ \emph {et~al.}(2004)\citenamefont {Boykin},
  \citenamefont {Klimeck}, \citenamefont {Eriksson}, \citenamefont {Friesen},
  \citenamefont {Coppersmith}, \citenamefont {von Allmen}, \citenamefont
  {Oyafuso},\ and\ \citenamefont {Lee}}]{Boykin2004}%
  \BibitemOpen
  \bibfield  {author} {\bibinfo {author} {\bibfnamefont {T.~B.}\ \bibnamefont
  {Boykin}}, \bibinfo {author} {\bibfnamefont {G.}~\bibnamefont {Klimeck}},
  \bibinfo {author} {\bibfnamefont {M.~A.}\ \bibnamefont {Eriksson}}, \bibinfo
  {author} {\bibfnamefont {M.}~\bibnamefont {Friesen}}, \bibinfo {author}
  {\bibfnamefont {S.~N.}\ \bibnamefont {Coppersmith}}, \bibinfo {author}
  {\bibfnamefont {P.}~\bibnamefont {von Allmen}}, \bibinfo {author}
  {\bibfnamefont {F.}~\bibnamefont {Oyafuso}}, \ and\ \bibinfo {author}
  {\bibfnamefont {S.}~\bibnamefont {Lee}},\ }\href {\doibase 10.1063/1.1637718}
  {\bibfield  {journal} {\bibinfo  {journal} {Appl. Phys. Lett.}\ }\textbf
  {\bibinfo {volume} {84}},\ \bibinfo {pages} {115} (\bibinfo {year}
  {2004})}\BibitemShut {NoStop}%
\bibitem [{\citenamefont {Yu}\ \emph {et~al.}(2023)\citenamefont {Yu},
  \citenamefont {Zihlmann}, \citenamefont {Abadillo-Uriel}, \citenamefont
  {Michal}, \citenamefont {Rambal}, \citenamefont {Niebojewski}, \citenamefont
  {Bedecarrats}, \citenamefont {Vinet}, \citenamefont {Dumur}, \citenamefont
  {Filippone}, \citenamefont {Bertrand}, \citenamefont {Franceschi},
  \citenamefont {Niquet},\ and\ \citenamefont {Maurand}}]{Yu2023}%
  \BibitemOpen
  \bibfield  {author} {\bibinfo {author} {\bibfnamefont {C.~X.}\ \bibnamefont
  {Yu}}, \bibinfo {author} {\bibfnamefont {S.}~\bibnamefont {Zihlmann}},
  \bibinfo {author} {\bibfnamefont {J.~C.}\ \bibnamefont {Abadillo-Uriel}},
  \bibinfo {author} {\bibfnamefont {V.~P.}\ \bibnamefont {Michal}}, \bibinfo
  {author} {\bibfnamefont {N.}~\bibnamefont {Rambal}}, \bibinfo {author}
  {\bibfnamefont {H.}~\bibnamefont {Niebojewski}}, \bibinfo {author}
  {\bibfnamefont {T.}~\bibnamefont {Bedecarrats}}, \bibinfo {author}
  {\bibfnamefont {M.}~\bibnamefont {Vinet}}, \bibinfo {author} {\bibfnamefont
  {{\'{E}}.}~\bibnamefont {Dumur}}, \bibinfo {author} {\bibfnamefont
  {M.}~\bibnamefont {Filippone}}, \bibinfo {author} {\bibfnamefont
  {B.}~\bibnamefont {Bertrand}}, \bibinfo {author} {\bibfnamefont {S.~D.}\
  \bibnamefont {Franceschi}}, \bibinfo {author} {\bibfnamefont {Y.-M.}\
  \bibnamefont {Niquet}}, \ and\ \bibinfo {author} {\bibfnamefont
  {R.}~\bibnamefont {Maurand}},\ }\href {\doibase 10.1038/s41565-023-01332-3}
  {\bibfield  {journal} {\bibinfo  {journal} {Nat. Nanotechnol.}\ }\textbf
  {\bibinfo {volume} {18}},\ \bibinfo {pages} {741} (\bibinfo {year}
  {2023})}\BibitemShut {NoStop}%
\bibitem [{\citenamefont {Gonzalez-Zalba}\ \emph {et~al.}(2021)\citenamefont
  {Gonzalez-Zalba}, \citenamefont {de~Franceschi}, \citenamefont {Charbon},
  \citenamefont {Meunier}, \citenamefont {Vinet},\ and\ \citenamefont
  {Dzurak}}]{GonzalezZalba2021}%
  \BibitemOpen
  \bibfield  {author} {\bibinfo {author} {\bibfnamefont {M.~F.}\ \bibnamefont
  {Gonzalez-Zalba}}, \bibinfo {author} {\bibfnamefont {S.}~\bibnamefont
  {de~Franceschi}}, \bibinfo {author} {\bibfnamefont {E.}~\bibnamefont
  {Charbon}}, \bibinfo {author} {\bibfnamefont {T.}~\bibnamefont {Meunier}},
  \bibinfo {author} {\bibfnamefont {M.}~\bibnamefont {Vinet}}, \ and\ \bibinfo
  {author} {\bibfnamefont {A.~S.}\ \bibnamefont {Dzurak}},\ }\href {\doibase
  10.1038/s41928-021-00681-y} {\bibfield  {journal} {\bibinfo  {journal} {Nat.
  Electron.}\ }\textbf {\bibinfo {volume} {4}},\ \bibinfo {pages} {872}
  (\bibinfo {year} {2021})}\BibitemShut {NoStop}%
\bibitem [{\citenamefont {Borjans}\ \emph {et~al.}(2019)\citenamefont
  {Borjans}, \citenamefont {Zajac}, \citenamefont {Hazard},\ and\ \citenamefont
  {Petta}}]{Borjans2019}%
  \BibitemOpen
  \bibfield  {author} {\bibinfo {author} {\bibfnamefont {F.}~\bibnamefont
  {Borjans}}, \bibinfo {author} {\bibfnamefont {D.~M.}\ \bibnamefont {Zajac}},
  \bibinfo {author} {\bibfnamefont {T.~M.}\ \bibnamefont {Hazard}}, \ and\
  \bibinfo {author} {\bibfnamefont {J.~R.}\ \bibnamefont {Petta}},\ }\href
  {\doibase 10.1103/physrevapplied.11.044063} {\bibfield  {journal} {\bibinfo
  {journal} {Phys. Rev. Appl.}\ }\textbf {\bibinfo {volume} {11}},\ \bibinfo
  {pages} {044063} (\bibinfo {year} {2019})}\BibitemShut {NoStop}%
\bibitem [{\citenamefont {Asaad}\ \emph {et~al.}(2020)\citenamefont {Asaad},
  \citenamefont {Mourik}, \citenamefont {Joecker}, \citenamefont {Johnson},
  \citenamefont {Baczewski}, \citenamefont {Firgau}, \citenamefont {Mądzik},
  \citenamefont {Schmitt}, \citenamefont {Pla}, \citenamefont {Hudson},
  \citenamefont {Itoh}, \citenamefont {McCallum}, \citenamefont {Dzurak},
  \citenamefont {Laucht},\ and\ \citenamefont {Morello}}]{Asaad2020}%
  \BibitemOpen
  \bibfield  {author} {\bibinfo {author} {\bibfnamefont {S.}~\bibnamefont
  {Asaad}}, \bibinfo {author} {\bibfnamefont {V.}~\bibnamefont {Mourik}},
  \bibinfo {author} {\bibfnamefont {B.}~\bibnamefont {Joecker}}, \bibinfo
  {author} {\bibfnamefont {M.~A.~I.}\ \bibnamefont {Johnson}}, \bibinfo
  {author} {\bibfnamefont {A.~D.}\ \bibnamefont {Baczewski}}, \bibinfo {author}
  {\bibfnamefont {H.~R.}\ \bibnamefont {Firgau}}, \bibinfo {author}
  {\bibfnamefont {M.~T.}\ \bibnamefont {Mądzik}}, \bibinfo {author}
  {\bibfnamefont {V.}~\bibnamefont {Schmitt}}, \bibinfo {author} {\bibfnamefont
  {J.~J.}\ \bibnamefont {Pla}}, \bibinfo {author} {\bibfnamefont {F.~E.}\
  \bibnamefont {Hudson}}, \bibinfo {author} {\bibfnamefont {K.~M.}\
  \bibnamefont {Itoh}}, \bibinfo {author} {\bibfnamefont {J.~C.}\ \bibnamefont
  {McCallum}}, \bibinfo {author} {\bibfnamefont {A.~S.}\ \bibnamefont
  {Dzurak}}, \bibinfo {author} {\bibfnamefont {A.}~\bibnamefont {Laucht}}, \
  and\ \bibinfo {author} {\bibfnamefont {A.}~\bibnamefont {Morello}},\ }\href
  {\doibase 10.1038/s41586-020-2057-7} {\bibfield  {journal} {\bibinfo
  {journal} {Nature}\ }\textbf {\bibinfo {volume} {579}},\ \bibinfo {pages}
  {205} (\bibinfo {year} {2020})}\BibitemShut {NoStop}%
\bibitem [{\citenamefont {Abadillo-Uriel}\ \emph {et~al.}(2023)\citenamefont
  {Abadillo-Uriel}, \citenamefont {Rodriguez-Mena}, \citenamefont {Martinez},\
  and\ \citenamefont {Niquet}}]{AbadilloUriel2023}%
  \BibitemOpen
  \bibfield  {author} {\bibinfo {author} {\bibfnamefont {J.~C.}\ \bibnamefont
  {Abadillo-Uriel}}, \bibinfo {author} {\bibfnamefont {E.~A.}\ \bibnamefont
  {Rodriguez-Mena}}, \bibinfo {author} {\bibfnamefont {B.}~\bibnamefont
  {Martinez}}, \ and\ \bibinfo {author} {\bibfnamefont {Y.-M.}\ \bibnamefont
  {Niquet}},\ }\href {\doibase 10.1103/physrevlett.131.097002} {\bibfield
  {journal} {\bibinfo  {journal} {Phys. Rev. Lett.}\ }\textbf {\bibinfo
  {volume} {131}},\ \bibinfo {pages} {097002} (\bibinfo {year}
  {2023})}\BibitemShut {NoStop}%
\bibitem [{\citenamefont {Mengistu}\ and\ \citenamefont
  {Garc{\'{\i}}a-Crist{\'{o}}bal}(2016)}]{Mengistu2016}%
  \BibitemOpen
  \bibfield  {author} {\bibinfo {author} {\bibfnamefont {H.~T.}\ \bibnamefont
  {Mengistu}}\ and\ \bibinfo {author} {\bibfnamefont {A.}~\bibnamefont
  {Garc{\'{\i}}a-Crist{\'{o}}bal}},\ }\href {\doibase
  10.1016/j.ijsolstr.2016.08.022} {\bibfield  {journal} {\bibinfo  {journal}
  {Int. J. Solids Struct.}\ }\textbf {\bibinfo {volume} {100-101}},\ \bibinfo
  {pages} {257} (\bibinfo {year} {2016})}\BibitemShut {NoStop}%
\bibitem [{\citenamefont {McSkimin}\ and\ \citenamefont
  {Andreatch}(1964)}]{McSkimin1964}%
  \BibitemOpen
  \bibfield  {author} {\bibinfo {author} {\bibfnamefont {H.~J.}\ \bibnamefont
  {McSkimin}}\ and\ \bibinfo {author} {\bibfnamefont {P.}~\bibnamefont
  {Andreatch}},\ }\href {\doibase 10.1063/1.1702809} {\bibfield  {journal}
  {\bibinfo  {journal} {J. Appl. Phys.}\ }\textbf {\bibinfo {volume} {35}},\
  \bibinfo {pages} {2161} (\bibinfo {year} {1964})}\BibitemShut {NoStop}%
\bibitem [{\citenamefont {McSkimin}\ and\ \citenamefont
  {Andreatch}(1963)}]{McSkimin1963}%
  \BibitemOpen
  \bibfield  {author} {\bibinfo {author} {\bibfnamefont {H.~J.}\ \bibnamefont
  {McSkimin}}\ and\ \bibinfo {author} {\bibfnamefont {P.}~\bibnamefont
  {Andreatch}},\ }\href {\doibase 10.1063/1.1729323} {\bibfield  {journal}
  {\bibinfo  {journal} {J. Appl. Phys.}\ }\textbf {\bibinfo {volume} {34}},\
  \bibinfo {pages} {651} (\bibinfo {year} {1963})}\BibitemShut {NoStop}%
\end{thebibliography}%

\clearpage
\newpage
\mbox{~}

\onecolumngrid

\setcounter{equation}{0}
\setcounter{figure}{0}
\setcounter{table}{0}
\setcounter{section}{0}

\renewcommand{\theequation}{S\arabic{equation}}
\renewcommand{\thefigure}{S\arabic{figure}}
\renewcommand{\thesection}{S\arabic{section}}

\begin{center}
	\textbf{\large Supplemental Material to  \\
		``Valley-Free Silicon Fins Caused by Shear Strain''}\\[.2cm]
	Christoph Adelsberger, Stefano Bosco, Jelena Klinovaja, and Daniel Loss\\[.1cm]
	{\itshape Department of Physics, University of Basel, Klingelbergstrasse 82, 4056 Basel, Switzerland}\\
\end{center}

	\title{Supplemental Material of \\
	Valley-free silicon fins by shear strain
	}

\author{Christoph Adelsberger}
\author{Stefano Bosco}
\author{Jelena Klinovaja}
\author{Daniel Loss}
\affiliation{Department of Physics, University of Basel, Klingelbergstrasse 82, CH-4056 Basel, Switzerland}


\section*{Abstract}

In the Supplemental Material we provide more details on the simulation of the strain tensor in our devices via the finite element method and continuum elasticity theory. We show results on the strain tensor components that cause the localization of the electron wave function in certain areas of the device cross section. We also provide a more detailed analysis of the electric field dependence of the valley splitting which demonstrates that a precisely aligned electric field is not required. Furthermore, we analyze the effect of atomistic disorder at the interface between the Si and the SiGe alloy and demonstrate that in contrast to planar heterostructures, disorder has no effect on our fins. Also a modification of the cross section of the triangular fin do not spoil the large valley splitting predicted in the main text. Additionally, we present simulations of fins with other cross sections, different from the one showed in the main text. In these systems, we find a valley splitting similar to the one for the setups discussed in the main text, thus corroborating our claim that fine-tuning of the cross section shape is not required to reach large values of the valley splitting.

\section{Pseudomorphic strain}
\label{sm:strain}

We consider strained Si/SiGe devices. Because of the mismatch of lattice constant between the materials,  a force develops at their interfaces, resulting in a displacement field $\vect{u}(\vect{r})$ for the atoms. Consequently, in equilibrium the lattice constants of the two materials match at the interface; this is referred to as the pseudomorphic condition.  

In linear elasticity theory~\citeS{Kosevich1986sm} the change of lengths in a deformed body is given by the strain tensor
\begin{align}
	\varepsilon_{ij} = \frac{1}{2} \left(\frac{\partial u_i}{\partial x_j} + \frac{\partial u_j}{\partial x_i}\right) \ .
\end{align}
The strain tensor elements $\varepsilon_{ij}$ are related to the stress  tensor elements $\sigma_{ij}$ by the material-dependent elastic stiffness tensor $C_{ijkl}$:
\begin{align}
	\sigma_{ij} = C_{ijkl} \varepsilon_{kl} \ ,
\end{align}
implying Einstein summation.
In the presence of a force $f_j$ that deforms the body, the stress tensor satisfies the equilibrium condition
\begin{align}
	\frac{\partial \sigma_{ij}}{\partial x_i} = - f_j \ .
\end{align}
Therefore, we calculate $\varepsilon_{ij}$ in the presence of a given force $f_j^0$ by solving the partial differential equation
\begin{align}
	\frac{\partial [  C_{ijkl}\varepsilon_{kl}]}{\partial x_i} = - f^0_j \ . 
\end{align}
Our system comprising two materials with different lattice constants $a_i(\vect{r})$ and two different elastic stiffness tensors $C_{ijkl}(\vect{r})$ can be simulated by linear elasticity theory by introducing the equivalent body force~\citeS{Mengistu2016sm}
\begin{align}
	f_i^0 = \frac{\partial}{\partial x_j} \left[C_{ijkl}(\vect{r}) \varepsilon_{kl}^0(\vect{r})\right] \ , \label{eqn:PDEstrain}
\end{align}
where the strain from the lattice constant mismatch is given by
\begin{align}
	\varepsilon_{kl}^0(\vect{r}) \delta_{kl} = \frac{a_k^{(\mathrm{ref})} - a_k(\vect{r})}{a_k(\vect{r})} \ .
\end{align}
Here, $a_k^{(\mathrm{ref})}$ is a reference lattice constant that can be chosen to be the lattice constant of one of the two materials without loss of generality. The elastic stiffness tensor in Eq.~\eqref{eqn:PDEstrain}   for crystals with cubic symmetry can be written as~\citeS{Mengistu2016sm,Kloeffel2014sm}
\begin{align}
	\underline{\vect{C}} = \begin{pmatrix}
		C_{11} & C_{12} & C_{12} & 0 & 0 & 0\\
		C_{12} & C_{11} & C_{12} & 0 & 0 & 0\\
		C_{12} & C_{12} & C_{11} & 0 & 0 & 0\\
		0 & 0 & 0 &  C_{44}   & 0 & 0 \\
		0 & 0 & 0 & 0 & C_{44} & 0\\
		0 & 0 & 0 & 0 & 0 & C_{44}
	\end{pmatrix} \ ,
\end{align}
where we use the Voigt notation. For our simulations the tensor is rotated such that it agrees with the $[110]$ growth direction of the fin considered in our system. For the finite-element method (FEM) simulations we assume free boundary conditions at the outer boundaries of the devices.

\begin{figure}[]
	\includegraphics[width=\columnwidth]{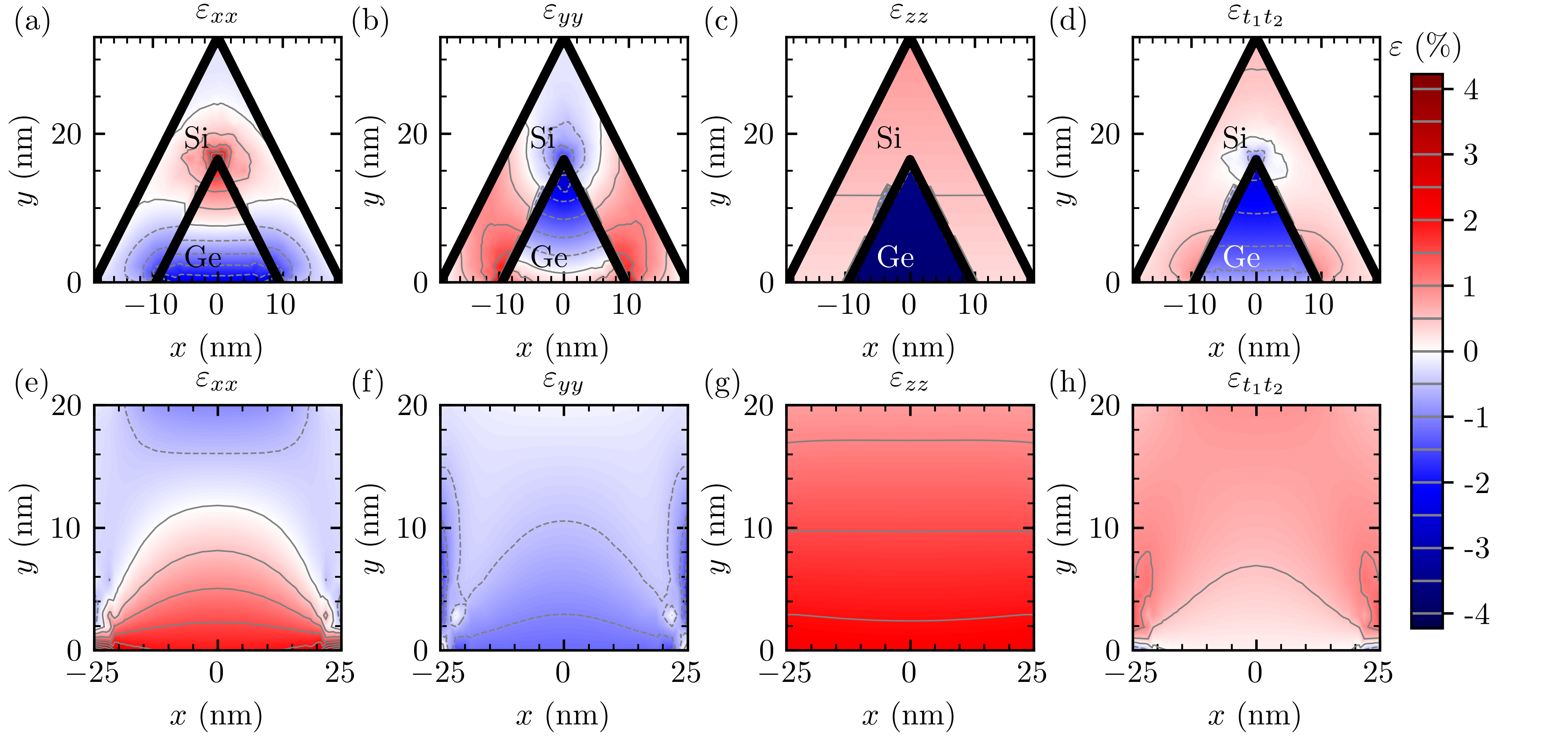}
	\caption[]{Strain tensor components $\varepsilon_{xx}$, $\varepsilon_{yy}$, $\varepsilon_{zz}$, and $\varepsilon_{t_1t_2} = (\varepsilon_{zz}-\varepsilon_{xx})/2$ simulated with the FEM in the devices analyzed in the main text. Here we consider pure Ge; in Si$_x$Ge$_{1-x}$ alloys $\underline{\vect{\varepsilon}}_{\mathrm{Si/Si}_x\mathrm{Ge}_{1-x}}=(1-x)\underline{\vect{\varepsilon}}_\mathrm{Si/Ge}$ is rescaled linearly by the concentration of Ge. (a-d) Triangular fin: $\varepsilon_{yy}$ has a minimum above the tip of the Ge fin localizing the electron wave function there. (e-h) Rectangular fin: $\varepsilon_{yy}$ is minimal at the left and right sides of the Si slab. Thus, the wave function is localized at these sides. Note that the deviation from perfect symmetry in the plots comes from numerical inaccuracies. The effects of the shear strain component $\varepsilon_{t_1t_2} = (\varepsilon_{zz}-\varepsilon_{xx})/2$ are discussed in the main text. We used $C_{11}^\text{Si} = \SI{168}{\giga\pascal}$, $C_{12}^\text{Si} = \SI{65.0}{\giga\pascal}$, $C_{44}^\text{Si} = \SI{80.4}{\giga\pascal}$~\citeS{McSkimin1964sm}, $C_{11}^\text{Ge} = \SI{131}{\giga\pascal}$, $C_{12}^\text{Ge} = \SI{49.2}{\giga\pascal}$, and  $C_{44}^\text{Ge} = \SI{68.2}{\giga\pascal}$~\citeS{McSkimin1963sm}. \label{fig:uaStrain} }
\end{figure}

To calculate the effect of strain due to lattice mismatch, we simulate the strain tensor elements $\varepsilon_{ij}$ in our devices by  solving the differential equation in Eq.~\eqref{eqn:PDEstrain} numerically. In particular, we use the FEM implemented in COMSOL Multiphysics\,\textsuperscript{\tiny\textregistered}~\citeS{COMSOLsm}. For the lifting of the valley degeneracy, we pay particular attention to the shear strain component which is the main source of the large valley gap $\Delta$ in our fins. By rotating the coordinate system such that the $z$ axis is aligned with the $[110]$ growth direction the strain tensor becomes
\begin{align}
		\underline{\vect{\varepsilon}} = \begin{pmatrix}
			\varepsilon_{t_1t_1} & \varepsilon_{t_1t_2} & \varepsilon_{t_1l} \\
			\varepsilon_{t_1t_2} & \varepsilon_{t_2t_2} & \varepsilon_{t_2l} \\
			\varepsilon_{lt_1} & \varepsilon_{lt_2} & \varepsilon_{ll} 
		\end{pmatrix} = 
		\begin{pmatrix}
			\frac{\varepsilon_{xx}-2\varepsilon_{xy}+\varepsilon_{zz}}{2} & \frac{\varepsilon_{zz}-\varepsilon_{xx}}{2} & \frac{-\varepsilon_{xy}+ \varepsilon_{yz}}{\sqrt{2}}\\
			 \frac{\varepsilon_{zz}-\varepsilon_{xx}}{2} & \frac{\varepsilon_{xx}+2\varepsilon_{xy}+\varepsilon_{zz}}{2} & \frac{\varepsilon_{xy}+ \varepsilon_{yz}}{\sqrt{2}}\\
			 \frac{-\varepsilon_{xy}+ \varepsilon_{yz}}{\sqrt{2}} & \frac{\varepsilon_{xy}+ \varepsilon_{yz}}{\sqrt{2}}& \varepsilon_{yy}
		\end{pmatrix}.
\end{align}
For the lowest-energy valleys the relevant shear strain component is $\varepsilon_{t_1t_2} = (\varepsilon_{zz}-\varepsilon_{xx})/2$ and the only uniaxial strain component entering the valley Hamiltonian is $\varepsilon_{ll} = \varepsilon_{yy}$.

In the main text, we show the simulated strain tensor component $\varepsilon_{t_1t_2}$ in the cross sections of the two devices analyzed. The finite value of shear strain above the tip of the inner Ge fin as well as at the sides of the Si slab explain the large values for $\Delta$ shown in the main text. In Fig.~\ref{fig:uaStrain}, we show the strain tensor components $\varepsilon_{xx}$, $\varepsilon_{yy}$, $\varepsilon_{zz}$, and $\varepsilon_{t_1t_2}$ for both devices. In the triangular fin the uniaxial component $\varepsilon_{yy}$ is negative at the region of interest above the tip of the Ge fin [see Fig.~\ref{fig:uaStrain}(b)]. Thus, the electron wave function is localized above the tip of the Ge fin. In the Si slab $\varepsilon_{yy}$ has the lowest negative values at the sides, localizing the wave function there [see Fig.~\ref{fig:uaStrain}(d)]. The $\varepsilon_{xx}$ and $\varepsilon_{zz}$ components are irrelevant for the localization of the electron and only cause the valley splitting. 

\section{Electric field dependence of the valley splitting}
\label{sm:efield}
\begin{figure}[]
	\includegraphics[width=0.8\textwidth]{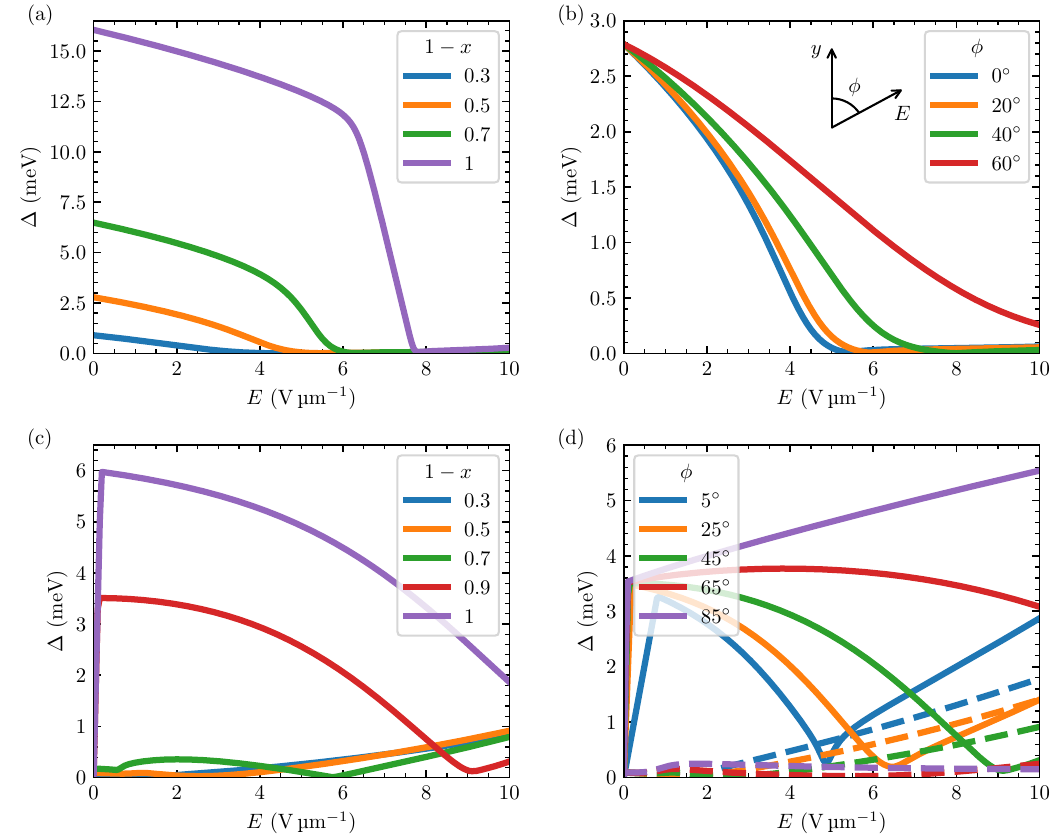}
	\caption{
		Valley splitting as a function of the electric field for (a) the triangular fin for different concentrations of Ge, (b) the triangular fin for misaligned electric field, (c) the rectangular fin for different concentrations of Ge, and (d) the rectangular fin for misaligned electric field. In (b) $1-x = 0.5$ and in (d) $1-x= 0.9$ for the solid lines and $1-x=0.5$ for the dashed lines. The inset in (b) defines the angle $\phi$ for both panels (b) and (d). Here, $\phi=0$ corresponds to the electric field direction used in the main text for triangular fin and $\phi=45^\circ$ to the direction used for the rectangular fin. All other parameters are the same as in the main text. \label{fig:Efield}}
\end{figure}

To get a clearer picture of the electric field dependence of the valley splitting, we explore how the valley splitting depends on Ge concentration. Moreover, we analyze the effects of a misaligned electric field from directions chosen in the main text. The results are shown in Fig.~\ref{fig:Efield}. 

In Fig.~\ref{fig:Efield}(a), we demonstrate that the qualitative electric field dependence of the valley splitting in the triangular fin does not depend on the concentration of Ge. For weak electric field the valley splitting decreases slightly with increasing $E$, followed by a drop of $\Delta$ to nearly zero. The value of $E$ at which the valley splitting nearly disappears  is larger for larger concentration of Ge. In Fig.~\ref{fig:Efield}(b), we present the valley splitting for electric fields applied at different angles $\phi$. The larger the deviation of the angle from the one chosen in the main text is the larger must the electric field be to push the electron wave function to the tip of the Si shell. Thus, for $\phi\neq 0$ the valley splitting drops to zero at larger electric field than for $\phi=0$.

The valley splitting in the rectangular fin increases abruptly from $\Delta = 0$ at $E=0$ to a large value at finite electric field if the Ge concentration is large ($1-x \gtrsim 0.8$) as shown in Fig.~\ref{fig:Efield}(c). Later, with increasing electric field, $\Delta$ decreases. For smaller Ge concentration, the valley splitting stays at a low value for weak electric field. Depending on the Ge concentration, $\Delta$ increases linearly starting from some electric field value and reaches $\Delta \approx \SI{1}{\milli\electronvolt}$ at $E = \SI{10}{\volt\per\micro\meter}$. In Fig.~\ref{fig:Efield}(d), we plot the valley splitting for electric fields that are misaligned from the direction toward the upper right corner of the Si slab. If $1-x = 0.9$ (solid lines), a more horizontal electric field direction is beneficial for large valley splitting. In the case of $1-x = 0.5$ (dashed lines), the largest values for $\Delta$ are obtained for almost vertical electric field. 

The analysis of the electric field direction shows that it is not required to precisely align the electric field as in the main text to obtain large valley splitting. In fact, a misalignment does  not change the qualitative electric field dependence and can even be beneficial for the size of the valley splitting.  

\section{Atomistic disorder at the $\text{Si}$/$\text{SiGe}$ interfaces and angle disorder}
\label{sm:steps}

\begin{figure}[]
	\includegraphics[width=0.8\textwidth]{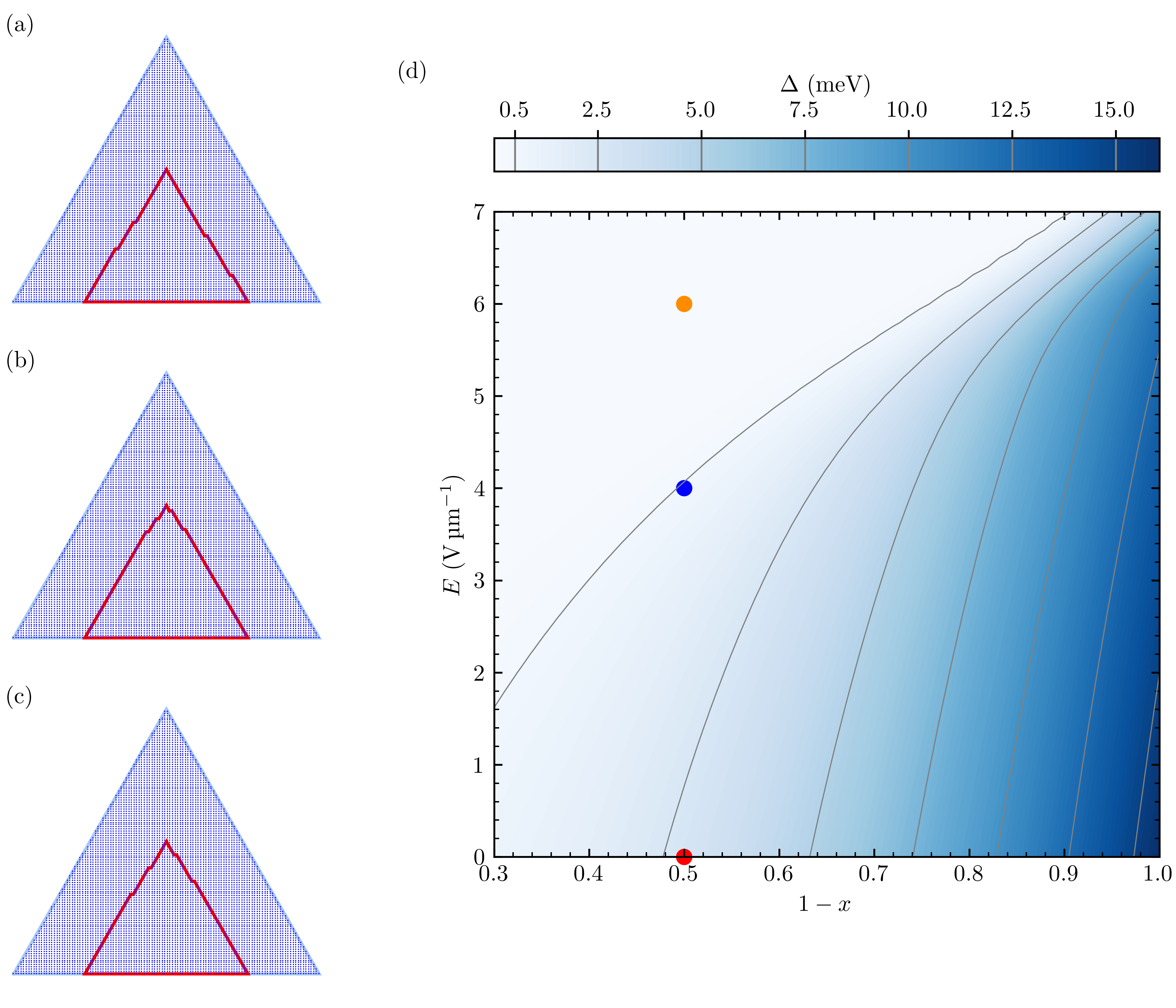}
	\caption{
Atomistic disorder in our Si/SiGe fins.	
	(a-c) Plot of the cross section of the triangular fin including three different disorder configurations at the Si/SiGe interfaces. The blue dots depict the discrete lattice points used for the numerical diagonalization of the Hamiltonian in the main text and the red lines mark the interface between the inner SiGe fin and the outer Si shell including atomic size steps at random positions. (d) Valley splitting $\Delta$ plotted against the Ge concentration $1-x$ and electric field $E$ (pointing along $y$ direction) in the triangular fin. The result is for all three configurations exactly the same as for the triangular fin without interface steps. This is not surprising since the electron wave function is located far away from the interface steps. \label{fig:steps} }
\end{figure}

In the main text, we analyze triangular and rectangular  Si/SiGe fin structures and we argue that, in contrast to planar heterostructures, in our fins atomistic disorder does not affect the valley splitting $\Delta$.
Here we show the results of simulations including atomic steps at the interfaces of the two materials, as shown in Fig.~\ref{fig:steps}(a-c). This kind of disorder is known to strongly affect the planar structures~\citeS{Friesen2007sm,Chutia2008sm,Saraiva2009sm,Hosseinkhani2020sm,Hosseinkhani2021sm,Lima2023sm,Wuetz2022sm} however we find that, as expected, it does not affect our fins.  

In particular, we diagonalize the Hamiltonian in the main text discretized on the lattice shown in Fig.~\ref{fig:steps}(a-c) with values of strain simulated in the triangular fin device without interface steps. The results are shown in  Fig.~\ref{fig:steps}(d) and comparing with the results in the main text we observe perfect agreement, thus corroborating our claim.  This result can be understood because in contrast to planar heterostructures, the electron is localized away from the interface steps as discussed in the main text. Moreover, the large valley splitting remains upon smoothening the interface between the Ge fin and the Si shell as we will discuss in the following section. 

\begin{figure}[]
	\includegraphics[width=0.8\textwidth]{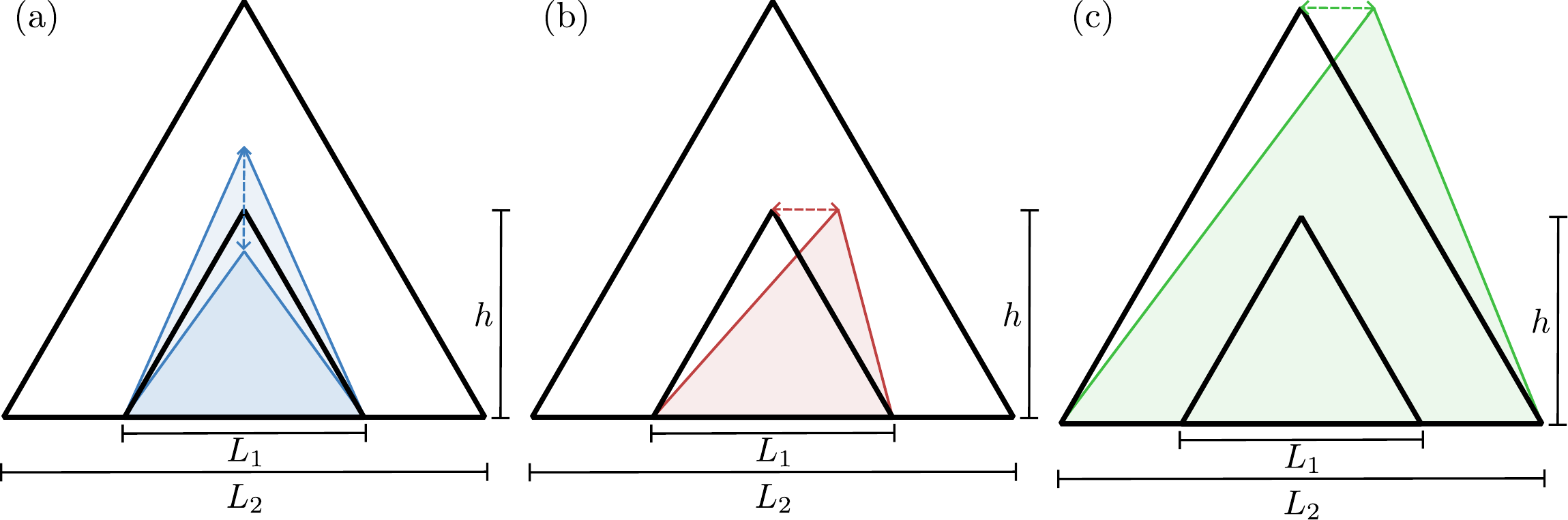}
	\caption{
		Modification of the triangular cross section. (a) Variation of the height $h$ of the inner triangle. The height of the inner triangle in the main text is $h_0 = \sqrt{3} L_1/2$. (b) Shift of the tip of the inner triangle ($x_\text{tip1}$) to the right. (c) Shift of the tip of the outer triangle ($x_\text{tip2}$) to the right.\label{fig:finAngles} }
		\end{figure}

Another type of disorder is different angles of the inner and outer triangles of the fin cross section or laterally shifted tips of the triangles. In the following we present results for the valley splitting of fins with modified triangular cross section. The modifications we consider are defined in Fig.~\ref{fig:finAngles}. In particular, we analyze different heights $h$ of the inner triangles [see Fig.~\ref{fig:finAngles}(a)], a lateral shift of the tip of the inner triangle [see Fig.~\ref{fig:finAngles}(b)], and a shift of the tip of the outer triangle to the right [see Fig.~\ref{fig:finAngles}(c)]. We denote the height of the inner triangle used in the main text by $h_0 = \sqrt{3} L_1/2$ and the $x$ coordinate of the position of the inner and outer tip by $x_\text{tip1}$ and $x_\text{tip2}$, respectively. The positions of the tips of the cross section in the main text are $x_\text{tip1}=0$ and $x_\text{tip2}=0$.

\begin{figure}[]
	\includegraphics[width=0.9\textwidth]{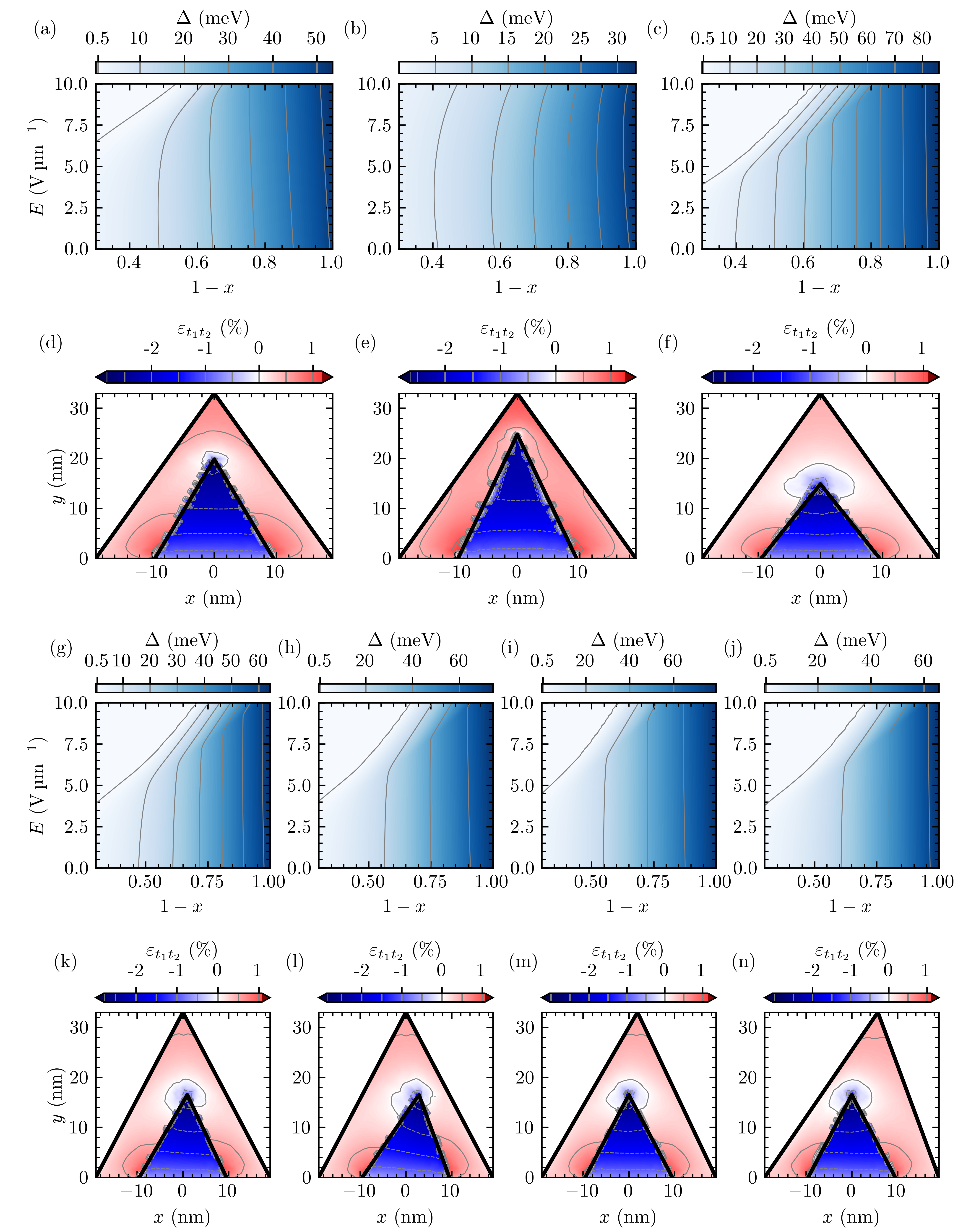}
	\caption{
		Valley splitting $\Delta$ and shear strain $\varepsilon_{t_1t_2}$ for the modified triangular cross section as introduced in Fig.~\ref{fig:finAngles}. (a-f) correspond to modifications of the height as shown in Fig.~\ref{fig:finAngles}(a), (g, h, k, l) to shifts of $x_\text{tip1}$ as in Fig.~\ref{fig:finAngles}(b), and (i, j, m, n) to shifts of $x_\text{tip2}$ as in Fig.~\ref{fig:finAngles}(c). (a,d) $h= 1.2 h_0$. (b, e) $h=1.5 h_0$. (c,f) $h = 0.9 h_0$. (g, k) $x_\text{tip1} = 0.1 L_1$. (h, l) $x_\text{tip1} = 0.3 L_1$. (i, m) $x_\text{tip2} = 0.1 L_2$. (j, n) $x_\text{tip2} = 0.3 L_2$. Here, $h_0$ is the height of the equilateral inner triangle considered in the main text and $x_\text{tip1}$ and $x_\text{tip2}$ are the $x$ coordinates of the positions of the tips of the inner and outer triangles, respectively. All other parameters are the same as in the main text.
		\label{fig:VS_angleDis} }
\end{figure}

The valley splitting $\Delta$ and the shear strain component $\varepsilon_{t_1t_2}$ are presented in Fig.~\ref{fig:VS_angleDis}. In Fig.~\ref{fig:VS_angleDis}(a-f), we demonstrate the results for different heights $h$ of the inner triangle;  in Fig.~\ref{fig:VS_angleDis}(g,h,k,l), for the tip of the inner triangle  being shifted to the right; in Fig.~\ref{fig:VS_angleDis}(i,j,m,n), for the tip of the outer triangle being shifted to the right. By symmetry a shift to the left has the same effect. In general, the qualitative dependence of the valley splitting on the electric field and the concentration of Ge is the same as for the triangular fin analyzed in the main text. Interestingly, breaking the symmetry in any of the considered ways results in an enhancement of the maximal absolute value of shear strain above the tip of the inner triangle, and thus a larger valley splitting of up to $\Delta = \SI{86}{\milli\electronvolt}$. As in the main text, also in the here considered situations the uniaxial strain causes a localization of the wave function between the tips of inner and outer triangle which can be controlled by the external electric field. Except for the case presented in Fig.~\ref{fig:VS_angleDis}(b), the valley splitting drops below $\Delta=\SI{0.5}{\milli\electronvolt}$ at strong electric field and low Ge concentration. When the height of the inner triangle is $1.5$ times the height of the equilateral triangle considered in the main text, i.e. $h = 1.5 h_0$, the valley splitting is large even at $1-x = 0.3$ and strong electric field simultaneously [see Fig.~\ref{fig:VS_angleDis}(b)]. This analysis demonstrates that our results in the main text do not depend on a certain symmetry and imperfections from fabrication do not spoil the effect of the lifting of the valley degeneracy.
		
\section{Semicylindrical $\text{Si}$/$\text{SiGe}$ device}
\label{sm:geometries}

\begin{figure}[]
	\includegraphics[width=\textwidth]{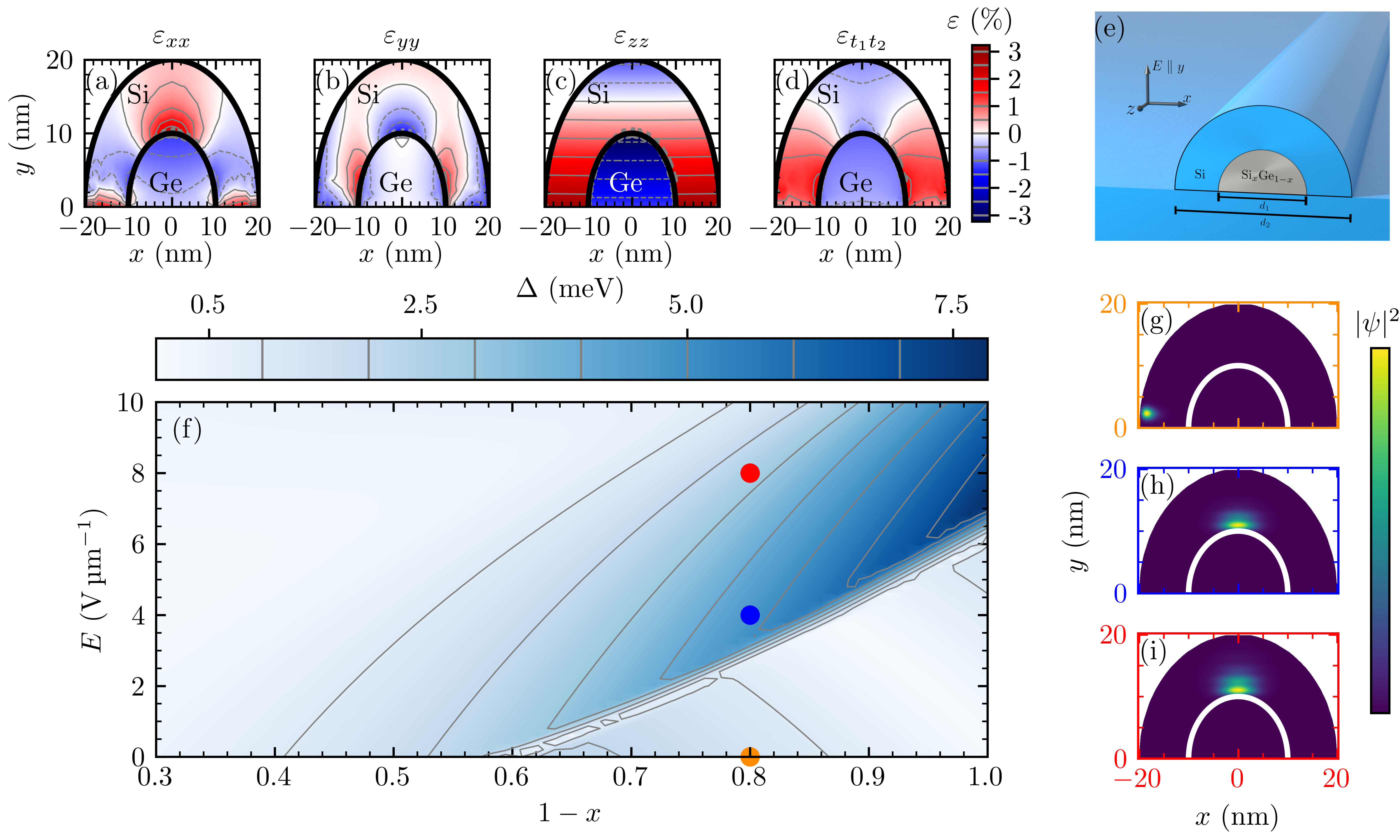}
	\caption{Simulation of a semicircular Si/SiGe fin.
		(a-d) Strain tensor components simulated with the FEM. Large shear strain appears at the bottom of the Si shell close to the substrate and at the top of the cross section. The material parameters are the same as in Fig.~\ref{fig:uaStrain}. (e) Sketch of the semicircular fin with inner diameter $d_1$ and outer diameter $d_2$. The electric field $E$ points along the $y$ direction.
		(f) Valley splitting $\Delta$ against the Ge concentration $1-x$  and the electric field $E$.
		(g-i) Probability densities $\abs{\psi}^2$ of the electrons at $1-x=0.8$ and (g) $E=0$, (h) $E=\SI{4}{\volt\per\micro\meter}$, and (i) $E=\SI{8}{\volt\per\micro\meter}$. These points are marked in (f). In analogy to the triangular fin analyzed in the main text, we obtain $\Delta > \SI{0.5}{\milli\electronvolt}$ for a wide experimentally-relevant range of parameters. This result proves that the sharp tip of the triangular fin is not required to enable a large valley splitting. (g) Interestingly, for a large Ge concentration ($1-x\gtrsim 0.6$) and strong strain, the electron is localized at the bottom of the device, close to the surface. This effect is caused by the uniaxial strain $\varepsilon_{yy}$ as shown in (b). (h, i) An electric field $E$ pushes the electron to the top of the device. Note that for smaller Ge concentration $1-x$, the electron is localized at the top even at $E=0$; this also occurs for triangular fins. We used $d_1= \SI{20}{\nano\meter}$, $d_2 = \SI{40}{\nano\meter}$, $a_x = \SI{0.34}{\nano\meter}$, and $a_y = \SI{0.17}{\nano\meter}$.  We chose the size of the system such that the Si and Si/Ge parts of the cross section cover the same areas as their counterparts in the triangular fin device in the main text. 
		\label{fig:NW_WFhalf}}
\end{figure}

In the main text we argue that Si/SiGe fins with different shapes have a similar valley splitting $\Delta$ and thus our results are largely independent of the fin shape. 
We support these claims here by simulating a semicircular fin, see Fig.~\ref{fig:NW_WFhalf}. This fin comprises a Ge semicircle on top of a Si substrate with a Si outer shell, where the electron is localized. This fin resembles the triangular fin discussed in the main text if the triangle has a round tip. In practice, a device without sharp tips is more realistic to be grown since the large strain values in the triangular device might cause instabilities of the structure or dislocations.

Following the same procedure as before, we simulate first the strain tensor. In Figs.~\ref{fig:NW_WFhalf}(a-d) we show the uniaxial and shear strain components. We observe a qualitatively similar trend as in the triangular fin, see main text. 

The local shear strain $\varepsilon_{t_1t_2}$ explains the trend of the valley splitting $\Delta$ simulated in Fig.~\ref{fig:NW_WFhalf}(f). At small concentration of Ge, with $1-x\lesssim 0.6$, the valley splitting decreases with increasing $E$.  At higher values of $1-x$, where strain is larger, we  observe a more interesting dependence of $\Delta$ on $E$. This dependence can be understood by looking at the localization of the electron at $1-x=0.8$ for different values of $E$, see Fig.~\ref{fig:NW_WFhalf}(g-h). At $E=0$, the wave function is localized at one of the lower corners of the fin, where $\varepsilon_{t_1t_2}$ is finite, thus resulting in a significant value of $\Delta$. 
As the electric field increases, the electron is pushed toward the tip of the fin, and in particular at  $E=\SI{4}{\volt\per\micro\meter}$, the electron is localized at the bottom of the tip, where $\varepsilon_{t_1t_2}$ is the largest, and thus resulting in a large value of $\Delta$. As $E$ is further increased the electron moves toward the topmost part of the fin, and  $\Delta$ decreases due to the weaker shear strain [see Fig.~\ref{fig:NW_WFhalf}(d)]. This trend is consistent with the simulation of the triangular fin discussed in the main text.
Consequently, we conclude that large values of $\Delta$ in Si/SiGe fins can be reached independently of the sharpness of the tip of the fin.

\bibliographystyleS{apsrev4-1}
\bibliographyS{LiteratureCQW}

\end{document}